\documentclass[fleqn,usenatbib,useAMS]{mnras}
 \usepackage{graphicx}
 \usepackage{amsmath}
 \usepackage{amssymb}
 \usepackage[T1]{fontenc}
 \usepackage{ae,aecompl}
 \usepackage{txfonts}
 \usepackage{enumerate}

 \title[YORP break-up events in Earth co-orbital space]
       {Using Mars co-orbitals to estimate the importance of rotation-induced YORP 
        break-up events in Earth co-orbital space} 

 \author[C. de la Fuente Marcos and R. de la Fuente Marcos]
        {C.~de~la~Fuente~Marcos$^{1}$\thanks{E-mail: nbplanet@ucm.es}
         and
         R. de la Fuente Marcos$^{2}$ \\
         $^1$Universidad Complutense de Madrid,
             Ciudad Universitaria, E-28040 Madrid, Spain \\
         $^2$AEGORA Research Group,
             Facultad de Ciencias Matem\'aticas,
             Universidad Complutense de Madrid,
             Ciudad Universitaria, E-28040 Madrid, Spain}
 \date{Accepted 2021 January 4.
       Received 2020 December 20;
       in original form 2020 August 24}
 \pubyear{2021}
 
\begin{document}
  \label{firstpage}
  \pagerange{\pageref{firstpage}--\pageref{lastpage}}
  \maketitle

  \begin{abstract}
     Both Earth and Mars host populations of co-orbital minor bodies. A 
     large number of present-day Mars co-orbitals is probably associated 
     with the fission of the parent body of Mars Trojan 5261~Eureka 
     (1990~MB) during a rotation-induced 
     Yarkovsky--O'Keefe--Radzievskii--Paddack (YORP) break-up event. Here, 
     we use the statistical distributions of the Tisserand parameter and 
     the relative mean longitude of Mars co-orbitals with eccentricity 
     below 0.2 to estimate the importance of rotation-induced YORP break-up 
     events in Martian co-orbital space. Machine-learning techniques 
     ($k$-means++ and agglomerative hierarchical clustering algorithms) are 
     applied to assess our findings. Our statistical analysis identified 
     three new Mars Trojans: 2009~SE, 2018~EC$_{4}$ and 2018~FC$_{4}$. Two 
     of them, 2018~EC$_{4}$ and 2018~FC$_{4}$, are probably linked to 
     Eureka but we argue that 2009~SE may have been captured, so it is not 
     related to Eureka. We also suggest that 2020~VT$_{1}$, a recent 
     discovery, is a transient Martian co-orbital of the horseshoe type. 
     When applied to Earth co-orbital candidates with eccentricity below 
     0.2, our approach led us to identify some clustering, perhaps linked 
     to fission events. The cluster with most members could be associated 
     with Earth quasi-satellite 469219~Kamo`oalewa (2016~HO$_{3}$) that is 
     a fast rotator. Our statistical analysis identified two new Earth 
     co-orbitals: 2020~PN$_{1}$, which follows a horseshoe path, and 
     2020~PP$_{1}$, a quasi-satellite that is dynamically similar to 
     Kamo`oalewa. For both Mars and Earth co-orbitals, we found pairs of 
     objects whose values of the Tisserand parameter differ by very small 
     amounts, perhaps hinting at recent disruption events. Clustering 
     algorithms and numerical simulations both suggest that 2020~KZ$_{2}$ 
     and Kamo`oalewa could be related. 
  \end{abstract}


  \section{Introduction}
     Asteroids can split into two or more components due to the effect of solar-radiation-driven forces on their spin rate. The 
     Yarkovsky--O'Keefe--Radzievskii--Paddack (YORP) mechanism (see e.g. \citealt{2006AREPS..34..157B}) can secularly increase the rotation 
     rate of asteroids and make them reach the critical fission frequency, triggering break-up events \citep{2008Natur.454..188W} that may 
     lead to the formation of unbound asteroid pairs \citep{2008AJ....136..280V,2010Natur.466.1085P,2011Icar..214..161J,2018Icar..304..183S,
     2019Icar..333..429P} and clusters \citep{2018Icar..304..110P,2020MNRAS.493.2556C,2020NatAs...4...83C,2020Icar..33813554F}.

     Mars hosts a group of Trojan asteroids that may have formed during a rotation-induced YORP break-up event \citep{2020Icar..33513370C}. 
     The so-called Eureka cluster or family -- after its largest member 5261~Eureka (1990~MB), which is a 2-km binary asteroid -- includes 
     at least nine members \citep{2020Icar..33513370C}, see Table~\ref{eureka}. Spectroscopic analysis has shown that some members of this 
     cluster exhibit a distinctive olivine-dominated composition that lends further support to the common origin scenario 
     \citep{2017MNRAS.466..489B}. On the other hand, most known Mars Trojans are believed to be primordial objects that may have remained in 
     their current dynamical state since Mars was formed \citep{2013Icar..224..144C,2013MNRAS.432L..31D,2017Icar..293..243C}. The stability 
     of Mars Trojans in general and of relevant known objects in particular has been studied by e.g. \citet{1994AJ....107.1879M},
     \citet{2005Icar..175..397S}, and \citet{2012CeMDA.113...23S}. Different strategies to find more Mars Trojans have been discussed by
     e.g. \citet{2012MNRAS.424..372T,2014MNRAS.437.4019T}.

     Although so far only one Earth Trojan has been discovered, 2010~TK$_{7}$ \citep{2011Natur.475..481C}, and it is not long-term stable 
     due to its chaotic orbit (see e.g. \citealt{2012A&A...541A.127D,2019A&A...622A..97Z}), there are dozens of known near-Earth asteroids 
     (NEAs) that move co-orbital to our planet. Strategies to discover additional Earth co-orbitals have been discussed by e.g. 
     \citet{2012MNRAS.420L..28T}. On the other hand, a number of NEA pairs are suspected to come from YORP-induced rotational fissions 
     \citep{2019MNRAS.483L..37D,2019Icar..333..165M} and Earth co-orbitals may also be experiencing these processes (see e.g. 
     \citealt{2018MNRAS.473.3434D}). Here, we use the statistical distributions of the Tisserand parameter and the relative mean longitude 
     of Mars co-orbitals to estimate the importance of YORP break-up events in Martian co-orbital space, then we apply the same method to 
     study a group of objects co-orbital to Earth. This paper is organized as follows. In Section~2, we discuss data and methods. The 
     analysis of Mars co-orbitals is presented in Section~3. Section~4 focuses on Earth co-orbitals. Machine-learning techniques are applied 
     in Section~5 to evaluate the significance of our findings. Our results are discussed in Section~6 and our conclusions are summarized in 
     Section~7.
%
%
     \begin{table}
        \centering
        \fontsize{8}{11pt}\selectfont
        \tabcolsep 0.15truecm
        \caption{Tisserand parameter with respect to Mars, $T_{\rm Mars}$, and relative mean longitude, $\lambda_{\rm r}$, for Eureka and 
                 confirmed family members with ${\Delta}T_{\rm Mars}$ relative to Eureka included. Data are referred to epoch 2459000.5, 
                 31-May-2020 00:00:00.0 TDB (J2000.0 ecliptic and equinox). Source: JPL's SBDB.  
               }
        \begin{tabular}{rccc}
           \hline
              Object                  & $T_{\rm Mars}$ & $\lambda_{\rm r}$ (\degr) & ${\Delta}T_{\rm Mars}$ \\
           \hline
              5261 Eureka (1990 MB)   & 2.872035959    & $-$57.98691               &   --                   \\   
              311999 (2007 NS$_{2}$)  & 2.892520748    & $-$55.03168               & 0.020484788            \\
              385250 (2001 DH$_{47}$) & 2.820265267    & $-$56.66416               & 0.051770692            \\
                      2011 SC$_{191}$ & 2.892095964    & $-$68.63854               & 0.020060004            \\
                      2011 SL$_{25}$  & 2.848673098    & $-$65.67817               & 0.023362862            \\
                      2011 SP$_{189}$ & 2.879069290    & $-$68.85747               & 0.007033331            \\
                      2011 UB$_{256}$ & 2.818159354    & $-$55.40139               & 0.053876606            \\
                      2011 UN$_{63}$  & 2.871098466    & $-$53.84016               & 0.000937494            \\
                      2016 CP$_{31}$  & 2.836038634    & $-$56.61722               & 0.035997325            \\
           \hline
        \end{tabular}
        \label{eureka}
     \end{table}
%
%

  \section{Data and methods}
     Here, we use publicly available data from Jet Propulsion Laboratory's (JPL) Small-Body Database 
     (SBDB)\footnote{\href{https://ssd.jpl.nasa.gov/sbdb.cgi}{https://ssd.jpl.nasa.gov/sbdb.cgi}} and HORIZONS on-line solar system data and 
     ephemeris computation service,\footnote{\href{https://ssd.jpl.nasa.gov/?horizons}{https://ssd.jpl.nasa.gov/?horizons}} both provided by 
     the Solar System Dynamics Group \citep{2011jsrs.conf...87G,2015IAUGA..2256293G}. The data are referred to epoch 2459000.5 Barycentric 
     Dynamical Time (TDB) and any computed parameters are referred to the same epoch, that is also the origin of time in the calculations. 
     Data as of 2020 December 20 have been retrieved from JPL's SBDB and HORIZONS using tools provided by the \textsc{Python} package 
     \textsc{Astroquery} \citep{2019AJ....157...98G}. Here, we focus on two groups of asteroids, co-orbitals of Mars and those of Earth with 
     eccentricity $e<0.2$, and on the Tisserand parameter \citep{Tisserand1896,1999ssd..book.....M}. In the following, we summarize some 
     concepts that will be used later to shape our analysis.

     Co-orbital minor bodies are trapped in the 1:1 mean-motion resonance with a planet; therefore, they go around the Sun in almost exactly 
     one planetary orbital period. Although their orbits resemble that of their host planet in terms of orbital period, they could be very 
     eccentric and/or inclined \citep{2002Icar..160....1M}. The key parameter to classify co-orbitals is the value of the relative mean 
     longitude, which is the difference between the mean longitude of the minor body (asteroid or comet) and that of its host planet. The 
     mean longitude is given by $\lambda=\Omega+\omega+M$, where $\Omega$ is the longitude of the ascending node, $\omega$ is the argument 
     of perihelion, and $M$ is the mean anomaly. Therefore, the critical angle is $\lambda_{\rm r}=\lambda-\lambda_{\rm P}$. If the value of 
     $\lambda_{\rm r}$ oscillates over time about 0{\degr}, the object is a quasi-satellite (or retrograde satellite, although it is not 
     gravitationally bound) to the planet (see e.g. \citealt{2006MNRAS.369...15M,2014CeMDA.120..131S}), if it librates around 60{\degr}, the 
     object is called an L$_4$ Trojan and leads the planet in its orbit, when it librates around $-$60{\degr} (or 300{\degr}), it is an 
     L$_5$ Trojan and it trails the planet, if the libration amplitude is wider than 180{\degr}, the minor body follows a horseshoe path 
     (see e.g. \citealt{1999ssd..book.....M}). The Eureka cluster occupies the West Lagrange point or L$_5$ with respect to Mars (that moves 
     around the Sun in a direct or eastward direction). Hybrids of these elementary co-orbital configurations are possible 
     \citep{1999PhRvL..83.2506N,2000CeMDA..76..131N}, as well as transitions between the various co-orbital states 
     \citep{1999Icar..137..293N,2000CeMDA..76..131N}, elementary or hybrid. Trojans can jump from librating around L$_4$ to librating around 
     L$_5$ and vice versa (see e.g. \citealt{2000A&A...354.1091T}; Connors et al. \citeyear{2011Natur.475..481C}; 
     \citealt{2018CeMDA.130...67S}). If $\lambda_{\rm r}$ does not oscillate around a certain value, we have a passing object whose position 
     relative to the host planet is not controlled by the 1:1 mean-motion resonance \citep{1999MNRAS.303..806C}. 

     Numerical simulations are required to confirm any putative resonant behaviour because, sometimes, an object may seem co-orbital in 
     terms of orbital period, but $\lambda_{\rm r}$ does not librate with time. If necessary to confirm the behaviour of $\lambda_{\rm r}$, 
     full $N$-body calculations have been carried out as described by \citet{2012MNRAS.427..728D} using software developed by 
     \citet{2003gnbs.book.....A}\footnote{\href{https://www.ast.cam.ac.uk/\%7esverre/web/pages/nbody.htm}
     {http://www.ast.cam.ac.uk/$\sim$sverre/web/pages/nbody.htm}} that implements the Hermite integration scheme described by 
     \citet{1991ApJ...369..200M}. The physical model includes the perturbations from the eight major planets, the Moon, the barycentre of 
     the Pluto-Charon system, and the three largest asteroids. When integrating the equations of motion, non-gravitational forces, 
     relativistic or oblateness terms are not taken into account. Relativistic terms can be safely neglected for orbits like the ones 
     discussed here (see e.g. \citealt{2008CeMDA.101..289B}) and the same can be said about the oblateness terms (see e.g. 
     \citealt{2015P&SS..117..223D}).
     
     Tisserand's criterion \citep{Tisserand1896} is often applied to decide if two orbits computed at different epochs may perhaps belong to 
     the same minor body; in order to apply this criterion, the object must interact directly with a single planet that must follow a 
     low-eccentricity orbit. Assuming that the subsystem Sun-planet-minor-body can be approximated by a circular restricted three-body 
     problem, Jacobi's integral should be a quasi-invariant (see the discussion at the beginning of Section~6). In this case, the integral 
     can be written as the Tisserand parameter: 
     \begin{equation}
        T_{\rm P} = \frac{a}{a_{\rm P}} + 2 \ \cos{i} \ \sqrt{\frac{a_{\rm P}}{a} \ (1 - e^{2})} \,, \label{Tisserand}
     \end{equation}
     where $a$, $e$, and $i$ are the semimajor axis, eccentricity and inclination of the orbit of the minor body, and $a_{\rm P}$ is the 
     semimajor axis of the planet \citep{1999ssd..book.....M}. Our initial hypothesis is that two co-orbital (to a certain planet) minor 
     bodies resulting from a YORP break-up event (or any less-than-gentle disruption episode for that matter) should have approximately the 
     same values of $T_{\rm P}$, particularly if the split took place relatively recently. Co-orbital bodies with moderate values of $e$ 
     only experience relatively distant flybys with their host planet and if the planet follows a nearly circular path, the scenario 
     originally envisioned by Tisserand is fulfilled. This situation is found for Mars co-orbitals and those of Earth with $e<0.2$. We would
     like to emphasize that the functional form of equation~(\ref{Tisserand}) makes it robust against large uncertainties of the orbital 
     parameters, which helps its application to objects with relatively poor orbit determinations.

     Later on, we will study the distributions of $\lambda_{\rm r}$ and $T_{\rm P}$. In order to analyse the results, we produce histograms 
     using the \textsc{Matplotlib} library \citep{2007CSE.....9...90H} with sets of bins computed using \textsc{NumPy} 
     \citep{2011CSE....13b..22V} by applying the Freedman and Diaconis rule \citep{FD81}; instead of using frequency-based histograms, we 
     consider counts to form a probability density so the area under the histogram will sum to one. A critical assessment of our results 
     will be performed in Section~5 by comparing our findings with those derived from the application of machine-learning techniques 
     provided by the \textsc{Python} libraries \textsc{Scikit-learn} \citep{Scikit2011,Scikit2013} and \textsc{SciPy} \citep{SciPy2020}.

  \section{Mars co-orbitals}
     Mars' co-orbital zone goes in terms of semimajor axis from $\sim$1.51645~au to $\sim$1.53095~au (see e.g. 
     \citealt{2005P&SS...53..617C}). Using this constraint and considering that $e<0.2$, we have retrieved data for 44 objects from JPL's 
     SBDB, and computed $\lambda_{\rm r}$ and $T_{\rm Mars}$ referred to epoch 2459000.5 for each one of them. 
%
%
     \begin{figure}
        \centering
        \includegraphics[width=\linewidth]{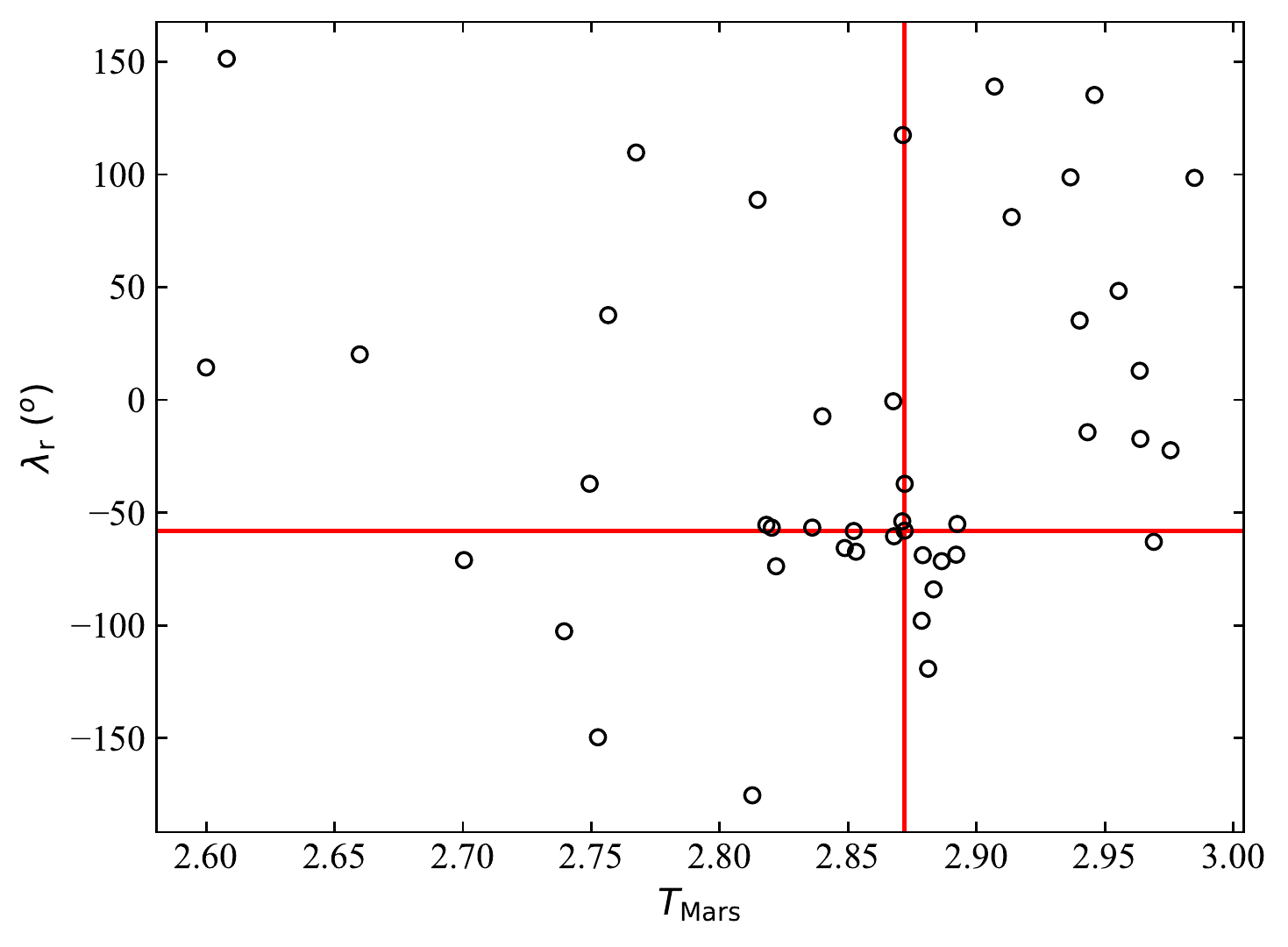}
        \caption{Distribution of relative mean longitudes, $\lambda_{\rm r}$, as a function of the Tisserand parameter, $T_{\rm Mars}$, for 
                 Mars co-orbital candidates (44 objects). The red lines mark the values for 5261~Eureka (1990~MB). 
                }
        \label{lambdarTM}
     \end{figure}
%
%

     Results are plotted in Fig.~\ref{lambdarTM}; the red lines mark the values for 5261~Eureka (1990~MB) that is the main object of the 
     cluster mentioned in Section~1. The values of $\lambda_{\rm r}$ and $T_{\rm Mars}$ for the members of the cluster discussed by 
     \citet{2020Icar..33513370C} are given in Table~\ref{eureka}. Following \citet{2020Icar..33513370C}, we will assume that these objects 
     are the result of one or more YORP break-up events. The absolute value of the difference between the value of $T_{\rm Mars}$ for the 
     object and that of Eureka is shown in Table~\ref{eureka} as ${\Delta}T_{\rm Mars}$. From these values, we find that 
     ${\Delta}T_{\rm Mars}<0.054$ could be compatible with a common origin under the assumptions made. On the other hand, we have that the 
     Martian Trojan with the value of the Tisserand parameter closest to that of Eureka is 2011~UN$_{63}$ followed by 2011~SP$_{189}$ (see 
     Table~\ref{eureka}). 

     In Fig.~\ref{lambdarTM}, we observe that there is a conspicuous cluster of objects around the position of Eureka; there is no 
     symmetrical cluster in the expected position of the L$_4$ Trojan cloud, which is consistent with previous findings 
     \citep{2018A&A...618A.178B,2020Icar..33513370C}. Either the event that led to the formation of the Eureka family was exclusive to 
     L$_5$ or there is a process that has been able to remove most objects from L$_4$ (see the analysis in \citealt{2020Icar..33513370C}). 
     In addition, we observe in Fig.~\ref{lambdarTM} that two objects have very low values of ${\Delta}T_{\rm Mars}$ with respect to Eureka 
     although they cannot be part of the L$_5$ Trojan cloud. The lowest is found for 2001~FG$_{24}$ with ${\Delta}T_{\rm Mars}=0.0000060$ 
     and $\lambda_{\rm r}=-37\fdg21$, that is not currently engaged in resonant behaviour, but simulations (not shown) reveal that it may 
     have been a transient horseshoe librator in the past and this situation may repeat in the future; the second lowest is 490718 
     (2010~RL$_{82}$) with ${\Delta}T_{\rm Mars}=0.0007294$ and $\lambda_{\rm r}=117\fdg45$, that shares dynamical status with 
     2001~FG$_{24}$ (their mutual ${\Delta}T_{\rm Mars}$ is 0.0007354). Fragments may leave the Trojan clouds via Yarkovsky-driven orbital 
     evolution as discussed by Scholl et al. (\citeyear{2005Icar..175..397S}), \citet{2015Icar..252..339C}, and \citet{2020Icar..33513370C}. 
     In general, asteroid families may evolve through a combination of YORP, Yarkovsky, and collisional events (see e.g. 
     \citealt{2020AJ....160..128M}). Two objects from Table~\ref{eureka}, 311999 (2007~NS$_{2}$) and 2011~SC$_{191}$, also have a very low 
     mutual ${\Delta}T_{\rm Mars}$, 0.0004248.     

\begin{figure}
       \centering
        \includegraphics[width=\linewidth]{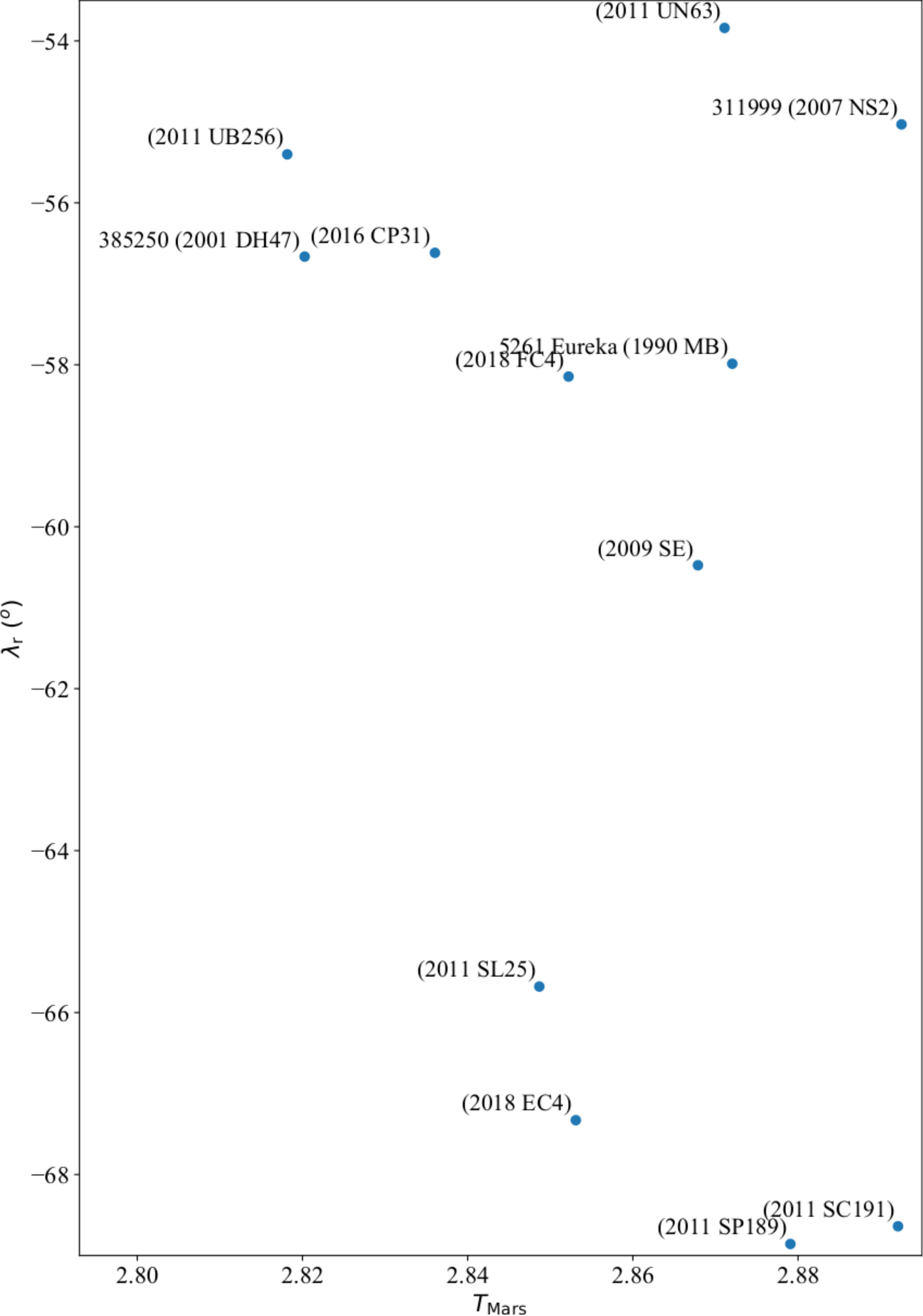}
        \caption{Relative positions in the $T_{\rm Mars}$-$\lambda_{\rm r}$ plane of the objects in Tables~\ref{eureka} and \ref{eureka+}. 
                 No objects have been removed from the displayed region, i.e. no known interlopers are present in the area of the 
                 $T_{\rm Mars}$-$\lambda_{\rm r}$ plane occupied by Mars' L$_5$ Trojan cloud.
                }
        \label{Eurekafamily}
     \end{figure}

\begin{figure}
       \centering
        \includegraphics[width=\linewidth]{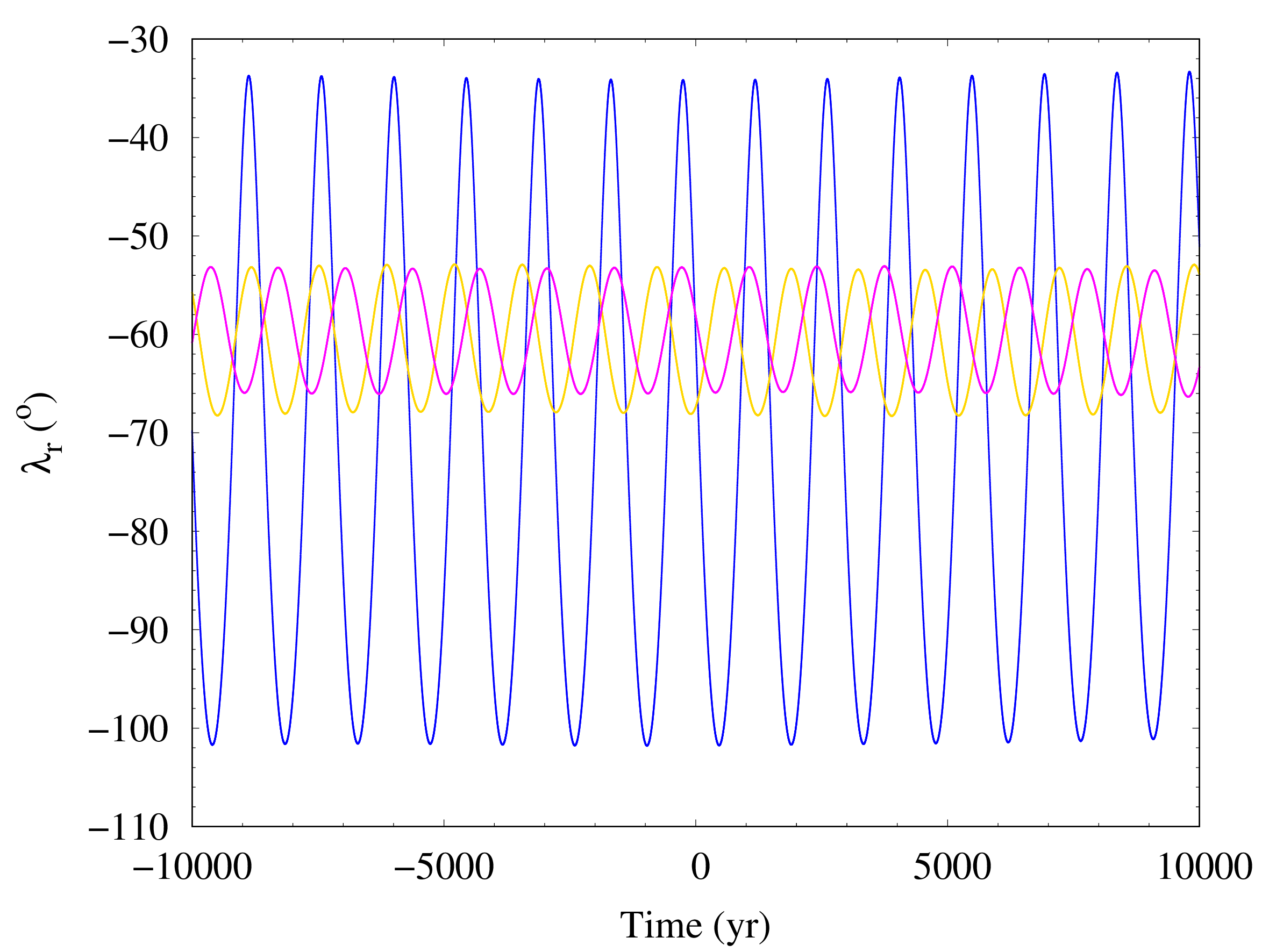}
        \includegraphics[width=\linewidth]{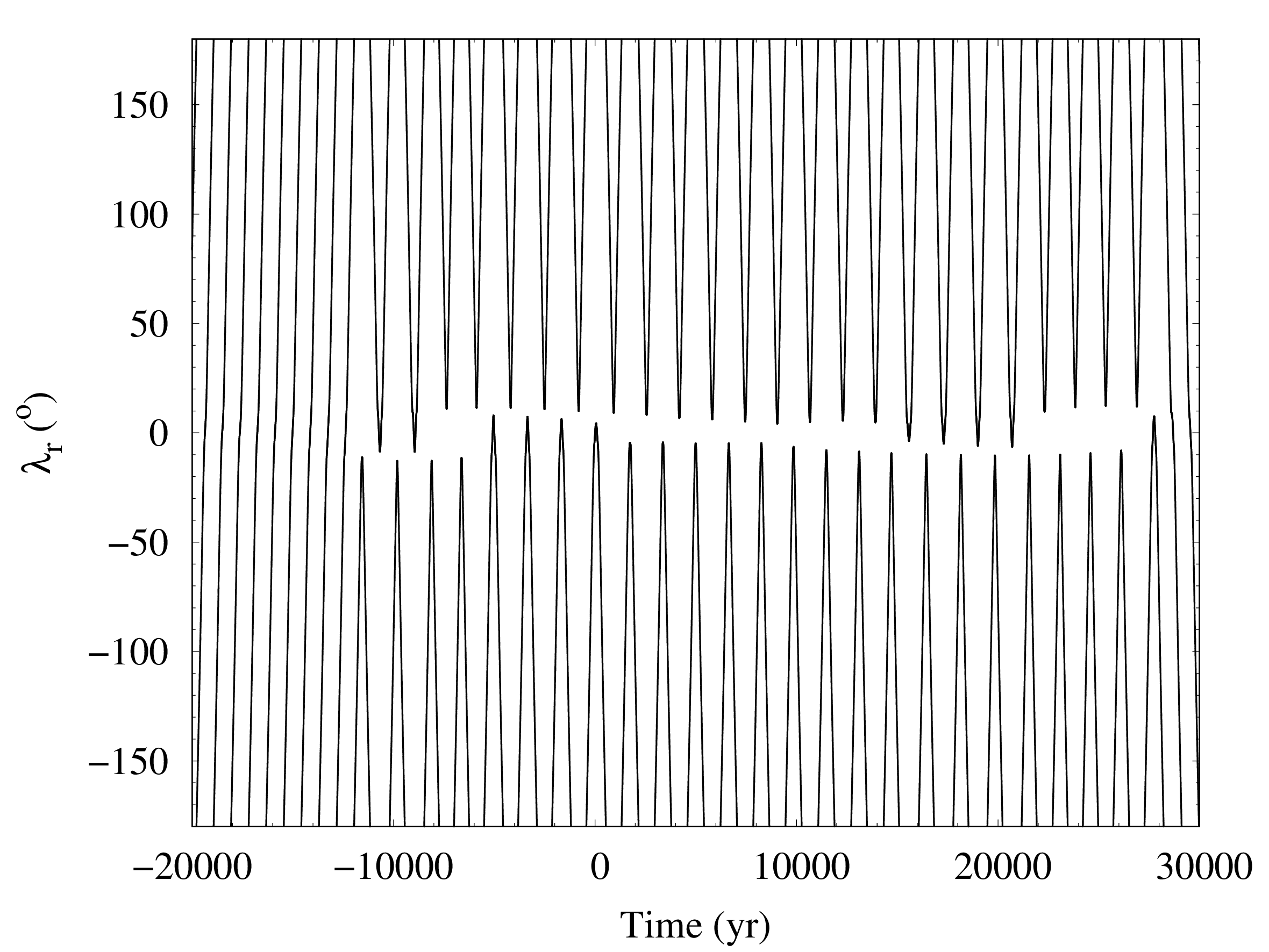}
        \caption{Evolution of the mean longitude difference of 2009~SE (blue), 2018~EC$_{4}$ (gold) and 2018~FC$_{4}$ (magenta) and Mars 
                 (top panel), and that of 2020~VT$_{1}$ (bottom panel). The output time-step size is 1~yr, the origin of time is epoch 
                 2459000.5 TDB, and only nominal orbits have been displayed (see Section~6.1 for a detailed discussion).
                }
        \label{newtrojans}
     \end{figure}

     Considering the boundaries defined by the data in Table~\ref{eureka} -- i.e. ${\Delta}T_{\rm Mars}<0.054$ with respect to Eureka and 
     $\lambda_{\rm r}\in(-68\fdg9, -53\fdg8)$, the locus of the Eureka cluster in the $T_{\rm Mars}$--$\lambda_{\rm r}$ plane -- at least 12 
     known Martian co-orbitals, including Eureka, could be Eureka family members (see Fig.~\ref{Eurekafamily}). The three new L$_5$ Trojans 
     -- 2009~SE, 2018~EC$_{4}$ and 2018~FC$_{4}$ -- are shown in Table~\ref{eureka+} and their status is confirmed in Fig.~\ref{newtrojans},
     top panel (but see the more detailed analysis in Section~6.1). These three objects can be regarded as robust L$_5$ Trojans and perhaps 
     members of the Eureka family (but see Section~6.1). Outside the locus of the cluster, we find 2020~VT$_{1}$, a recent discovery that 
     has $\lambda_{\rm r}=-0\fdg5983$ (but it is not a quasi-satellite, see Fig.~\ref{newtrojans}, bottom panel) and ${\Delta}T_{\rm Mars}$
     relative to Eureka of 0.0044492 (see Section~6.1 for additional details).
%
%
%
%
%
%
      \begin{table}
        \centering
        \fontsize{8}{11pt}\selectfont
        \tabcolsep 0.15truecm
        \caption{As Table~\ref{eureka} but for new Eureka family member candidates within the locus discussed in the text. The last column
                 gives the data-arc in days.
                }
        \begin{tabular}{rcccr}
          \hline
             Object                  & $T_{\rm Mars}$ & $\lambda_{\rm r}$ (\degr) & ${\Delta}T_{\rm Mars}$ & arc (d) \\
          \hline
             2009 SE                 & 2.867875826    & $-$60.47576               & 0.004160134            &  3133   \\
             2018 EC$_{4}$           & 2.853092552    & $-$67.32835               & 0.018943407            &  3131   \\
             2018 FC$_{4}$           & 2.852214411    & $-$58.14554               & 0.019821549            &   790   \\
          \hline
        \end{tabular}
        \label{eureka+}
      \end{table}
%
%
%
%
%
%

\begin{figure}
       \centering
        \includegraphics[width=\linewidth]{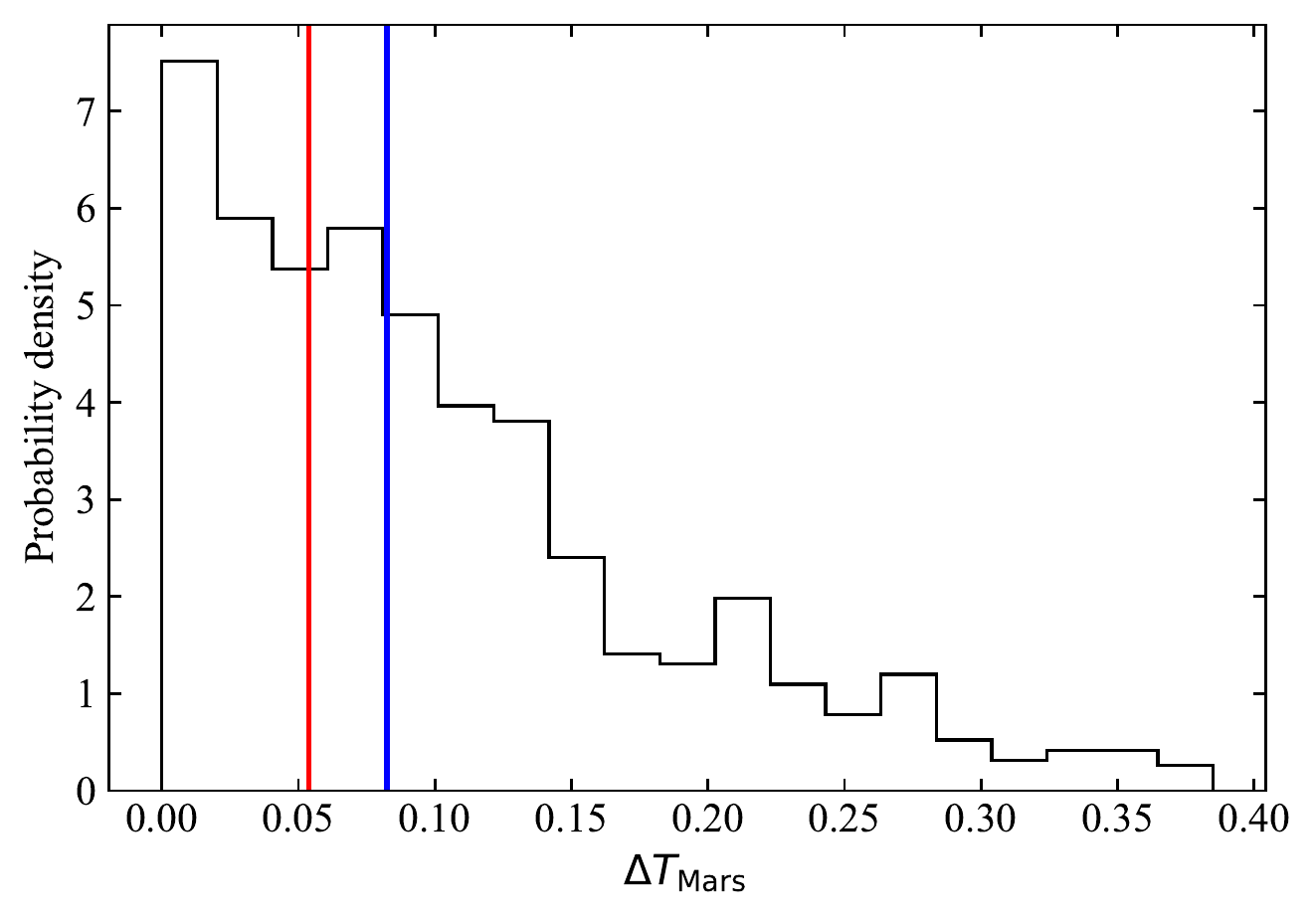}
        \caption{Distribution of ${\Delta}T_{\rm Mars}$ for all pairs of objects in our sample of Mars co-orbitals. The median value is 
                 shown in blue and the critical value of 0.054 mentioned in the text is shown in red. 
                }
        \label{dTM}
     \end{figure}

     In order to investigate the statistical significance of our results regarding ${\Delta}T_{\rm Mars}$, we have computed the mutual 
     differences between the values of $T_{\rm Mars}$ for all the objects in our sample. The resulting histogram is shown in Fig.~\ref{dTM}. 
     The mean value and standard deviation are 0.10$\pm$0.08; the median and 16th and 84th percentiles are 0.08$_{-0.06}^{+0.11}$. Although 
     the distributions are not normal, in a normal distribution a value that is one standard deviation above the mean is equivalent to the 
     84th percentile and a value that is one standard deviation below the mean is equivalent to the 16th percentile (see e.g. 
     \citealt{2012psa..book.....W}). By providing these values (standard deviation and the relevant percentiles), we wanted to quantify how 
     non-normal the distribution is. 

     The distribution in Fig.~\ref{dTM} shows that the probability of having two objects with ${\Delta}T_{\rm Mars}<0.054$ is, 
     $P(<0.054)=0.3404$; we also have $P(<0.005)=0.0402$ and $P(<0.0001)=0.0011$. Therefore, the case of Eureka and 2001~FG$_{24}$ pointed 
     out above is indeed unusual. In terms of probability, it may be possible that 2001~FG$_{24}$ (that has a data-arc of 7119~d) was part 
     of Eureka (a fast rotator) in the past. However, this cannot have happened in the recent past because we have performed extensive 
     calculations backwards in time for these two objects during 10$^{4}$~yr and their closest flyby might have been at about 0.008~au. 
     Similar calculations for the pair Eureka--2011~UN$_{63}$ give a distance of closest approach of 0.0008~au with a relative velocity of 
     3.5~km~s$^{-1}$. These values are not typical of pairs resulting from a rotation-induced YORP break-up event (see e.g. 
     \citealt{2010Natur.466.1085P}). On the other hand, every object in our sample of 44 has at least one other object with mutual 
     ${\Delta}T_{\rm Mars}<0.054$. The fact that no known interlopers are present in the area of the $T_{\rm Mars}$-$\lambda_{\rm r}$ plane 
     occupied by Mars' L$_5$ Trojan cloud in Fig.~\ref{Eurekafamily} gives some support to using this plane to find groups of minor bodies
     that may be related. 
%
%
%
%

  \section{Earth co-orbitals}
\begin{figure}
       \centering
        \includegraphics[width=\linewidth]{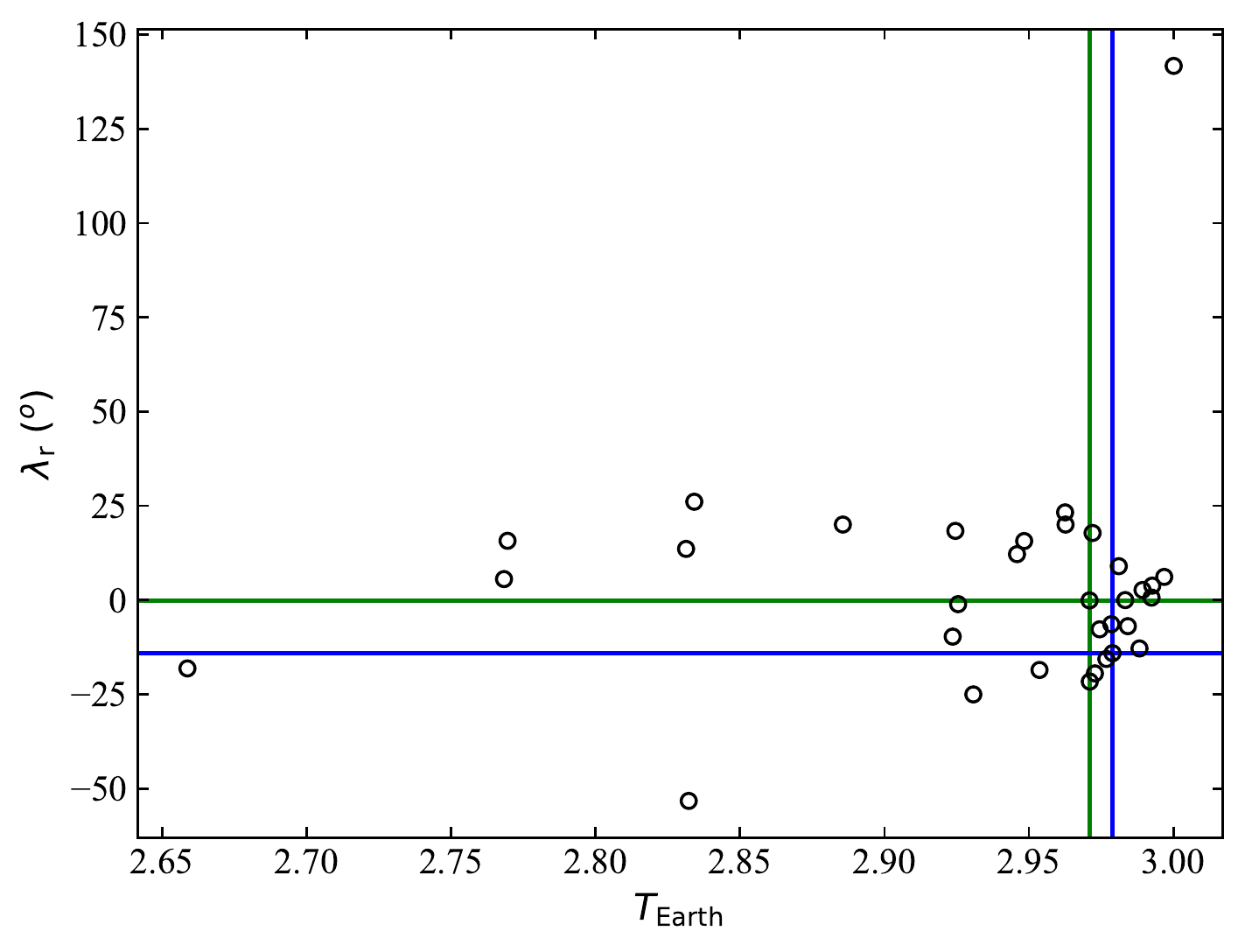}
        \caption{Distribution of relative mean longitudes, $\lambda_{\rm r}$, as a function of the Tisserand parameter, $T_{\rm Earth}$, for 
                 Earth co-orbital candidates (33 objects). The green lines mark the values for Earth quasi-satellite 469219~Kamo`oalewa and 
                 the blue ones those of 478784 (2012~UV$_{136}$).  
                }
        \label{lambdarTE}
     \end{figure}

\begin{figure}
       \centering
        \includegraphics[width=\linewidth]{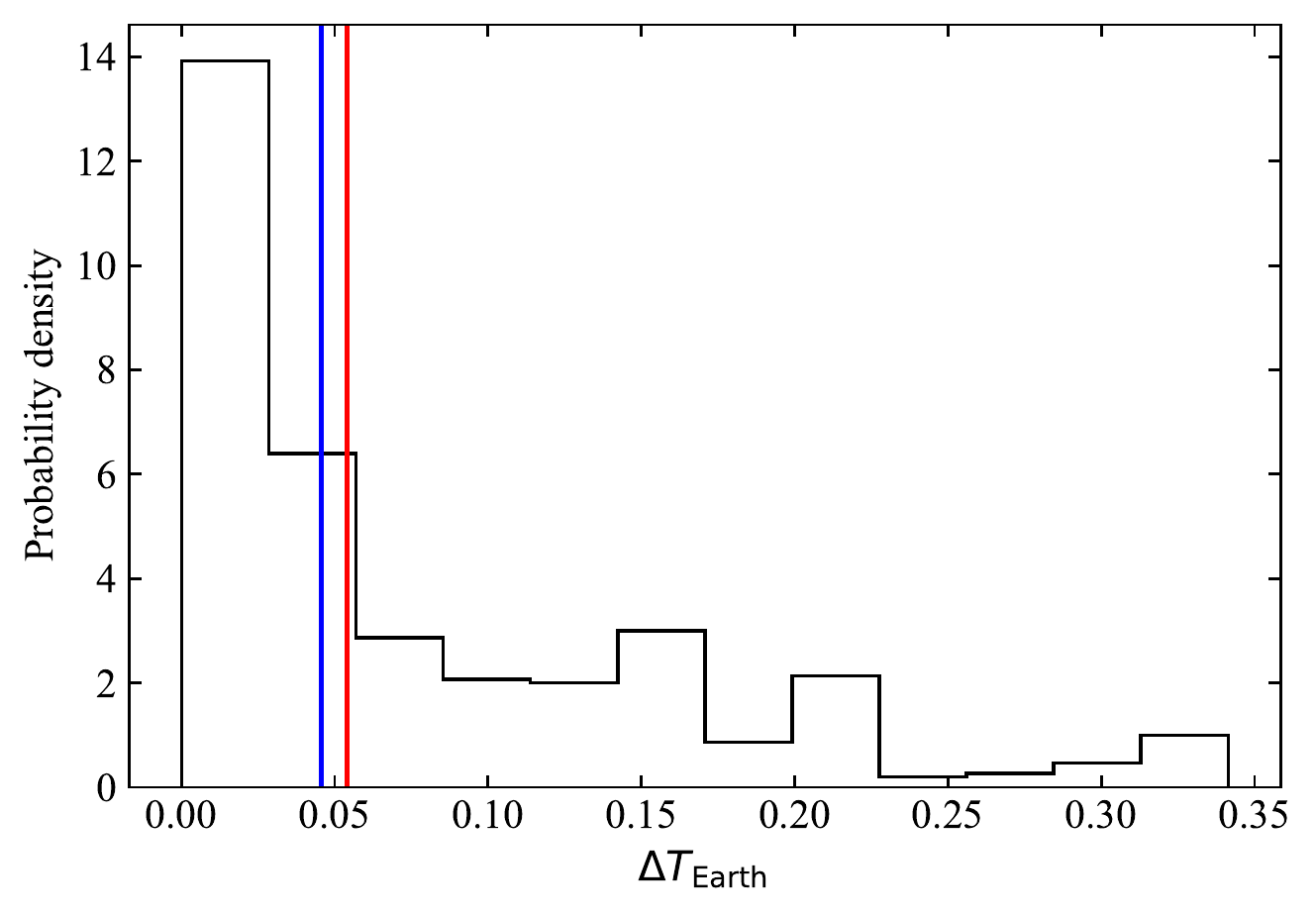}
        \caption{Distribution of ${\Delta}T_{\rm Earth}$ for all pairs of objects in our sample of Earth co-orbitals. The median value is
                 shown in blue and the critical value of 0.054 mentioned in the text is shown in red. 
                }
        \label{dTE}
     \end{figure}

     Earth's co-orbital zone goes from $\sim$0.994~au to $\sim$1.006~au (see e.g. \citealt{2018MNRAS.473.3434D}). Using this constraint and 
     considering that $e<0.2$, we have retrieved data for 33 objects from JPL's SBDB, and computed $\lambda_{\rm r}$ and $T_{\rm Earth}$ as
     we did in Section~3 in the case of Mars and its co-orbitals. Figure~\ref{lambdarTE} shows the resulting distribution and some 
     conspicuous clustering similar to the one present in Fig.~\ref{lambdarTM} is clearly visible, but now groups appear concentrated 
     towards $\lambda_{\rm r}\sim0$\degr. When we computed the mutual differences between the values of $T_{\rm Earth}$, 
     ${\Delta}T_{\rm Earth}$, for all the objects in our sample to investigate the statistical significance of the groupings, we obtained 
     the histogram in Fig.~\ref{dTE}. Now the median and 16th and 84th percentiles are 0.05$_{-0.04}^{+0.11}$ that is below the threshold of 
     0.054 found in Section~3 (mean and standard deviation, 0.08$\pm$0.08). The values of the probabilities analogues to those computed in 
     Section~3 are: $P(<0.054)=0.5644$, $P(<0.005)=0.0871$ and $P(<0.0001)=0.0019$. 
%
%
%
%
%
%
%
%

\begin{figure}
       \centering
        \includegraphics[width=\linewidth]{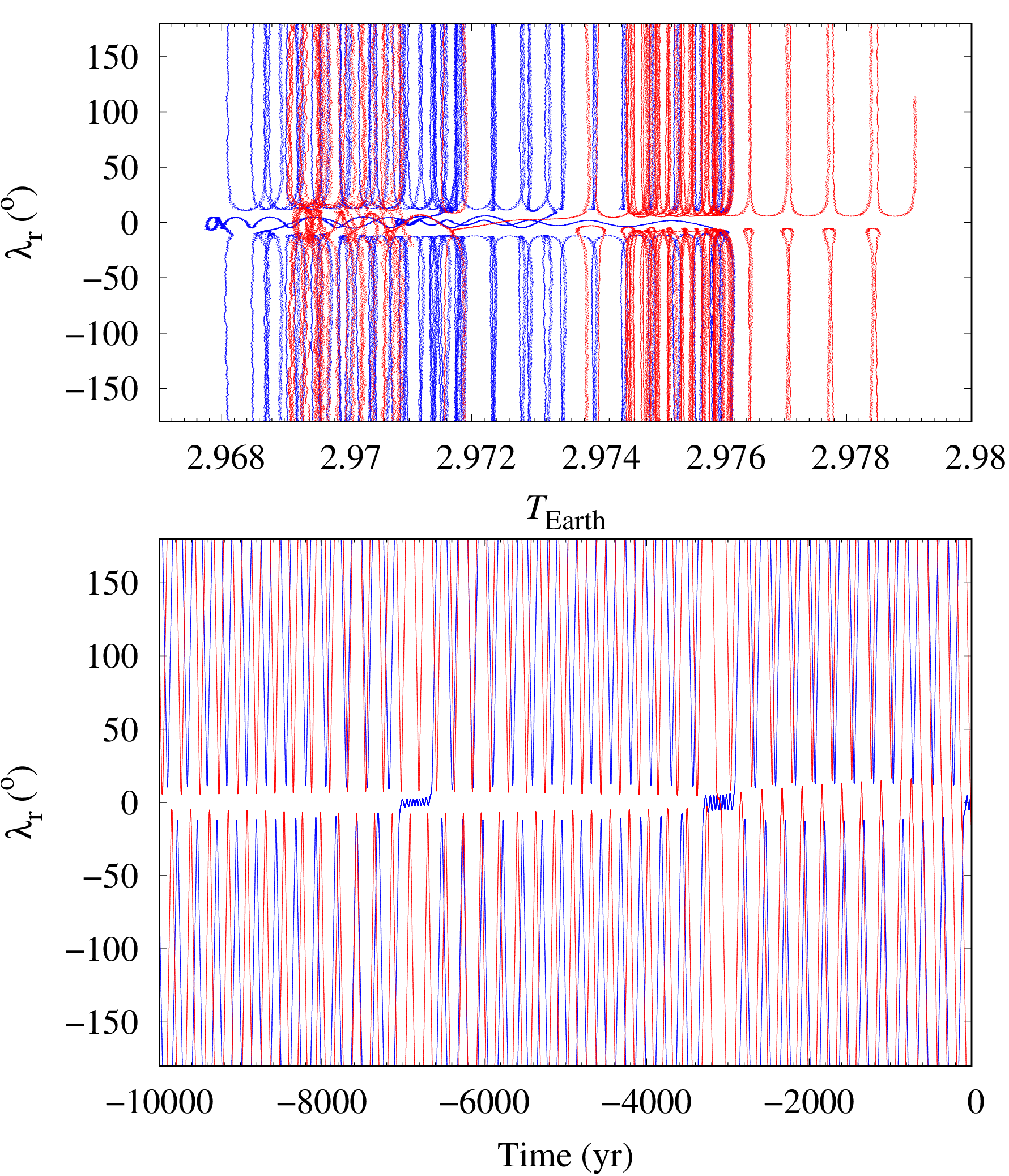}
        \caption{Concurrent past evolution over 10\,000~yr of 469219~Kamo`oalewa (in blue) and 2016~FU$_{12}$ (in red) in the 
                 $T_{\rm Earth}$-$\lambda_{\rm r}$ plane (top panel) and in $\lambda_{\rm r}$ (bottom panel). Kamo`oalewa switches 
                 multiple times between the quasi-satellite and horseshoe configurations, while 2016~FU$_{12}$ followed an asymmetric 
                 and irregular horseshoe path. The output time-step size is 0.002~yr, the origin of time is epoch 2459000.5 TDB, and only 
                 nominal orbits have been displayed.
                }
        \label{ho3fu12}
     \end{figure}

\begin{figure}
       \centering
        \includegraphics[width=\linewidth]{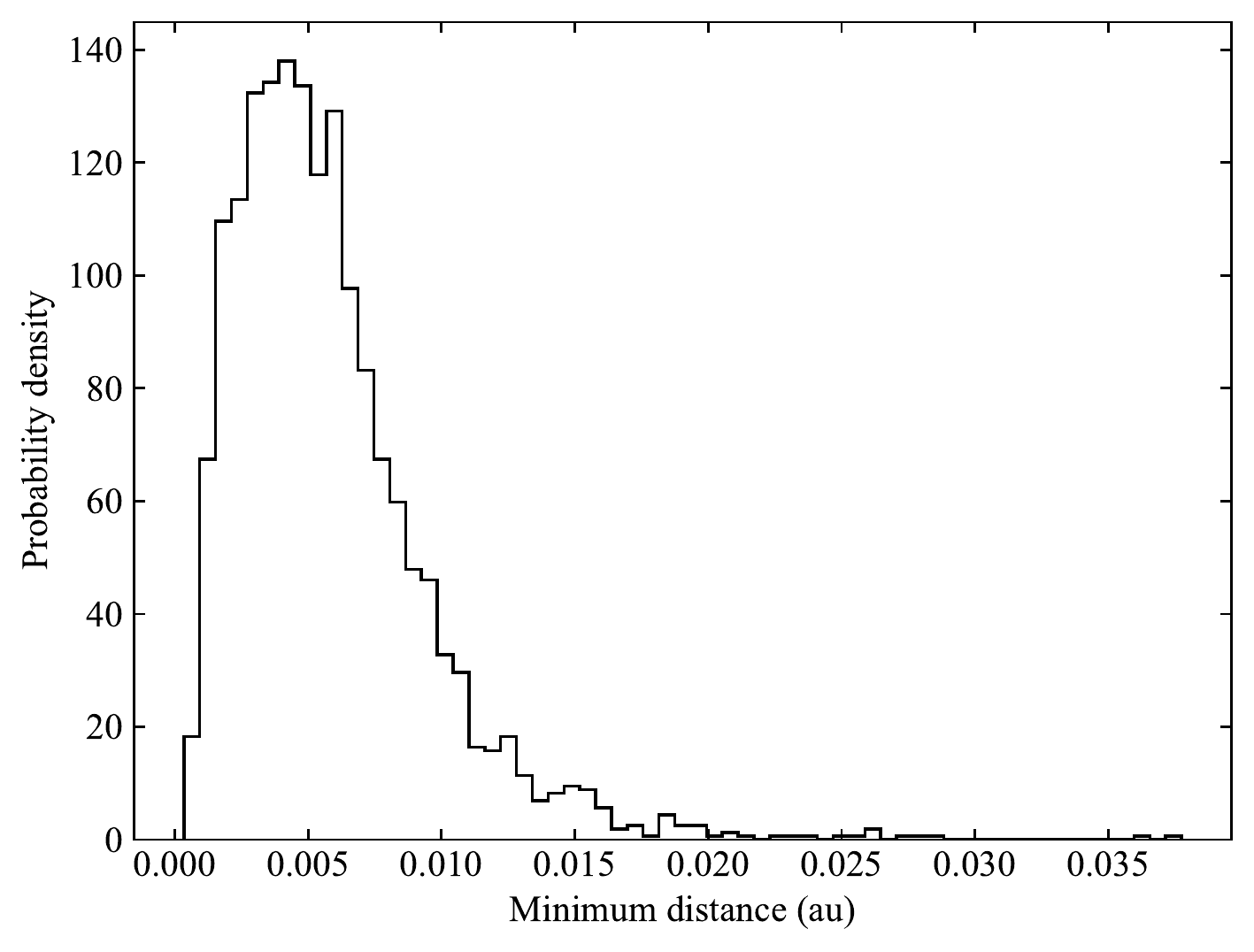}
        \includegraphics[width=\linewidth]{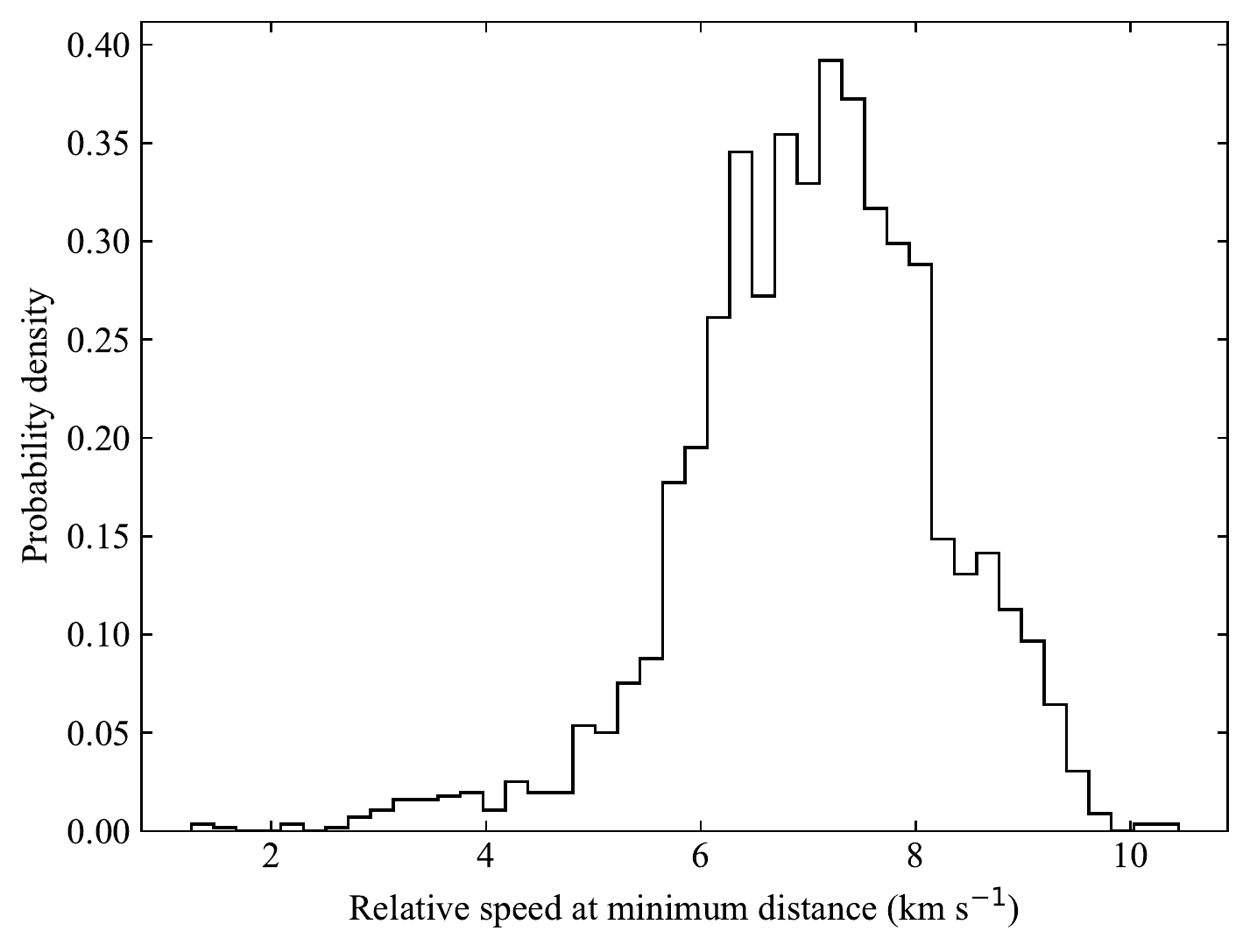}
        \includegraphics[width=\linewidth]{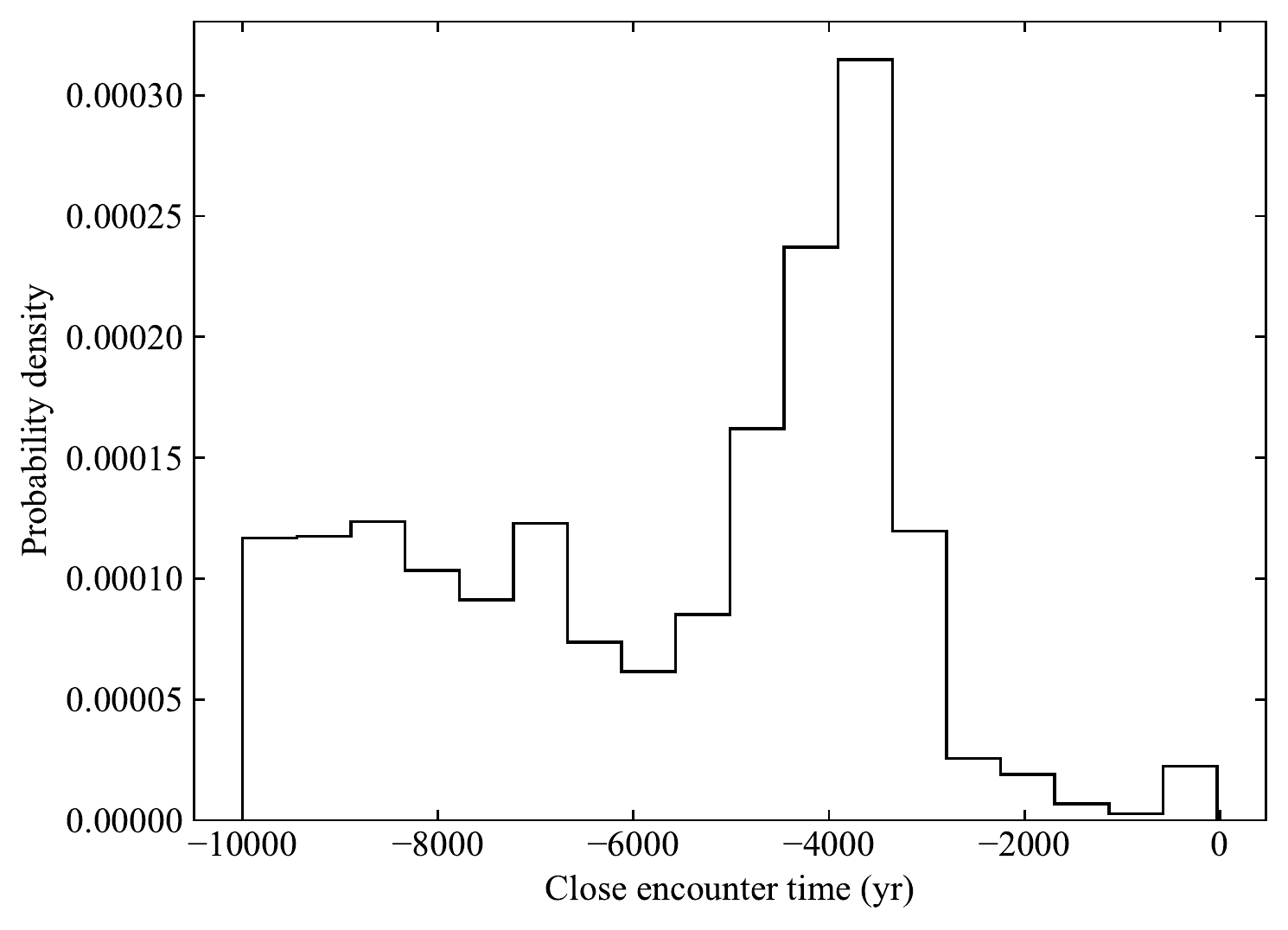}
        \caption{Distribution of minimum approach distance (top panel), relative speed at minimum distance (middle panel), and time at
                 minimum distance (bottom panel) for calculations including 469219 Kamo`oalewa (2016~HO$_{3}$) and 2016~FU$_{12}$. The 
                 origin of time in the bottom panel is epoch 2459000.5 TDB. Input values from Tables~\ref{vectorKamo} and 
                 \ref{vector2016FU12}.
                }
        \label{KamoFU12}
     \end{figure}

     The smallest value of ${\Delta}T_{\rm Earth}$, 0.00008836, is found for the pair made of 469219 Kamo`oalewa (2016~HO$_{3}$), a 
     quasi-satellite of Earth \citep{2016MNRAS.462.3441D}, and 2016~FU$_{12}$. Given the fact that Kamo`oalewa is an extremely fast rotator 
     with a period of 28.03~minutes \citep{2017DPS....4920407R}, it is not difficult to argue that 2016~FU$_{12}$ could be the by-product of 
     a YORP break-up event that affected the quasi-satellite in relatively recent times. A concurrent simulation into the past of the 
     nominal orbits of both objects appears to support such a scenario as shown in Fig.~\ref{ho3fu12}, bottom panel, where multiple 
     overlappings can be seen. However, extensive calculations backwards in time during 10$^{4}$~yr for these two objects show that their 
     closest flyby may have been at about 0.0003~au or close to 45\,000 km with a relative velocity of 6.2~km~s$^{-1}$. The summary of our 
     results for the recent flybys of this pair is shown in Fig.~\ref{KamoFU12} that includes the outcome of 2500 simulations that take into 
     account the uncertainties in the orbit determinations of both bodies (see Table~\ref{kamolike}). Most flybys occur when both objects 
     are horseshoe librators, compare the bottom panels of Figs~\ref{ho3fu12} and \ref{KamoFU12}. From these results, it is clear that close 
     approaches and perhaps collisions between Earth co-orbitals are possible. Putative collisions may take place at speeds (see 
     Fig.~\ref{KamoFU12}, middle panel) in excess of those typical in the main asteroid belt, 6$\pm$2~km~s$^{-1}$ (see e.g. 
     \citealt{1992Icar...97..111F}). As pointed out in the previous section for relevant pairs of Mars co-orbitals, these values are not 
     typical of pairs resulting from a rotation-induced YORP break-up event (see e.g. \citealt{2010Natur.466.1085P}).
%
%
%
%
%
%
         \begin{table*}
          \fontsize{8}{11pt}\selectfont
          \tabcolsep 0.10truecm
          \caption{Heliocentric Keplerian orbital elements of 469219~Kamo`oalewa (2016~HO$_{3}$), 2016~FU$_{12}$, and 2020~KZ$_{2}$ used in 
                   this study. The orbit determination of Kamo`oalewa is based on 307 observations spanning a data-arc of 5140~d or 14.07~yr 
                   (solution date, 2018-Apr-30 06:51:17 PDT), the one of 2016~FU$_{12}$ is based on 19 observations spanning a data-arc of 
                   4~d (solution date, 2017-Apr-06 08:27:11 PDT), and that of 2020~KZ$_{2}$ is based on 39 observations spanning a data-arc 
                   of 8~d (solution date, 2020-May-29 06:07:53 PDT). Values include the 1$\sigma$ uncertainty. The orbit determinations have 
                   been computed at epoch JD 2459000.5 that corresponds to 00:00:00.000 TDB on 2020 May 31 (J2000.0 ecliptic and equinox). 
                   Source: JPL's SBDB.
                  }
          \begin{tabular}{lcccc}
           \hline
            Orbital parameter                                 &   & 469219 Kamo`oalewa            & 2016~FU$_{12}$      & 2020~KZ$_{2}$           \\
           \hline
            Semimajor axis, $a$ (au)                          & = &   1.001247952$\pm$0.000000003 &   1.0050$\pm$0.0002 &   1.00530$\pm$0.00003   \\
            Eccentricity, $e$                                 & = &   0.1033008$\pm$0.0000002     &   0.166$\pm$0.002   &   0.028770$\pm$0.000004 \\
            Inclination, $i$ (\degr)                          & = &   7.78529$\pm$0.00002         &   2.12$\pm$0.02     &   7.231$\pm$0.011       \\
            Longitude of the ascending node, $\Omega$ (\degr) & = &  66.15571$\pm$0.00002         & 224.2$\pm$0.2       &  64.9562$\pm$0.0010     \\
            Argument of perihelion, $\omega$ (\degr)          & = & 306.18784$\pm$0.00002         & 198.3$\pm$0.3       &   7.498$\pm$0.011       \\
            Mean anomaly, $M$ (\degr)                         & = & 236.29311$\pm$0.00002         & 164.6$\pm$0.7       & 176.268$\pm$0.012       \\
            Perihelion, $q$ (au)                              & = &   0.8978182$\pm$0.0000002     &   0.8384$\pm$0.0015 &   0.97637$\pm$0.00002   \\
            Aphelion, $Q$ (au)                                & = &   1.104677713$\pm$0.000000003 &   1.1717$\pm$0.0002 &   1.03422$\pm$0.00003   \\
            Absolute magnitude, $H$ (mag)                     & = &  24.3$\pm$0.5                 &  26.9$\pm$0.5       &  27.7$\pm$0.4           \\
           \hline
          \end{tabular}
          \label{kamolike}
         \end{table*}
%
%
%
%
%
%

     No actual collisions between asteroids have ever been observed although the outcome of several of them may have been studied: 
     354P/LINEAR (see e.g. \citealt{2010Natur.467..814S}), (596) Scheila (see e.g. \citealt{2011ApJ...733L...3B}), (493) Griseldis 
     \citep{2015DPS....4741403T}, and P/2016~G1 (PANSTARRS) \citep{2016ApJ...826L..22M,2017AJ....154..248M,2019A&A...628A..48H}. The 
     observed fragments resulting from these events have been produced by excavation, leading to a low launch velocity of the ejecta from 
     the surface of the target body, a relatively large asteroid (larger than objects like Kamo`oalewa). Numerical simulations indicate that 
     velocities of ejecta relative to target may not be much higher than a few hundred m~s$^{-1}$ and often significantly lower than that 
     (see e.g. Appendix~E in \citealt{2017Icar..296..239S} or \citealt{2010Icar..207...54J}). However, we have detailed observations of at 
     least one high-speed collision of small bodies in our immediate neighbourhood, the Iridium--Cosmos collision event. On 2009 February 
     10, an inactive Russian communications satellite, Cosmos~2251, collided with an active commercial communications satellite, Iridium~33, 
     at a relative speed of nearly 10~km~s$^{-1}$ \citep{2009amos.confE...3K}. The two artificial satellites had similar masses 
     (Cosmos~2251, 900~kg and Iridium~33, 689~kg), their orbital planes were nearly perpendicular, and the resulting clouds of orbital 
     debris remained at almost right angles to each other, spreading along their original paths in a short time-scale; realistic orbital 
     evolution analyses suggest that most fragments will probably remain in orbit around Earth for decades \citep{2009amos.confE...3K,
     2009amos.confE...4V,2010amos.confE..37S}. Although the Low Earth Orbit dynamical environment and the one travelled by Earth co-orbitals 
     are rather different and we are speaking of manufactured not natural objects, the outcome of the high-speed collision between 
     Iridium~33 and Cosmos~2251 may teach us a valuable lesson, that catastrophic disruptions of small bodies due to high-speed impacts may 
     create clouds of debris that may eventually lead to objects with no genetic relationship experiencing close encounters (or even 
     impacts). There are no reasons to assume that Iridium--Cosmos collision-like events may not be happening in Earth co-orbital space 
     where mean-motion resonances may easily lead to intersecting orbits.  

\begin{figure}
       \centering
        \includegraphics[width=\linewidth]{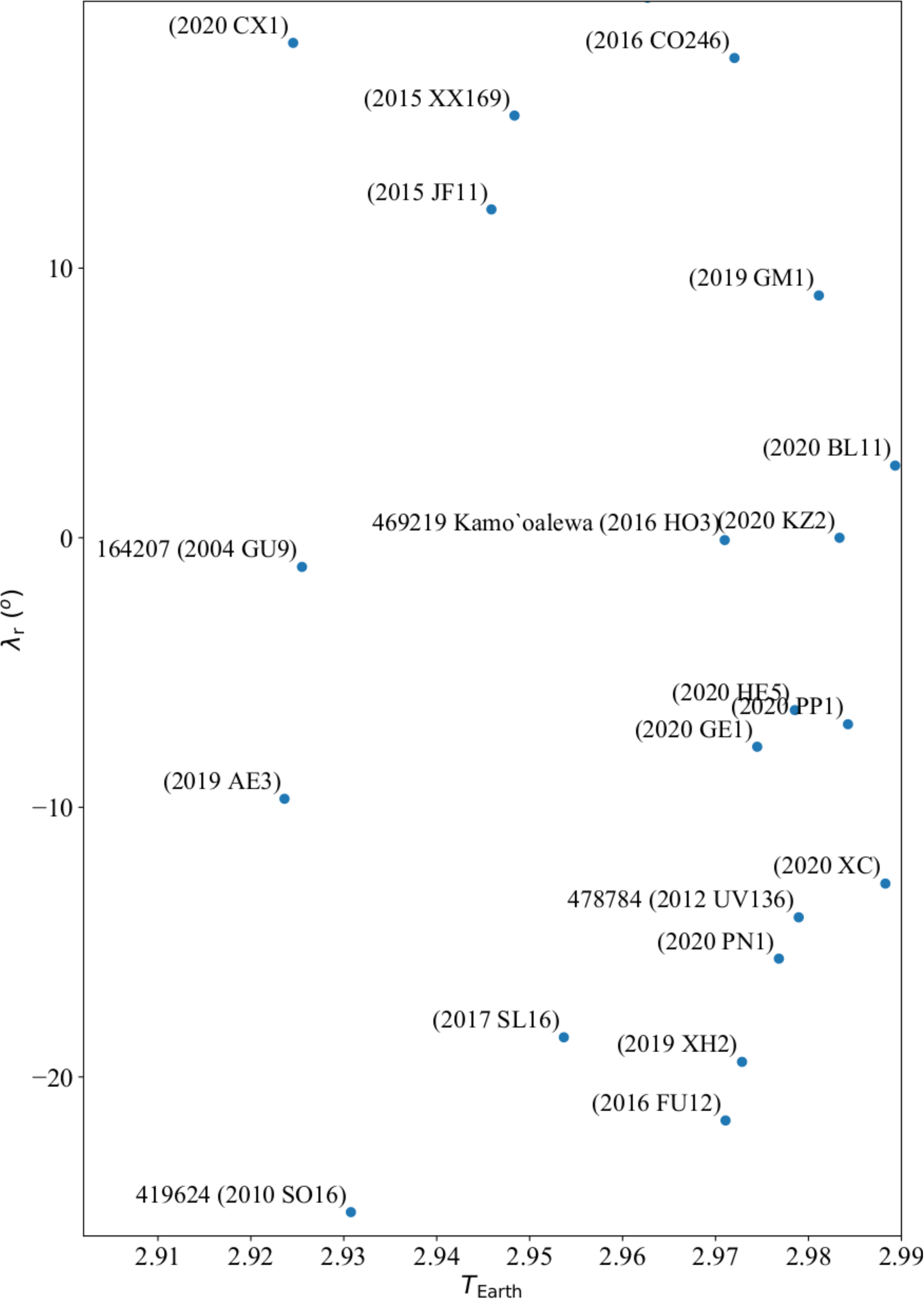}
        \caption{Relative positions of 20 Earth co-orbitals that include most of those cited in the text.
                }
        \label{EarthGroups}
     \end{figure}

     Additional objects that may be related to Kamo`oalewa are shown in Table~\ref{kamo} although only Kamo`oalewa and 2016~CO$_{246}$ have 
     robust orbits. Out of the various clusterings seen in Fig.~\ref{lambdarTE}, the ones with the most members are associated with 
     Kamo`oalewa and 478784 (2012~UV$_{136}$) that is not currently engaged in resonant behaviour (see Table~\ref{uv136} for additional 
     members). The relative positions in the $T_{\rm Earth}$-$\lambda_{\rm r}$ plane of many of the objects mentioned in our discussion are
     shown in Fig.~\ref{EarthGroups}.
%
%
     \begin{table}
        \centering
        \fontsize{8}{11pt}\selectfont
        \tabcolsep 0.10truecm
        \caption{Tisserand parameter with respect to Earth, $T_{\rm Earth}$, and relative mean longitude, $\lambda_{\rm r}$, for Earth 
                 quasi-satellite Kamo`oalewa and suspected family members with ${\Delta}T_{\rm Earth}$ relative to Kamo`oalewa included.  
                }
        \begin{tabular}{rccc}
          \hline
             Object                            & $T_{\rm Earth}$ & $\lambda_{\rm r}$ (\degr) & ${\Delta}T_{\rm Earth}$ \\
          \hline
            469219 Kamo`oalewa (2016~HO$_{3}$) & 2.97098699      &  $-$0.08584               &   --                    \\
                               2016~CO$_{246}$ & 2.97203544      &    17.79712               & 0.00104845              \\
                               2016~FU$_{12}$  & 2.97107535      & $-$21.61721               & 0.00008836              \\
                               2019~XH$_{2}$   & 2.97284981      & $-$19.44256               & 0.00186281              \\
                               2020~GE$_{1}$   & 2.97448273      &  $-$7.75249               & 0.00349574              \\ 
          \hline
        \end{tabular}
        \label{kamo}
     \end{table}
%
%
%
%
      \begin{table}
        \centering
        \fontsize{8}{11pt}\selectfont
        \tabcolsep 0.15truecm
        \caption{Similar to Table~\ref{kamo} but for 478784 (2012~UV$_{136}$).
                }
        \begin{tabular}{rccc}
          \hline
             Object                  & $T_{\rm Earth}$ & $\lambda_{\rm r}$ (\degr) & ${\Delta}T_{\rm Earth}$ \\
          \hline
            478784 (2012~UV$_{136}$) & 2.97894830      & $-$14.07993               &   --                    \\
                      2019~GM$_{1}$  & 2.98111706      &     8.98722               & 0.00216877              \\
                      2020~GE$_{1}$  & 2.97448273      &  $-$7.75249               & 0.00446557              \\
                      2020~HE$_{5}$  & 2.97853099      &  $-$6.39397               & 0.00041731              \\
                      2020~KZ$_{2}$  & 2.98334044      &  $-$0.00089               & 0.00439214              \\
                      2020~PN$_{1}$  & 2.97681846      & $-$15.61311               & 0.00212984              \\
          \hline
        \end{tabular}
        \label{uv136}
      \end{table}
%
%
%
%
%
%

  \section{Clustering validation}
     So far, we have spoken loosely of clusterings but there are a number of statistical tools that can be used to perform such a data 
     exploration in a rigorous way. Machine-learning techniques have sometimes been used to investigate the presence of coherent dynamical 
     groups or genetic families within the populations of asteroids of the main belt (see e.g. \citealt{1990AJ....100.2030Z,
     1994AJ....107..772Z,2013ApJ...770....7M,2017MNRAS.469.2024S,2019MNRAS.488.1377C,2019MsT.........30M,2020MNRAS.496..540C}). In this 
     section, we apply unsupervised machine-learning in the form of clustering algorithms to the data discussed above. Our objective is to
     validate the conclusions obtained in the previous sections and perhaps extract new ones that may lead us to understand the populations
     of Mars and Earth co-orbitals better.

     As part of the data preparation process, we have scaled the data set using standardization or Z-score normalization: found the mean and
     standard deviation for $T_{\rm P}$ and $\lambda_{\rm r}$, subtracted the relevant mean from each value, and then divided by its 
     corresponding standard deviation. This was carried out by applying the method {\tt fit\_transform} that is part of the 
     {\tt StandardScaler} class provided by the \textsc{Python} library \textsc{Scikit-learn} \citep{Scikit2011}. This rescaling is strongly 
     recommended when applying unsupervised machine-learning algorithms. Distance assignment between the objects in our sample assumes a 
     Euclidean metric.

     We have applied two different algorithms to evaluate data clustering: $k$-means++ and agglomerative hierarchical clustering. The 
     $k$-means++ algorithm \citep{Kmeans07} performs a centroid-based analysis and it is an improved version of the $k$-means algorithm
     (see e.g. \citealt{Kmeans57,Kmeans67,Kmeans82}). Here, we have used the implementation of $k$-means++ in the method {\tt fit} that is 
     part of the {\tt KMeans} class provided by the \textsc{Python} library \textsc{Scikit-learn} \citep{Scikit2011}. On the other hand, the 
     agglomerative hierarchical clustering algorithm (see e.g. \citealt{AHC63}) studies connectivity-based clustering aimed at building a 
     hierarchy of clusters where each observation or data point (in our case, each object) starts as an individual cluster and clusters are 
     merged by iteration. Here, we have used the implementation of the agglomerative hierarchical clustering algorithm included in the 
     {\tt hierarchy} module of the {\tt clustering} package that is part of the \textsc{Python} library \textsc{SciPy} \citep{SciPy2020}, 
     specifically the functions {\tt linkage} that performs hierarchical/agglomerative clustering analysis and {\tt dendrogram} that plots 
     the resulting hierarchical clustering as a dendrogram. The function {\tt linkage} has been invoked using the Ward variance minimization 
     algorithm \citep{AHC63}.

     For the agglomerative hierarchical clustering analysis, the dendrograms in our figures display the hierarchical merging process with 
     the horizontal axis showing the distance at which two given clusters were merged. We have considered 0.7 of the maximal merging 
     distance as the threshold to define the final clusters in the data set so the merging distance of objects within a cluster is 
     significantly shorter than the merging distance of the final clusters (see additional comments below). For $k$-means++, we have used 
     the elbow method to determine the optimal value of clusters, $k$; we invoke the method {\tt fit} with $k$ in the interval (1, 10) and 
     select the value of $k$ that minimizes the sum of the distances of all data points or objects to their respective cluster centres. 

     \subsection{Mars co-orbitals}
\begin{figure*}
       \centering
        \includegraphics[width=\linewidth]{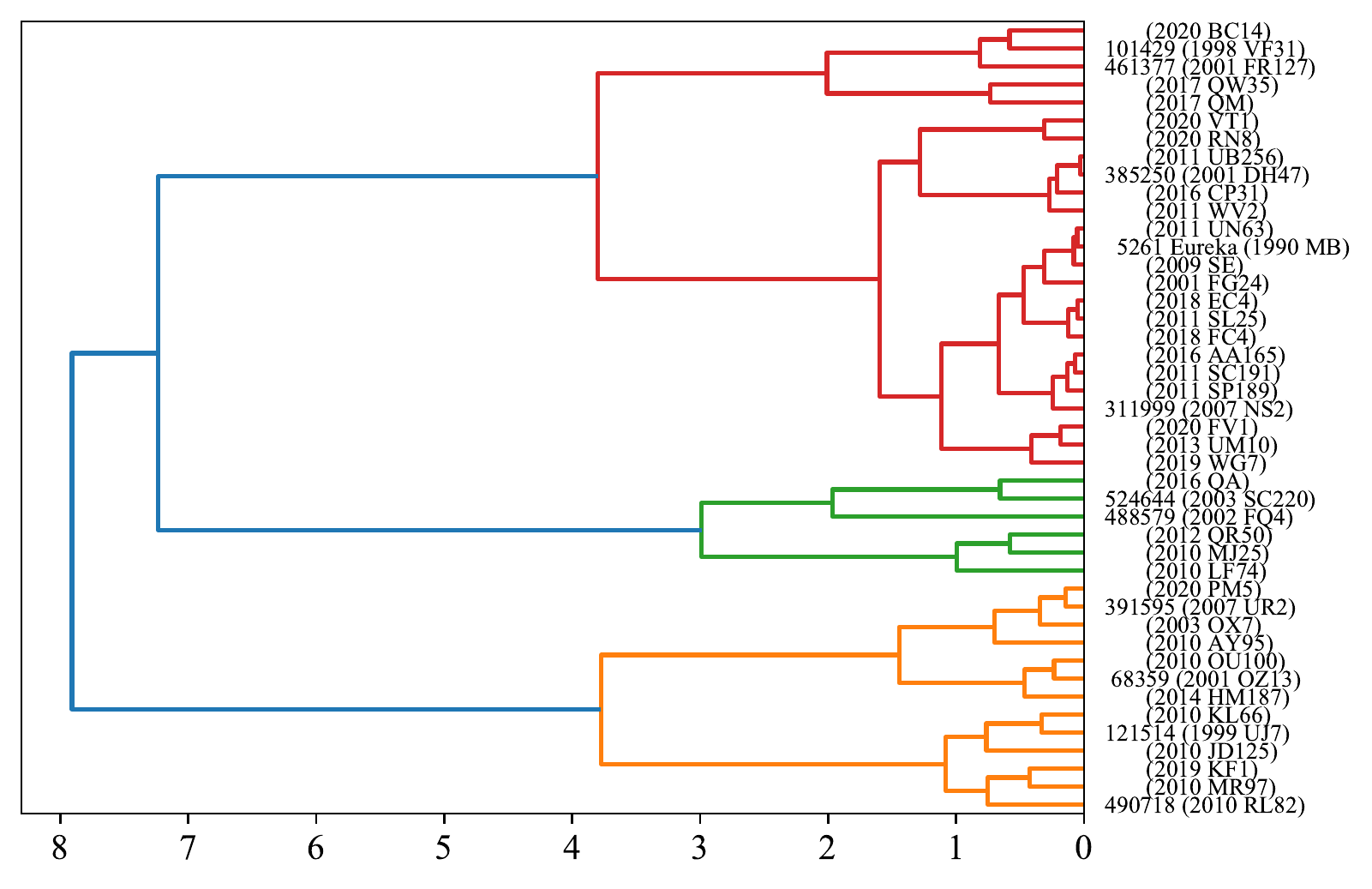}
        \caption{Dendrogram for the data set made of 44 Mars co-orbitals. The data set has been previously standardized by removing the mean
                 and scaling to unit variance using \textsc{Scikit-learn}'s {\tt StandardScaler} class (its method {\tt fit\_transform}). 
                 The $x$-axis shows the Euclidean distance between the clusters or merging distance and the $y$-axis displays the 
                 designations of the objects. This dendrogram has been computed using \textsc{SciPy}'s {\tt dendrogram} and the Ward 
                 variance minimization algorithm of the {\tt linkage} function.
                }
        \label{dendrogramMars}
     \end{figure*}

\begin{figure}
       \centering
        \includegraphics[width=\linewidth]{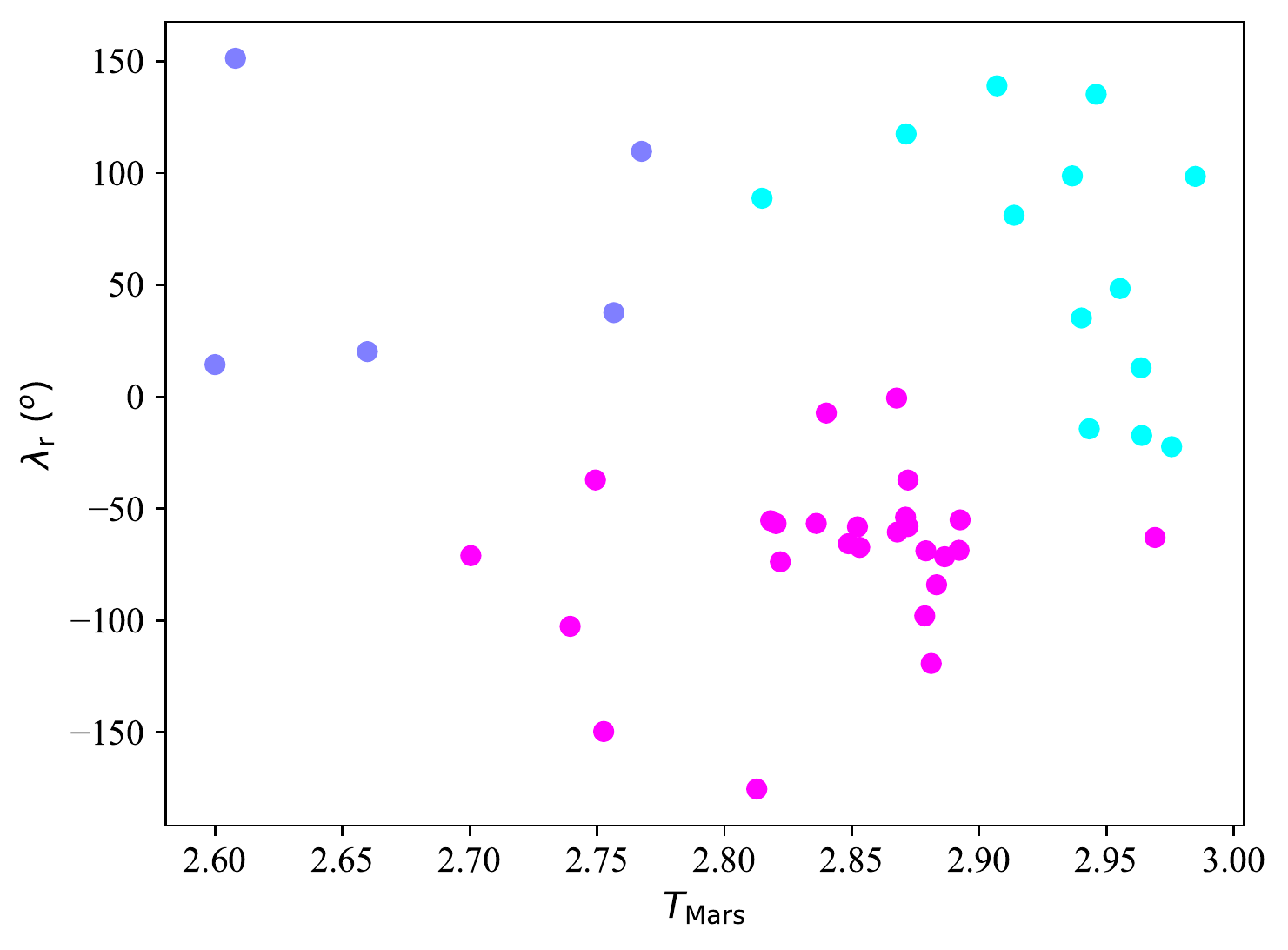}
        \caption{Colour-coded clusters generated by the $k$-means++ algorithm applied to the data set made of 44 Mars co-orbitals. The 
                 plum-coloured points correspond to the green cluster in Fig.~\ref{dendrogramMars}, the pink points correspond to the red 
                 cluster in Fig.~\ref{dendrogramMars}, and the cyan points belong in the orange cluster in Fig.~\ref{dendrogramMars}.  
                }
        \label{clustersMars}
     \end{figure}

        We have used the data set discussed in Section~3 as input to perform the clustering analyses as described above. First, we applied 
        agglomerative hierarchical clustering to obtain the dendrogram shown in Fig.~\ref{dendrogramMars}. The maximal merging distance is
        close to 8 so our threshold to define the final clusters is about 5.6 and we obtain three clusters. The top one, in red, includes all
        the members of the Eureka family in Table~\ref{eureka} and the three new Mars Trojans pointed out in Section~3. The additional L$_5$ 
        Mars Trojan, 101429 (1998~VF$_{31}$), also appears included in this cluster, but in a different, separate branch. The L$_4$ Mars 
        Trojan 121514 (1999~UJ$_7$) is part of a different cluster, the orange one. When applying the $k$-means++ algorithm and the elbow 
        method to the same data set we also obtain three clusters that are shown in Fig.~\ref{clustersMars}. The analysis presented in 
        Section~3 and the one carried out here are somewhat complementary as in Section~3 we discussed one-dimensional clustering via 
        ${\Delta}T_{\rm Mars}$ and here we consider two-dimensional clustering in the $T_{\rm Mars}$-$\lambda_{\rm r}$ plane.
%
%
%
%
%
%
%
%

     \subsection{Earth co-orbitals}
\begin{figure*}
       \centering
        \includegraphics[width=\linewidth]{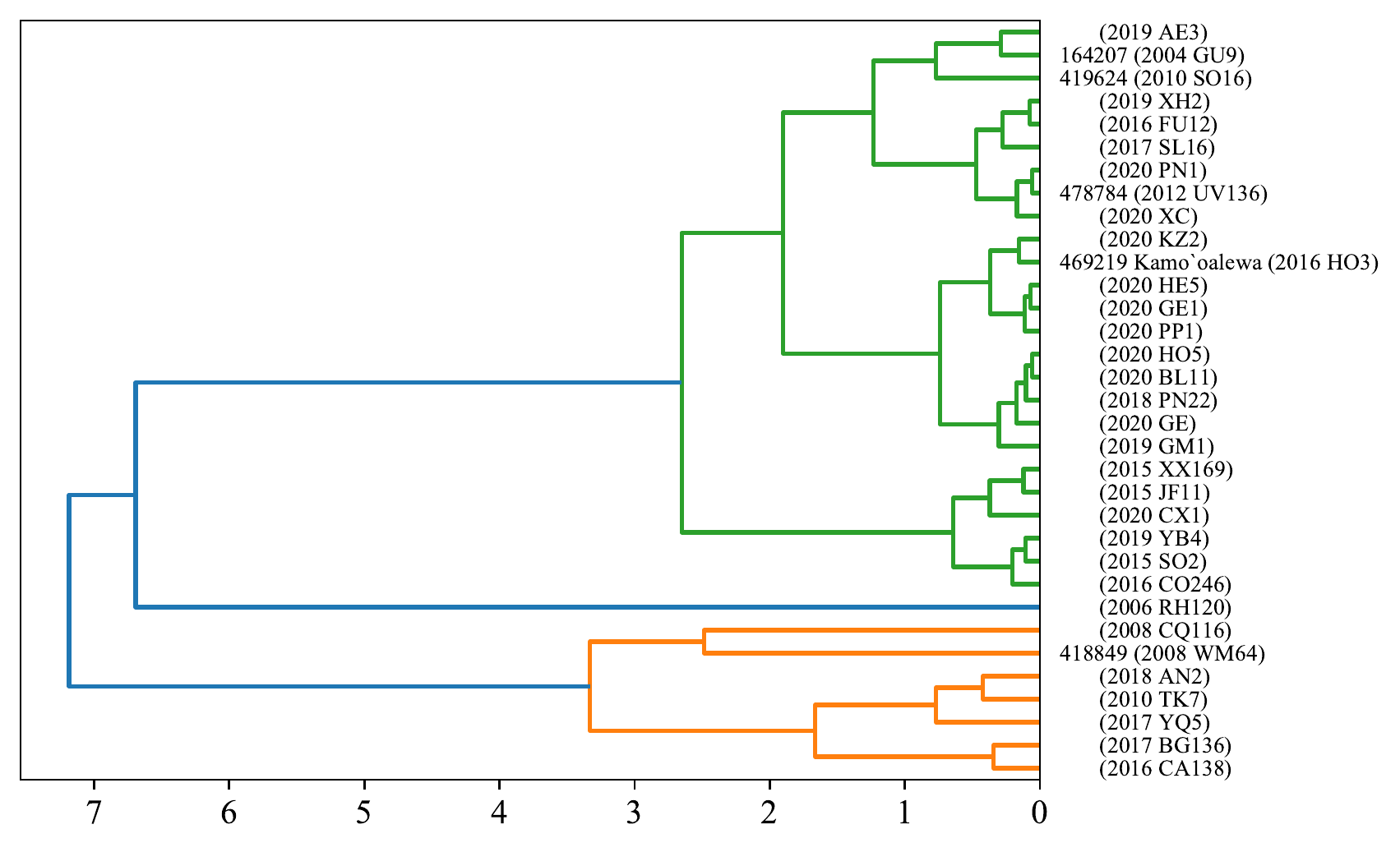}
        \caption{Same as Fig.~\ref{dendrogramMars} but for the data set made of 33 Earth co-orbitals. 
                }
        \label{dendrogramEarth}
     \end{figure*}

\begin{figure}
       \centering
        \includegraphics[width=\linewidth]{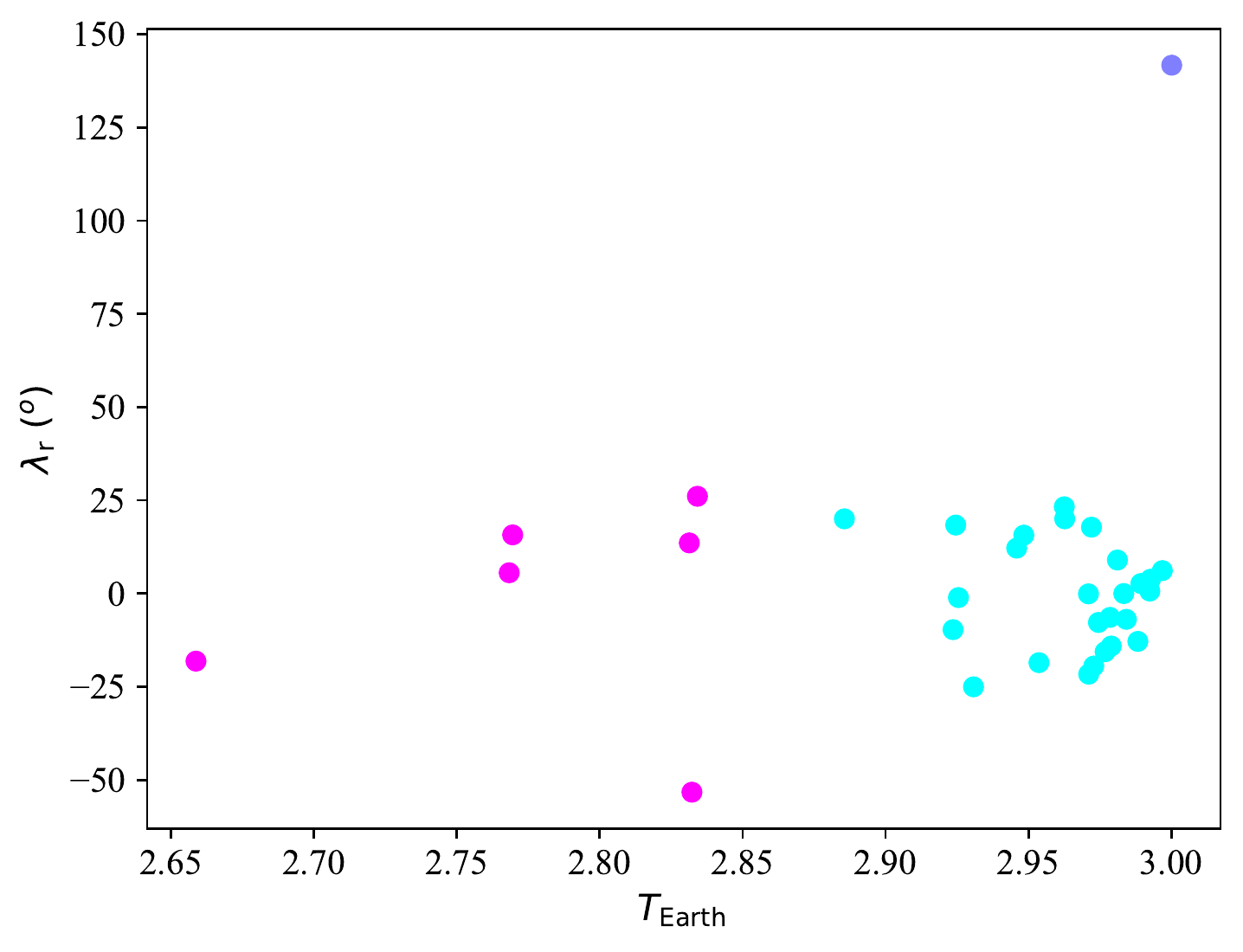}
        \caption{Same as Fig.~\ref{clustersMars} but for the data set made of 33 Earth co-orbitals. The plum-coloured point corresponds to 
                 the outlier 2006~RH$_{120}$ (see the text) that appears in blue in Fig.~\ref{dendrogramEarth}. The pink points correspond 
                 to the orange cluster in Fig.~\ref{dendrogramEarth} that includes the only known Earth Trojan, 2010~TK$_{7}$. The cyan
                 points are associated with the green cluster in Fig.~\ref{dendrogramEarth}.
                }
        \label{clustersEarth}
     \end{figure}

        Figure~\ref{dendrogramEarth} shows the result of the application of the agglomerative hierarchical clustering to the data set 
        discussed in Section~4. The maximal merging distance is slightly above 7 and the threshold to define the final clusters is about 5
        that yields three clusters. The green one includes the vast majority of the 33 objects and both 469219 Kamo`oalewa (2016~HO$_{3}$)
        and 478784 (2012~UV$_{136}$). The outlier in blue is 2006~RH$_{120}$ that is a former temporary satellite of Earth 
        \citep{2008MPEC....D...12B,2009A&A...495..967K} with a value of $T_{\rm Earth}$ very close to 3, which means that this object 
        follows the most Earth-like known orbit. The orange cluster includes the only known Earth Trojan, 2010~TK$_{7}$. The application of 
        the $k$-means++ algorithm to this data set also produces three clusters that are displayed in Fig.~\ref{clustersEarth}. Again, the 
        analysis presented in Section~4 and the one carried out here are somewhat complementary as in Section~4 we discussed one-dimensional 
        clustering via ${\Delta}T_{\rm Earth}$ and here we consider two-dimensional clustering in the $T_{\rm Earth}$-$\lambda_{\rm r}$ 
        plane.   
%
%
%
%
%
%
%
%

\begin{figure}
       \centering
        \includegraphics[width=\linewidth]{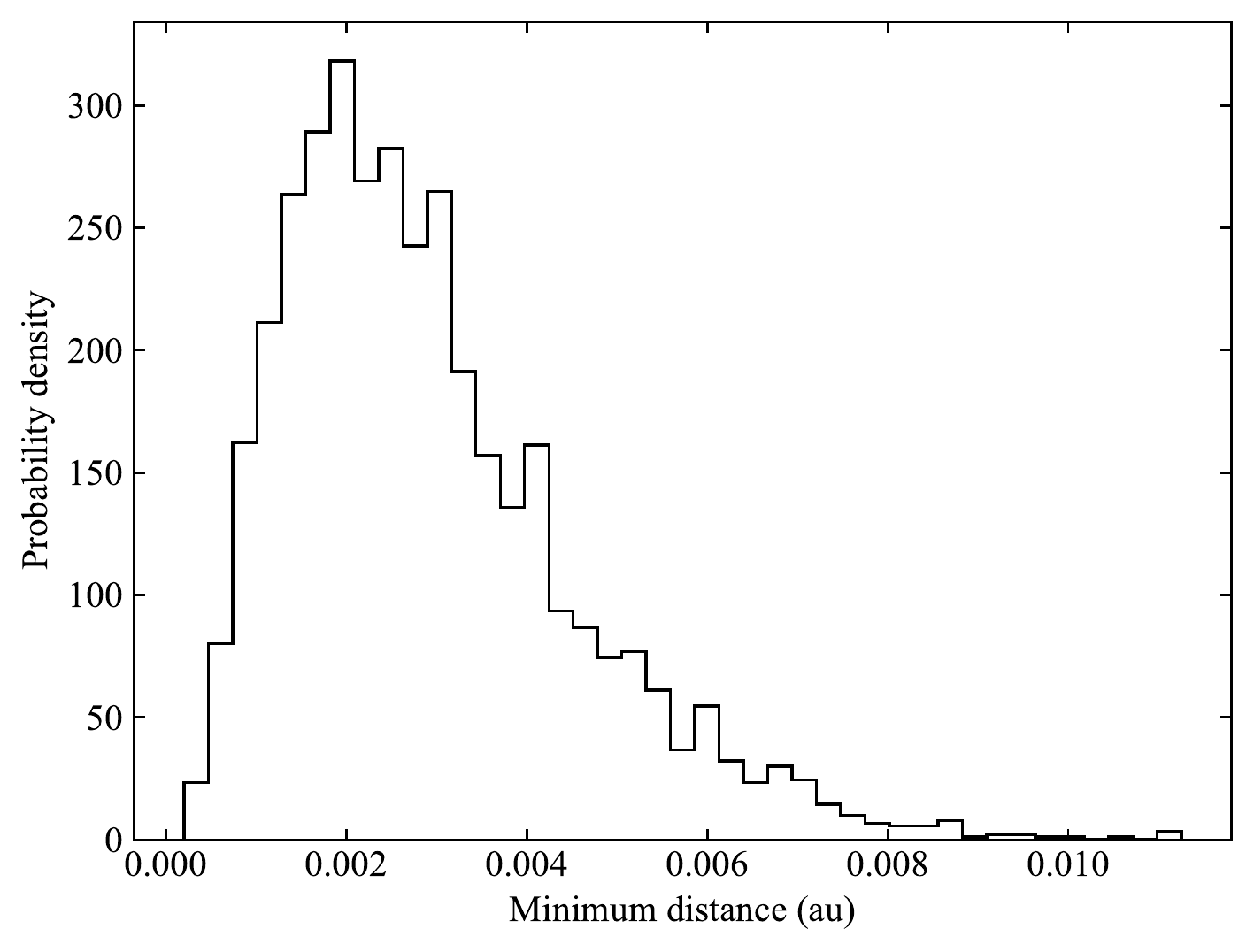}
        \includegraphics[width=\linewidth]{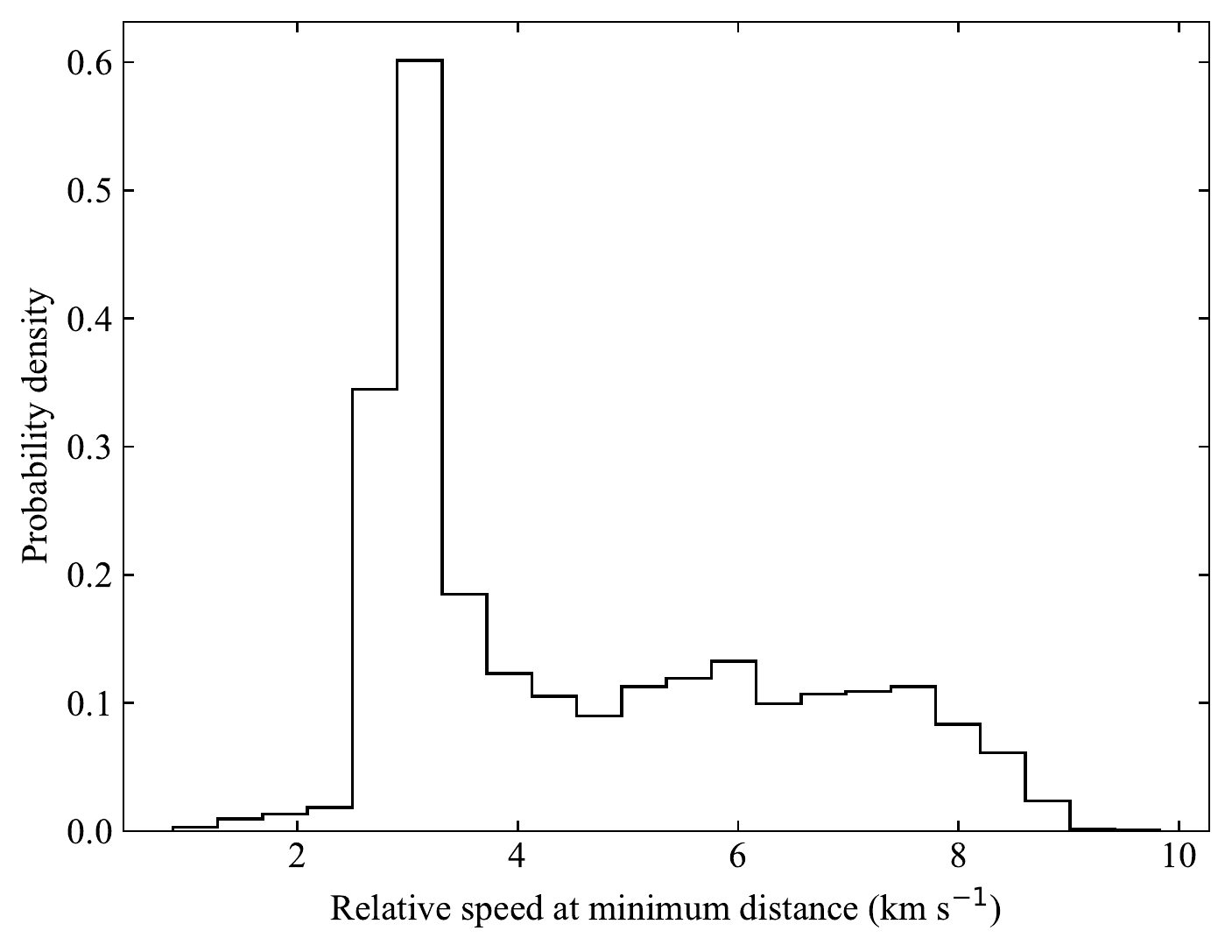}
        \includegraphics[width=\linewidth]{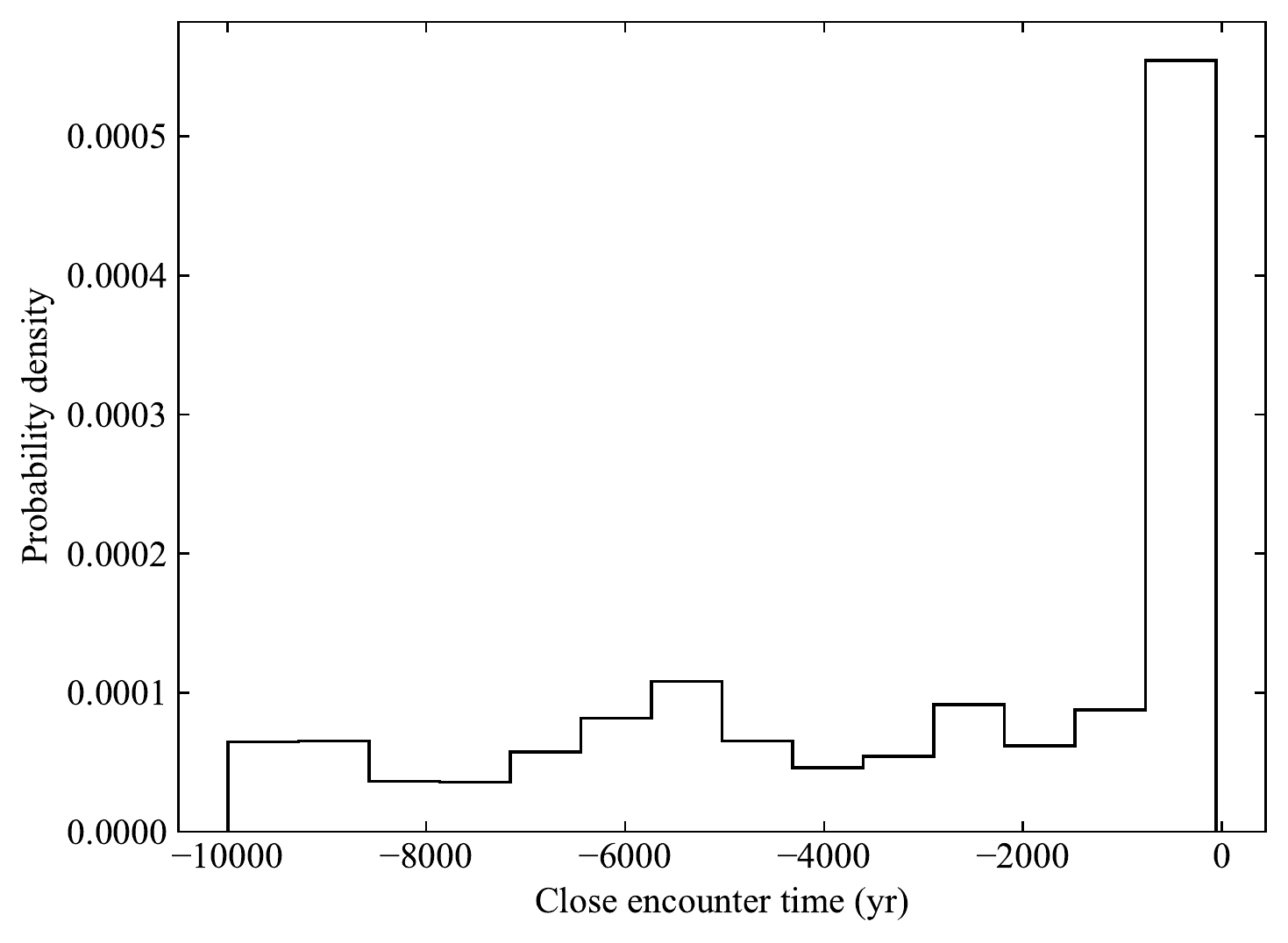}
        \caption{Distribution of minimum approach distance (top panel), relative speed at minimum distance (middle panel), and time at
                 minimum distance (bottom panel) for calculations including 469219 Kamo`oalewa (2016~HO$_{3}$) and 2020~KZ$_{2}$. The origin 
                 of time in the bottom panel is epoch 2459000.5 TDB. Input values from Tables~\ref{vectorKamo} and \ref{vector2020KZ2}.
                }
        \label{KamoKZ2}
     \end{figure}

        The dendrogram in Fig.~\ref{dendrogramEarth} indicates that the known Earth co-orbital closest to Kamo`oalewa in the 
        $T_{\rm Earth}$-$\lambda_{\rm r}$ plane is 2020~KZ$_{2}$. Their orbits are very similar (see Table~\ref{kamolike}, but not as 
        similar as those discussed in \citealt{2019MNRAS.483L..37D}) and if we perform a numerical study analogue to the one carried out for 
        the case of Kamo`oalewa and 2016~FU$_{12}$, we obtain Fig.~\ref{KamoKZ2} that summarizes the results of 3500 experiments. The 
        distributions in Figs~\ref{KamoFU12} and ~\ref{KamoKZ2} are quite different. Although the orbit determinations of both 
        2016~FU$_{12}$ and 2020~KZ$_{2}$ are not as robust as that of Kamo`oalewa (see Table~\ref{kamolike}), the inclusion of the 
        uncertainties in the calculations reveals that 2020~KZ$_{2}$ could be a very recent fragment of Kamo`oalewa: approaches at distances 
        as short as 30\,000~km are possible at relative velocities as low as 900~m~s$^{-1}$ during the last 1000~yr. Again and as pointed 
        out in the previous sections, these values are not typical of pairs resulting from a rotation-induced YORP break-up event (see e.g. 
        \citealt{2010Natur.466.1085P}). Our exploration in Section~4 suggested that 2020~KZ$_{2}$ could be related to 478784 but the 
        agglomerative hierarchical clustering analysis carried out here and summarized in Fig.~\ref{dendrogramEarth} together with the 
        results of the calculations shown in Fig.~\ref{KamoKZ2} support a dynamical connection between Kamo`oalewa and 2020~KZ$_{2}$. 
        However, if they are indeed related, it is unlikely that they formed during a gentle split; a more plausible scenario may be that of 
        two members of the population of Earth co-orbitals experiencing a collision at relatively low speed. Such a collision may initially 
        generate two clouds of fragments moving along the original paths that eventually could spread throughout the colliding orbits. In 
        this scenario (see also the second to last paragraph of Section~4), two fragments undergoing a flyby at a later time may have 
        different surface properties and their relative velocity at the distance of closest approach could be close to the original impact 
        speed as they have independent sources (the two original impactors). 
%
%
%
%

  \section{Discussion}
\begin{figure}
       \centering
        \includegraphics[width=\linewidth]{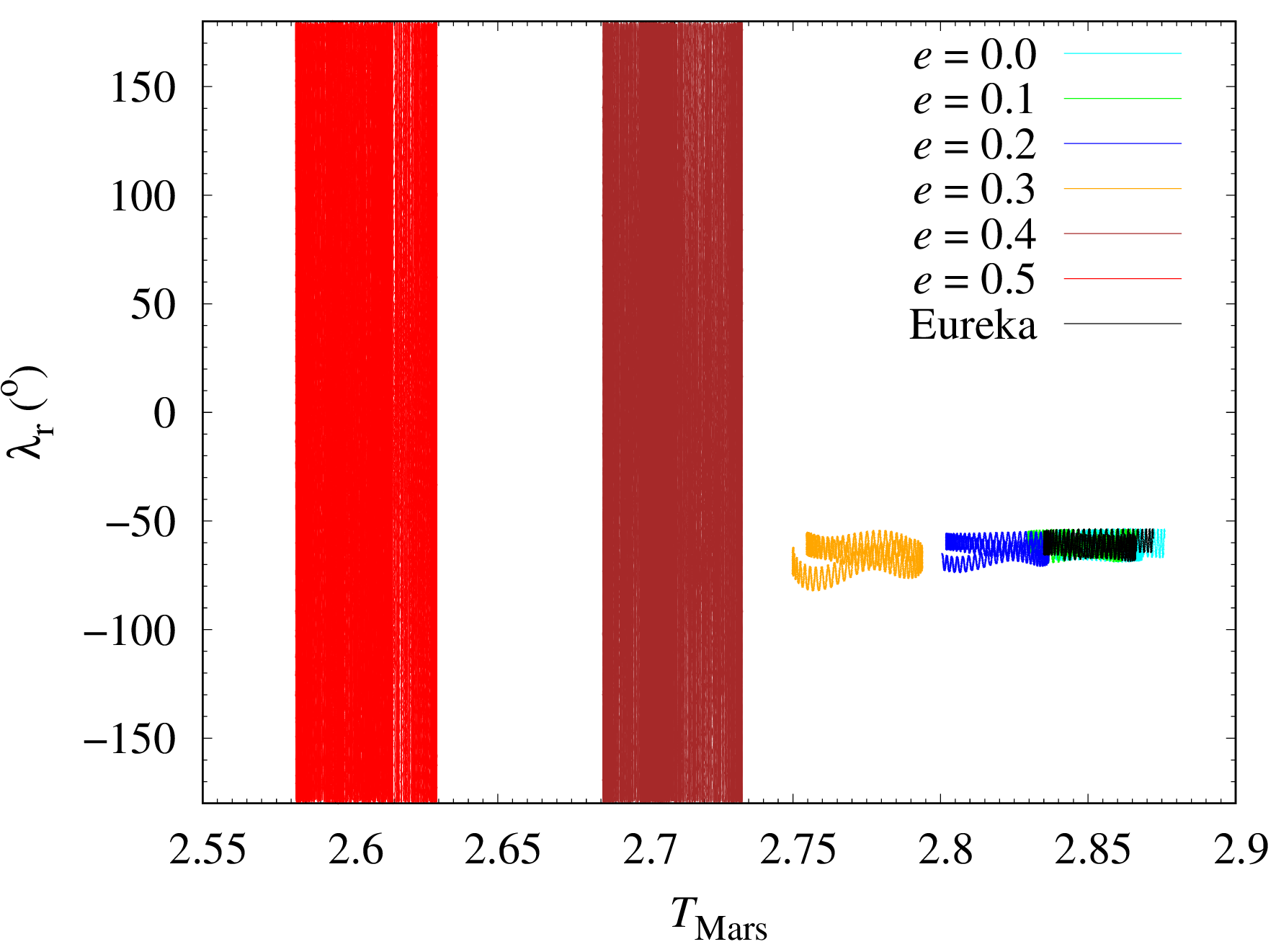}
        \includegraphics[width=\linewidth]{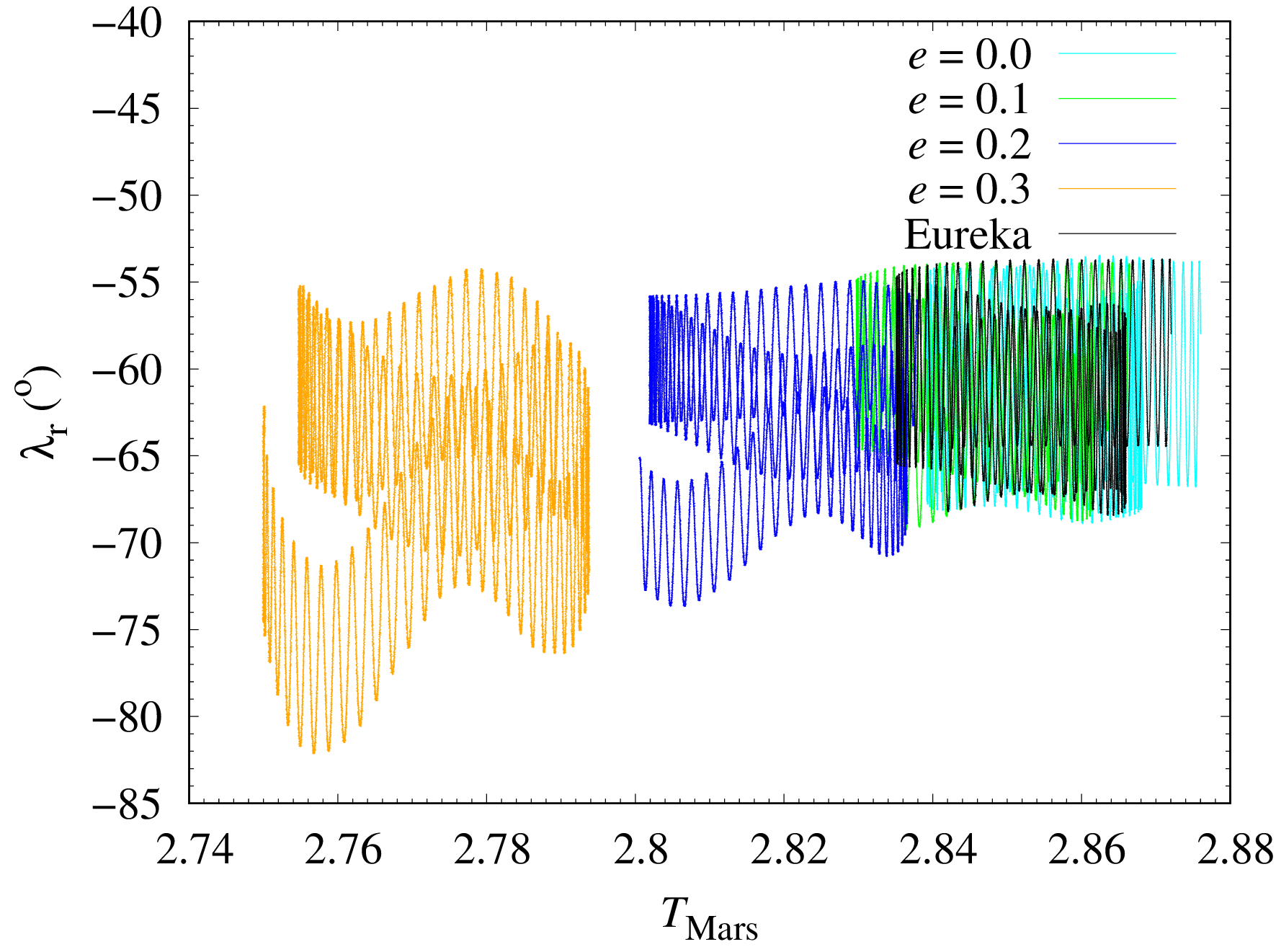}
        \caption{Effect of the eccentricity on the evolution of Eureka-like orbits. The nominal orbit of 5261~Eureka (1990~MB) and those of 
                 six other virtual test objects based on Eureka's nominal orbit with $e=0.0$, 0.1, 0.2, 0.3, 0.4 and 0.5 have been 
                 integrated for 10$^{5}$~yr forward in time. The evolution in the $T_{\rm Mars}$-$\lambda_{\rm r}$ plane is shown in the top 
                 panel. The bottom panel is a magnified version of the top one that focuses on Eureka and test orbits with $e=0.0$, 0.1, 0.2 
                 and 0.3. The output time-step size is 1~yr.}
        \label{ecceffect}
     \end{figure}

     Our analyses follow a novel approach based on the Tisserand parameter that has seldom been used within the context of asteroid family
     studies, although it has a clear potential to help in a shortlisting effort. Two main objections can be made regarding this use: that 
     the Tisserand parameter is a quasi-invariant that vary over a planetary orbital period and that we have restricted somewhat arbitrarily 
     the application to objects with $e<0.2$. In order to understand better the implications of these objections, we have used
     the orbit of 5261~Eureka (1990~MB) as a test case and integrated its nominal orbit for 10$^{5}$~yr forward in time together with five 
     other orbits similar to that of Eureka but with $e=0.0$, 0.1, 0.2, 0.3, 0.4 and 0.5 (Eureka has $e=0.065$). The results of these 
     integrations are displayed, in the $T_{\rm Mars}$-$\lambda_{\rm r}$ plane, in Fig.~\ref{ecceffect}. The top panel of the figure shows 
     that an originally confined path becomes unconfined as the value of the eccentricity increases, supporting our choice of $e<0.2$ to 
     select relevant objects. The Tisserand parameter is relevant to the circular restricted three-body problem that is an idealized 
     situation in which the Sun, a planet, and a massless body interact. Therefore, it assumes that the dynamics of the massless body is
     not significantly influenced by a fourth body. This is very nearly the case of Mars co-orbitals with $e<0.2$, but Earth co-orbitals
     evolve within the dynamical context of the Earth-Moon system although few known objects seem to be significantly affected by the Moon
     (see below). 

     The issue of the quasi-invariant nature of the Tisserand parameter becomes clear when considering the bottom 
     panel of Fig.~\ref{ecceffect}: although the osculating value (the one displayed) changes over time, the average value remains 
     reasonably constant over extended periods of time. The oscillations observed are the result of secular perturbations driven by planets
     other than Mars: the eccentricity fluctuates with a periodicity of about 200\,000~yr and the inclination oscillates on a time-scale of
     600\,000~yr \citep{1994CeMDA..58...53M}. Its eccentricity oscillates mainly due to secular resonances with the Earth and the 
     oscillation in inclination appears to be driven by secular resonances with Jupiter (see e.g. \citealt{2013MNRAS.432L..31D}). 
     Figure~\ref{ecceffect}, bottom panel, provides a good context to understand the clusterings discussed in the previous sections: any 
     co-orbital body that splits tends to have its resulting fragments confined within short distance in the $T_{\rm P}$-$\lambda_{\rm r}$ 
     plane. It is however possible that a similar arrangement may arise from a purely dynamical mechanism, without any fragmentation 
     involved, but it probably requires a very large source population, large enough to facilitate the process of capture inside the 
     co-orbital resonance with very similar values of $T_{\rm P}$ that have very low intrinsic probabilities (see the values in Sections~3 
     and 4).
%
%
%
%

     The dynamical sampling displayed in Fig.~\ref{ecceffect} throws some light on the origin of the overall distribution of objects in 
     Figs~\ref{lambdarTM} and ~\ref{lambdarTE}. New objects in the form of fragments may be formed with initially low values of the 
     eccentricity that may increase secularly leading to smaller values of $T_{\rm Mars}$ or $T_{\rm Earth}$, transforming co-orbitals into 
     passing objects whose position relative to the host planet is no longer controlled by the 1:1 mean-motion resonance. This evolutionary
     pathway appears particularly clear in the case of Mars co-orbitals (compare Figs~\ref{lambdarTM} and ~\ref{ecceffect}); unfortunately, 
     observational bias prevents a better assessment in the case of Earth co-orbitals (see Fig.~\ref{lambdarTE}) because the geometry of 
     ground-based observations precludes the discovery of objects with mean longitudes different from that of Earth. It is however possible 
     that unrelated objects or interlopers may also be present, creating pairs of objects, without any previous common dynamical history, 
     whose values of the Tisserand parameter differ by very small amounts just by chance. 

\begin{figure}
       \centering
        \includegraphics[width=\linewidth]{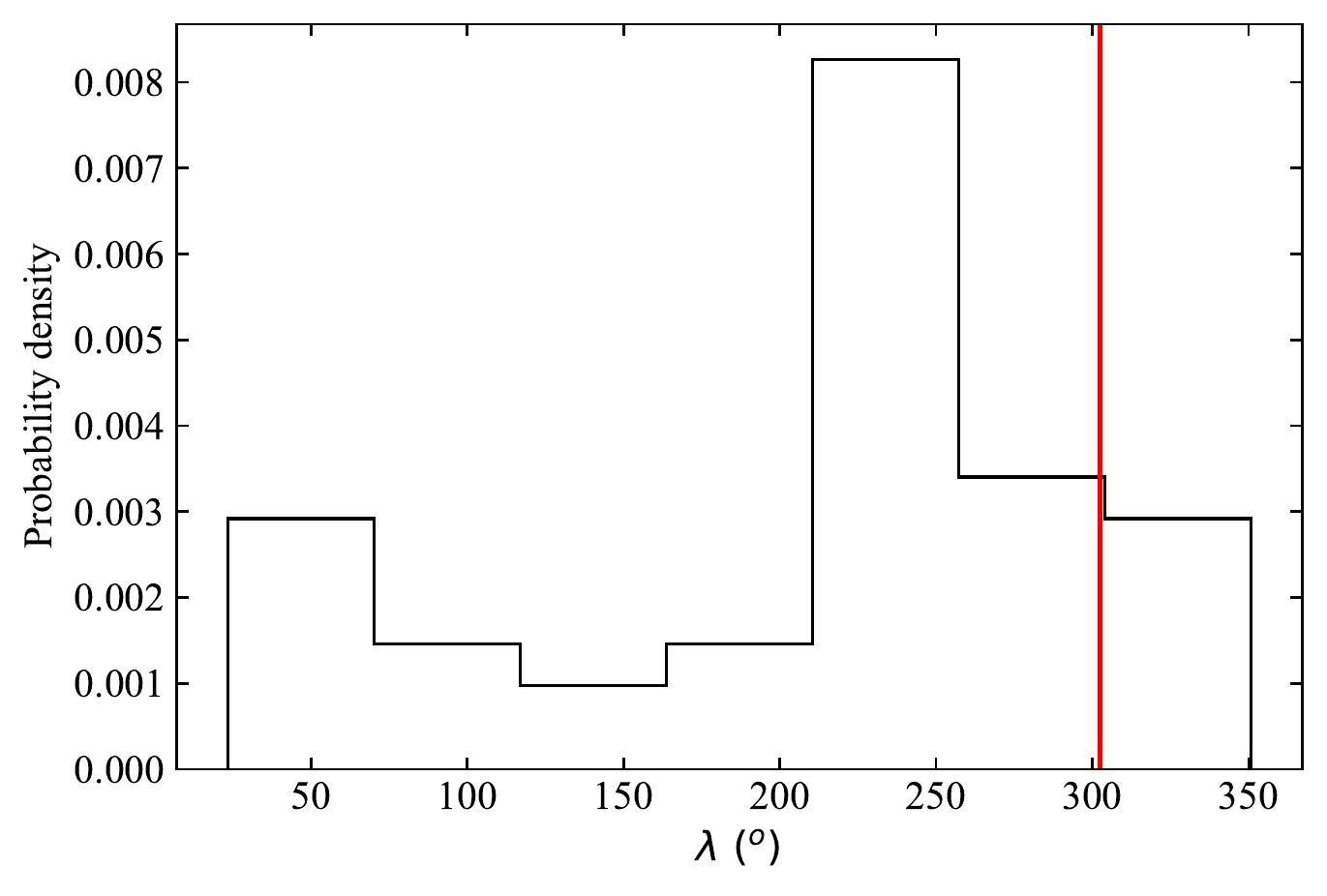}
        \includegraphics[width=\linewidth]{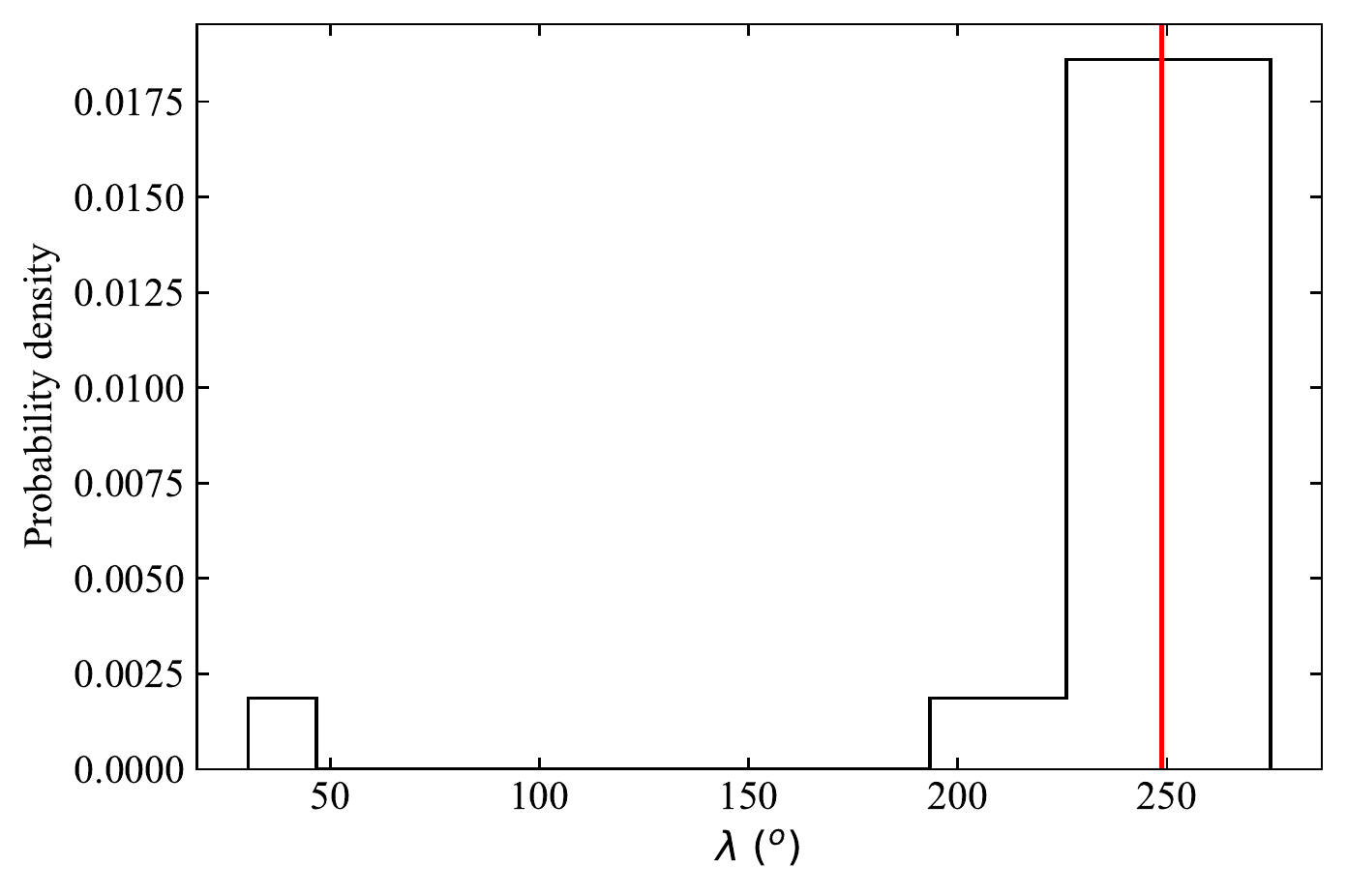}
        \caption{Distribution of mean longitudes for the samples of Mars co-orbitals (top panel) and Earth co-orbitals (bottom panel) 
                 discussed in this work. The mean longitudes of Mars and Earth are shown in red. The geometry of ground-based observations 
                 seems to make very difficult the discovery of objects with mean longitudes different from that of Earth. 
                }
        \label{lambdaM}
     \end{figure}

     The median value of ${\Delta}T_{\rm Earth}$ is lower than that of ${\Delta}T_{\rm Mars}$, which suggest that fragmentations may be more
     frequent in Earth co-orbital space. However, when we consider the histograms of mean longitudes in Fig.~\ref{lambdaM}, the sample of
     Earth co-orbitals appears to be biased in favour of those objects with mean longitudes close to that of Earth. This may however signal
     that fragmentations are mainly linked to quasi-satellites or asymmetric horseshoes or, more likely, to those objects that experience 
     recurrent transitions between the quasi-satellite and horseshoe resonant states (perhaps because they are intrinsically more numerous). 
     On the other hand, one object appears in both Table~\ref{kamo} and \ref{uv136} and this is more consistent with cascades of disruptions 
     than isolated events, a scenario favoured by Fatka et al. (\citeyear{2020Icar..33813554F}). However, the most simple interpretation is 
     that the data for the case of Earth may be strongly biased towards objects with low eccentricity and $\lambda_{\rm r}\sim0{\degr}$ as 
     these small bodies are physically the closest to the Earth and therefore more likely to be discovered. 
%
%
%
%

     \citet{2020Icar..33513370C} argued that fragments produced during YORP-induced break-up events should be called YORPlets because the 
     YORP mechanism produces them. Although 54509~YORP (2000~PH$_{5}$) itself, an irregular horseshoe to Earth that experiences the effects 
     of the YORP mechanism, has been left out of our analysis because it has $e>0.2$, applying the same approach as before, we obtain 
     Table~\ref{yorp} that includes 2015~JF$_{11}$ that has a very low ${\Delta}T_{\rm Earth}$ with respect to YORP.   
%
%
      \begin{table}
        \centering
        \fontsize{8}{11pt}\selectfont
        \tabcolsep 0.15truecm
        \caption{Similar to Table~\ref{kamo} but for YORP.
                }
        \begin{tabular}{rccc}
          \hline
             Object                    & $T_{\rm Earth}$ & $\lambda_{\rm r}$ (\degr) & ${\Delta}T_{\rm Earth}$ \\
          \hline
            54509~YORP (2000~PH$_{5}$) & 2.945839220     & $-$75.20226               &   --                    \\
                       2014~HL$_{199}$ & 2.945164475     &     5.57231               & 0.000674745             \\ 
                       2015~JF$_{11}$  & 2.945883357     &    12.18048               & 0.000044137             \\  
                       2015~XX$_{169}$ & 2.948366642     &    15.66149               & 0.002527422             \\ 
          \hline
        \end{tabular}
        \label{yorp}
      \end{table}
%
%
 
     One may contend that some objects mentioned above have poor orbit determinations: some data-arcs are as short as a few days. This is 
     indeed true for many Earth co-orbitals that are dim and can only be sparsely observed. However and given the functional form of the 
     Tisserand parameter in equation~(\ref{Tisserand}), even relatively poor orbit determinations may provide robust values of this 
     parameter. The new co-orbitals identified in this research are further studied below.

     \subsection{New Mars co-orbitals} 
        The three new Trojans follow tadpole orbits around Mars' L$_5$; the candidate horseshoe librator follows a compound orbit that also
        encompasses Mars on one side (see Fig.~\ref{newtrojans}, bottom panel). In addition to the Eureka family members, one additional 
        L$_5$ Mars Trojan, 101429 (1998~VF$_{31}$), and one L$_4$, 121514 (1999~UJ$_7$) are known. Trojan 101429 was identified by 
        \citet{1999ApJ...517L..63T} and it is not as stable as those part of the Eureka family (see e.g. fig.~5 in 
        \citealt{2013MNRAS.432L..31D}). In addition to having different orbital behaviour \citep{2013Icar..224..144C,2013MNRAS.432L..31D}, 
        spectroscopic studies have confirmed that the physical nature of 101429 is different from that of most Eureka family members 
        \citep{2007Icar..192..434R,2021Icar..35413994C}. The surface mineralogy of 121514 is also different \citep{2019EPSC...13.1254B}. The
        three new Mars Trojans identified in this work are mostly unstudied objects and the only data available about them are astrometric
        observations and their associated photometry. Small body 2020~VT$_{1}$ joins the subset of non-Trojan Mars co-orbitals discussed by
        \citet{2005P&SS...53..617C}, although their objects move in more eccentric orbits. As a very recent discovery, it remains unstudied.        
%
%
         \begin{table*}
          \fontsize{8}{11pt}\selectfont
          \tabcolsep 0.10truecm
          \caption{Heliocentric Keplerian orbital elements of 2009~SE, 2018~EC$_{4}$, 2018~FC$_{4}$, and 2020~VT$_{1}$ used in this study. 
                   The orbit determination of 2009~SE is based on 56 observations spanning a data-arc of 3133~d or 8.58~yr (solution date, 
                   2018-Apr-23 02:36:33 PDT), the one of 2018~EC$_{4}$ is based on 70 observations spanning a data-arc of 3131~d or 8.57~yr 
                   (solution date, 2020-May-30 05:28:03 PDT), and that of 2018~FC$_{4}$ is based on 35 observations spanning a data-arc of 
                   790~d (solution date, 2020-Nov-11 09:52:11 PST). The orbit determination of 2020~VT$_{1}$is based on 28 observations 
                   spanning a data-arc of 24~d (solution date, 2020-Nov-15 04:55:38 PST). Values include the 1$\sigma$ uncertainty. The 
                   orbit determinations have been computed at epoch JD 2459000.5 that corresponds to 00:00:00.000 TDB on 2020 May 31 
                   (J2000.0 ecliptic and equinox). Source: JPL's SBDB.
                  }
          \begin{tabular}{lccccc}
           \hline
            Orbital parameter                                 &   & 2009~SE                   & 2018~EC$_{4}$               & 2018~FC$_{4}$             &
                                                                    2020~VT$_{1}$  \\ 
           \hline
            Semimajor axis, $a$ (au)                          & = &   1.524472$\pm$0.000002   &   1.52357630$\pm$0.00000007 &   1.5238457$\pm$0.0000013 &
                                                                      1.5231$\pm$0.0003       \\
            Eccentricity, $e$                                 & = &   0.0650794$\pm$0.0000010 &   0.06052671$\pm$0.00000012 &   0.017077$\pm$0.000006   &
                                                                      0.16702$\pm$0.00010     \\
            Inclination, $i$ (\degr)                          & = &  20.6248$\pm$0.0002       &  21.835796$\pm$0.000013     &  22.1437$\pm$0.0002       &
                                                                     18.717$\pm$0.009         \\
            Longitude of the ascending node, $\Omega$ (\degr) & = &   6.82030$\pm$0.00005     &  47.371564$\pm$0.000010     & 187.55390$\pm$0.00003     &
                                                                     50.169$\pm$0.003         \\
            Argument of perihelion, $\omega$ (\degr)          & = & 354.156$\pm$0.010         & 344.1754$\pm$0.0004         &  52.009$\pm$0.007         &
                                                                    296.191$\pm$0.012         \\
            Mean anomaly, $M$ (\degr)                         & = & 240.916$\pm$0.012         & 203.4934$\pm$0.0005         &   4.660$\pm$0.007         &
                                                                    315.410$\pm$0.013         \\
            Perihelion, $q$ (au)                              & = &   1.425261$\pm$0.000002   &   1.43135923$\pm$0.00000015 &   1.497823$\pm$0.000009   &
                                                                      1.26868$\pm$0.00013     \\
            Aphelion, $Q$ (au)                                & = &   1.623684$\pm$0.000002   &   1.61579336$\pm$0.00000007 &   1.5498684$\pm$0.0000013 &
                                                                      1.7774$\pm$0.0004       \\
            Absolute magnitude, $H$ (mag)                     & = &  19.9$\pm$0.4             &  20.1$\pm$0.4               &  21.3$\pm$0.4             &
                                                                     22.9$\pm$0.3             \\
           \hline
          \end{tabular}
          \label{elementsnewtro}
         \end{table*}
%
%

\begin{figure}
       \centering
        \includegraphics[width=\linewidth]{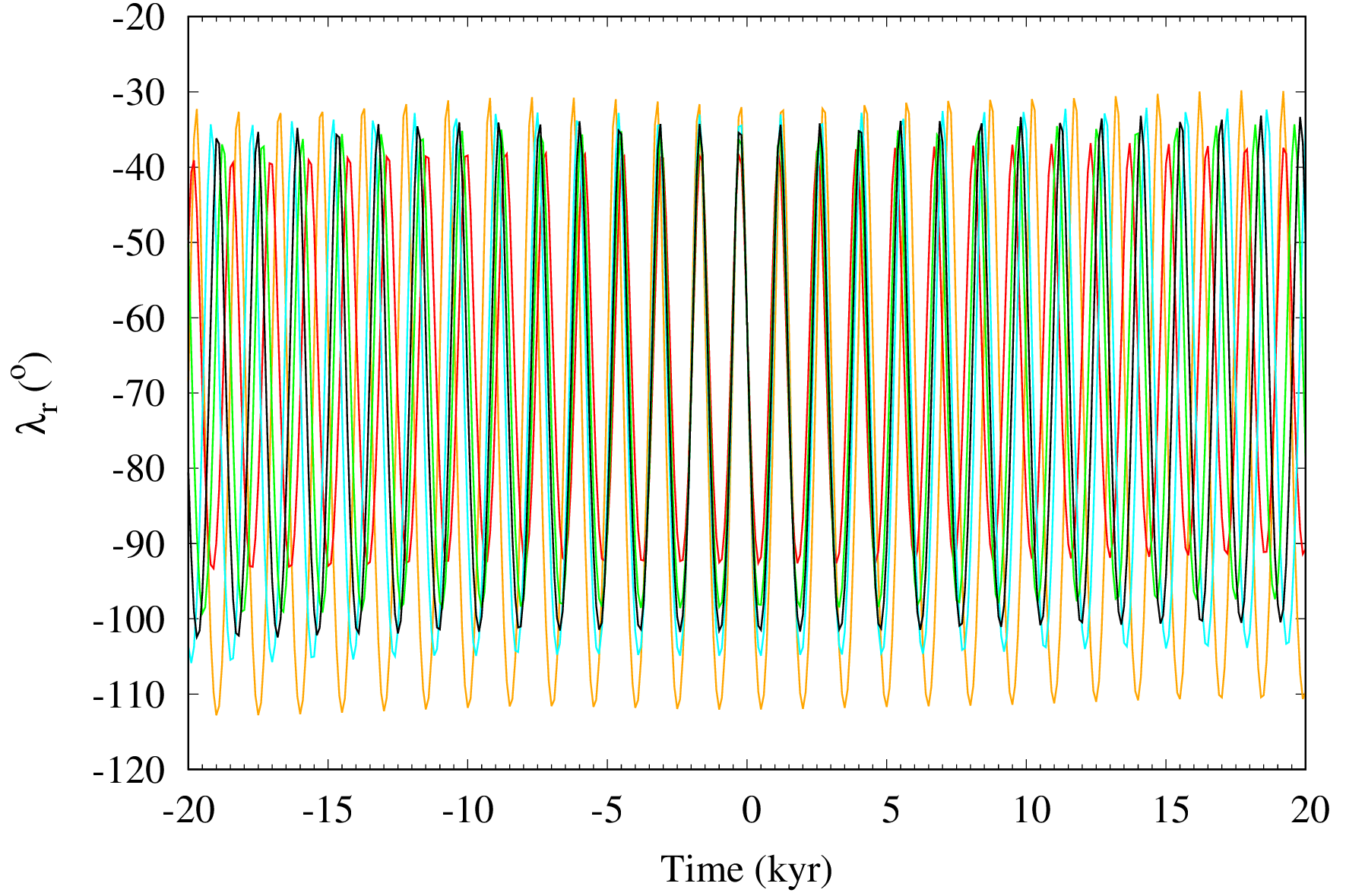}
        \includegraphics[width=\linewidth]{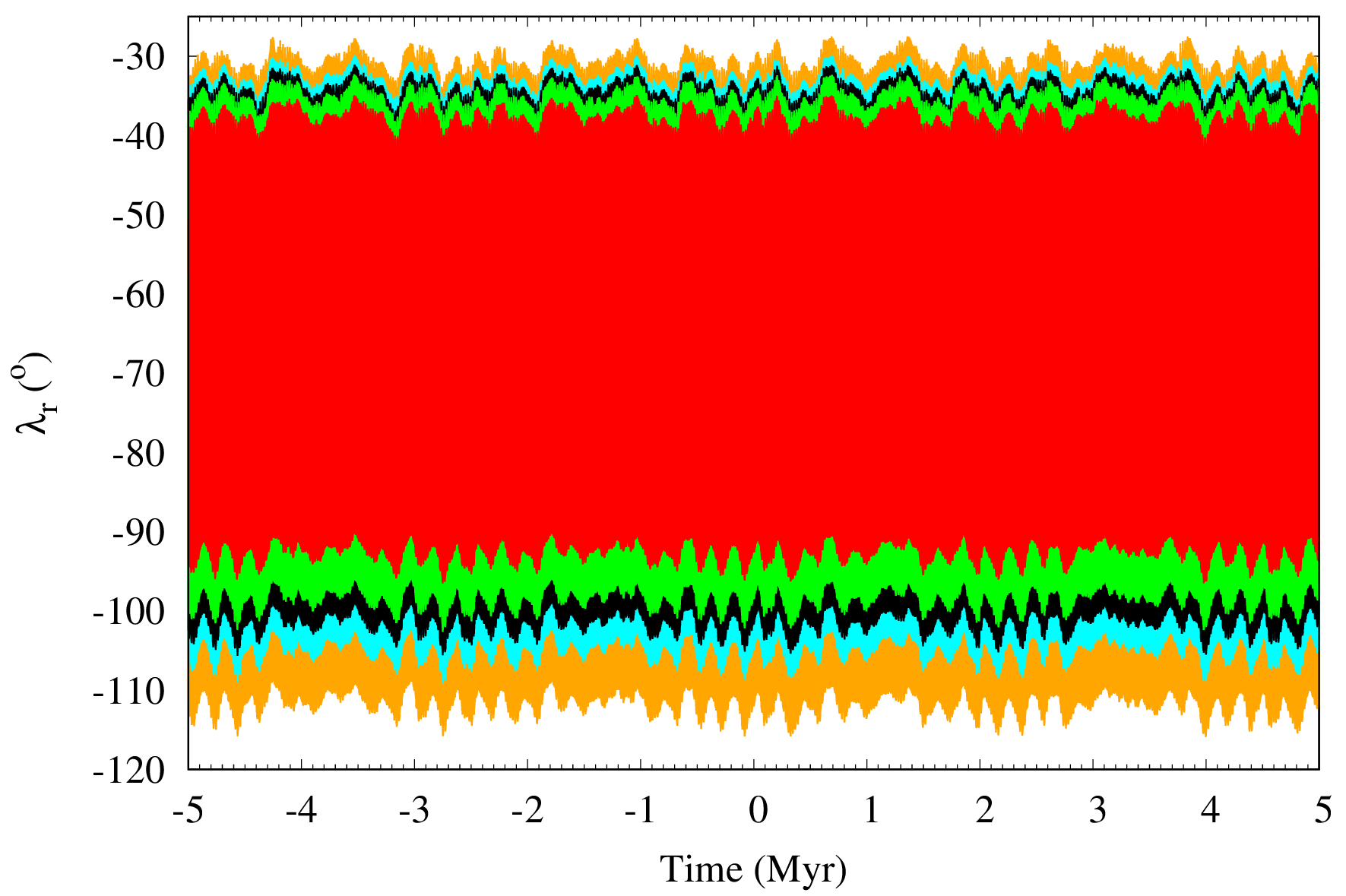}
        \caption{Evolution of the relative mean longitude with respect to Mars of 2009~SE. The time interval (-20, 20)~kyr is shown in the
                 top panel; the bottom panel displays the full calculations spanning 10~Myr. Our calculations include the nominal orbit (in 
                 black) as described by the orbit determination in Table~\ref{elementsnewtro} and those of control orbits or clones with 
                 Cartesian vectors separated $+$3$\sigma$ (in green), $-$3$\sigma$ (in cyan), $+$9$\sigma$ (in red), and $-$9$\sigma$ (in 
                 orange) from the nominal values in Table~\ref{vector2009SE}. The output time-step size is 100~yr.   
                 }
        \label{2009SE}
     \end{figure}

        {\bf 2009~SE}. It was discovered on 2009 September 16 by the Catalina Sky Survey 
        (CSS).\footnote{\href{http://www.lpl.arizona.edu/css/css\_facilities.html}{http://www.lpl.arizona.edu/css/css\_facilities.html}} 
        Its orbit determination is robust as it is based on 56 observations spanning a data-arc of 3133~d or 8.58~yr (see 
        Table~\ref{elementsnewtro}). The size, shape and orientation of its orbit is similar to those of known L$_5$ Mars Trojans. It is a 
        relatively bright object, suitable for future spectroscopic observations at an apparent visual magnitude of nearly 20. Its absolute 
        magnitude is $H$=19.9~mag (assumed $G=0.15$), which suggests a diameter in the range $\sim$150--1200~m for an assumed albedo in the 
        range 0.60--0.01. \citet{2018AJ....156...60M} have shown that albedos of $\sim$0.01 are possible but fairly rare; therefore, as 
        asteroid size calculations are more sensitive to the lower bound than to the upper bound -- because the albedo appears in the 
        denominator -- for a given range, smaller sizes are far more likely. Our calculations indicate that it is a stable L$_5$ Mars 
        Trojan, see Fig.~\ref{2009SE}. From the nominal orbit (in black), its relative mean longitude oscillates around $-$60{\degr} with an 
        amplitude of 70{\degr} and a libration period of 1430~yr, see Fig.~\ref{2009SE}, top panel. These values are comparable to those of 
        101429 and 121514 (see e.g. \citealt{2013MNRAS.432L..31D}). Figure~\ref{2009SE}, bottom panel, also shows the modulation with 
        periodicity of about 200\,000~yr pointed out originally by \citet{1994CeMDA..58...53M} in the case of Eureka. Although its orbital 
        evolution appears to be very stable on Myr time-scales (see Fig.~\ref{2009SE}, bottom panel), the libration amplitude is much larger 
        than that of Eureka and this fact may make its long-term evolution less stable. Such results strongly suggest that 2009~SE, like 
        101429 and 121514, is not a member of the Eureka family, although a spectroscopic confirmation is still required. $N$-body 
        simulations spanning Gyr time-scales show that both 101429 and 121514 may not be primordial Trojans but perhaps were captured about 
        4~Gyr ago (see e.g. \citealt{2013MNRAS.432L..31D}). Based on the analysis carried out here and what we already knew about the likely 
        past orbital evolution of 101429 and 121514, we argue that 2009~SE may also be a captured object, not related to Eureka. This 
        conclusion is robust when considering the relatively small uncertainties of the orbit determination in Table~\ref{elementsnewtro} 
        (see Fig.~\ref{2009SE}, bottom panel, evolution of control orbits with Cartesian vectors separated from the nominal one).
        \hfil\par
%
%
%
%

\begin{figure}
       \centering
        \includegraphics[width=\linewidth]{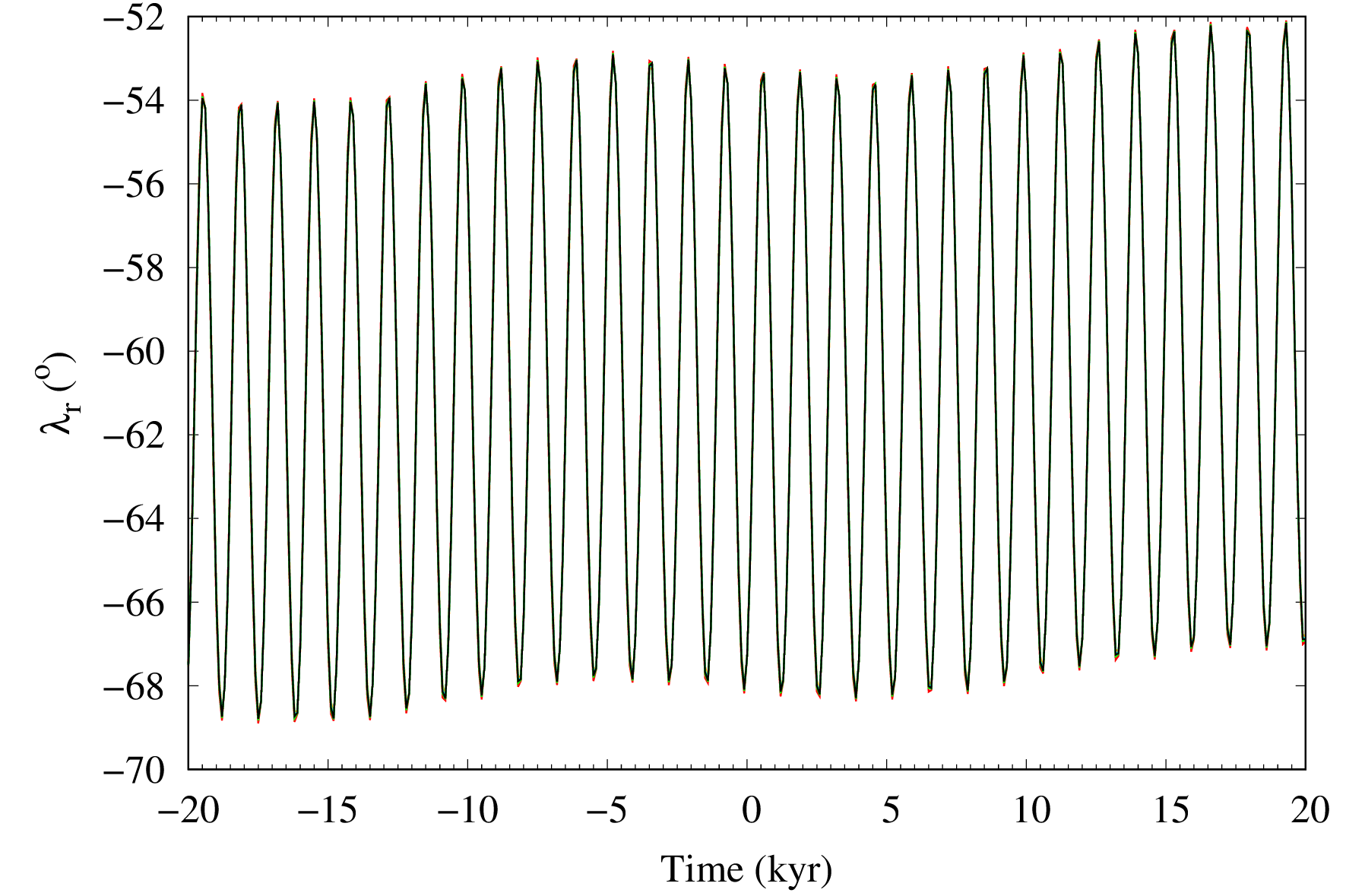}
        \includegraphics[width=\linewidth]{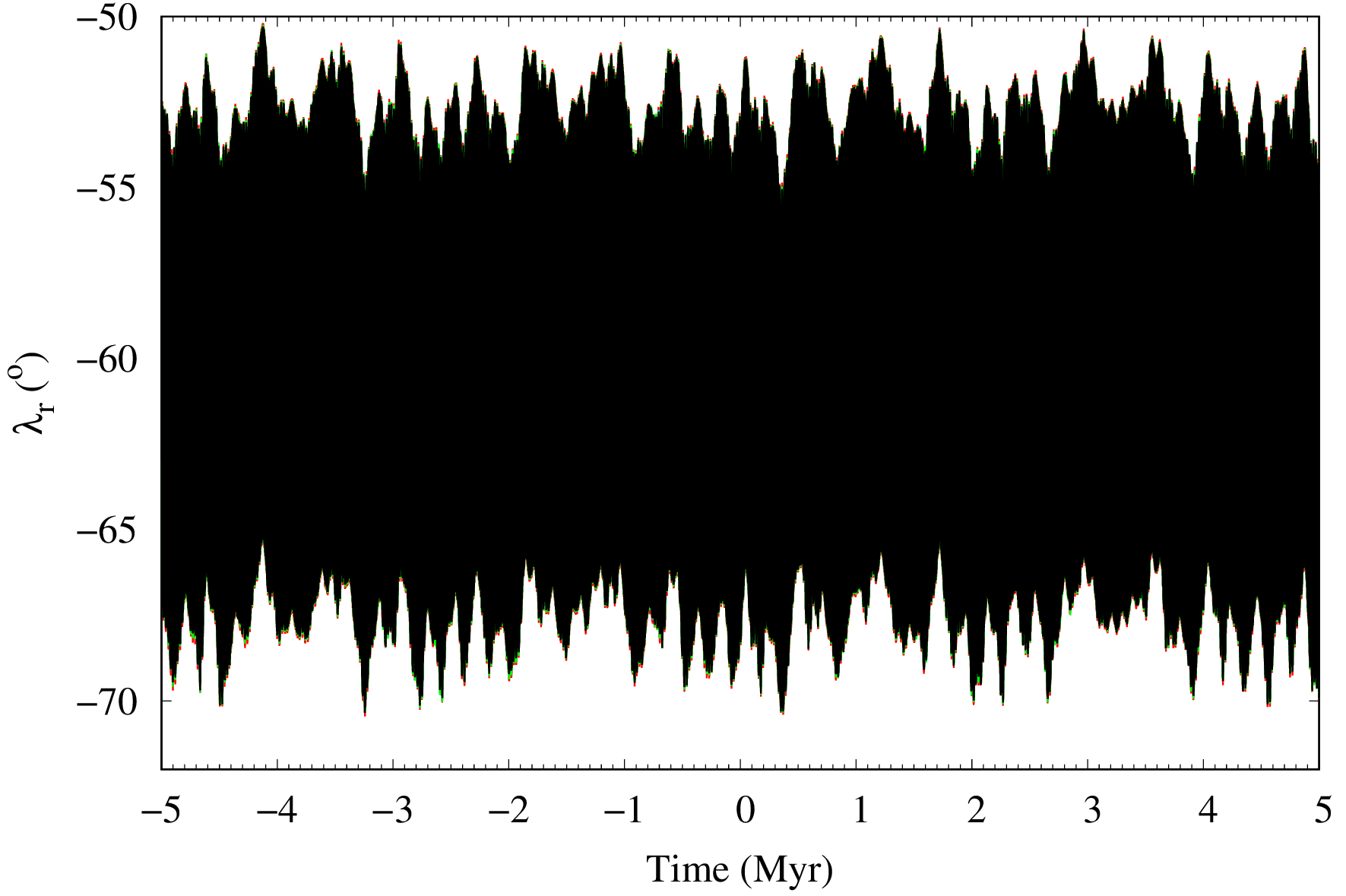}
        \caption{Same as Fig.~\ref{2009SE} but for 2018~EC$_{4}$. In this case, even orbits separated $\pm$9$\sigma$ from the nominal one
                 produce virtually identical evolution on Myr time-scales. Input values from Table~\ref{vector2018EC4}.
                 }
        \label{2018EC4}
     \end{figure}

        {\bf 2018~EC$_\mathbf{4}$}. Although this object was first observed on 2011 October 29 by the Pan-STARRS~1 telescope system at 
        Haleakala, it remained unidentified until it was discovered by the Mt. Lemmon Survey (which is part of CSS) on 2018 March 10. As in 
        the case of 2009~SE, its orbit determination is robust and based on 70 observations spanning a data-arc of 3131~d or 8.57~yr (see
        Table~\ref{elementsnewtro}). Its orbit as well as apparent brightness and size are similar to those of 2009~SE. Our calculations 
        also indicate that it is a stable L$_5$ Mars Trojan, see Fig.~\ref{2018EC4}. The evolution of the nominal orbit (in black) shows 
        that its relative mean longitude oscillates around $-$60{\degr} with an amplitude of 17{\degr} and a libration period of 1250~yr, 
        see Fig.~\ref{2018EC4}, top panel. These values are similar to those of 2011~SC$_{191}$ that may have been trapped at the Lagrangian 
        L$_5$ point of Mars since the formation of the Solar system (see e.g. \citealt{2013MNRAS.432L..31D}). {\'C}uk et al. 
        (\citeyear{2015Icar..252..339C}) have argued that 2011~SC$_{191}$ may have been the first fragment ejected from Eureka. If distance 
        from Eureka in the $T_{\rm Mars}$-$\lambda_{\rm r}$ plane (see Fig.~\ref{Eurekafamily}) could be considered as an indication of the 
        time passed since a given fragmentation event took place, then 2011~SC$_{191}$ is one of the most distant members of the Eureka 
        family together with 2011~SP$_{189}$ and 2018~EC$_{4}$ (but see also Fig.~\ref{dendrogramMars}) and they may have been produced 
        first within the hierarchy of disintegrations induced by multiple YORP-induced spin-ups. That 2018~EC$_{4}$ is a probable member of 
        the Eureka family is a robust conclusion when considering the small uncertainties of the orbit determination in 
        Table~\ref{elementsnewtro} (see Fig.~\ref{2018EC4}, bottom panel, evolution of control orbits with Cartesian vectors separated from 
        the nominal one) although a spectroscopic confirmation is still required. Figure~\ref{2018EC4}, bottom panel, also hints at the 
        modulation with periodicity of about 200\,000~yr pointed out originally by \citet{1994CeMDA..58...53M} in the case of Eureka.  
        \hfil\par
%
%
%
%

\begin{figure}
       \centering
        \includegraphics[width=\linewidth]{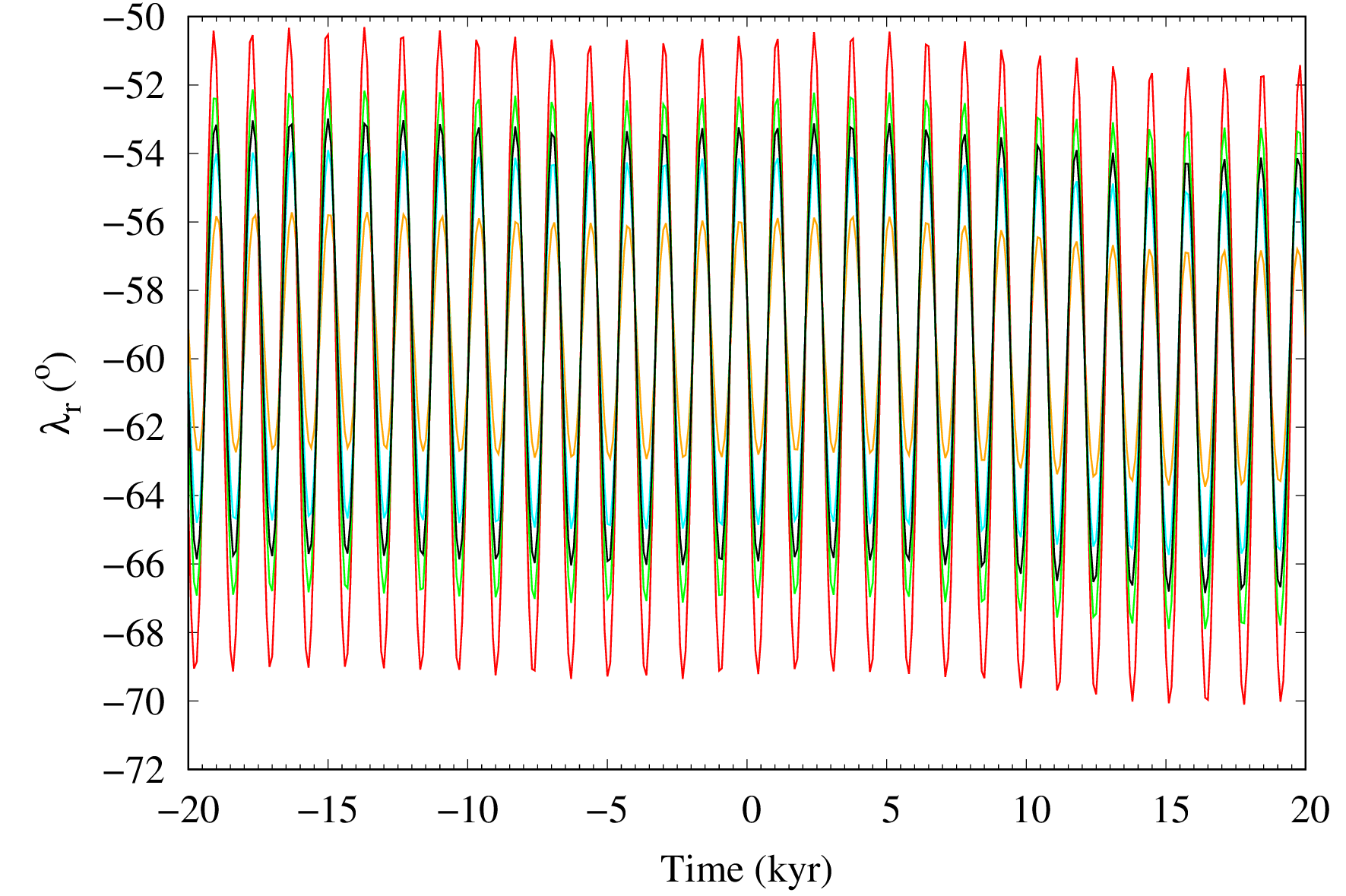}
        \includegraphics[width=\linewidth]{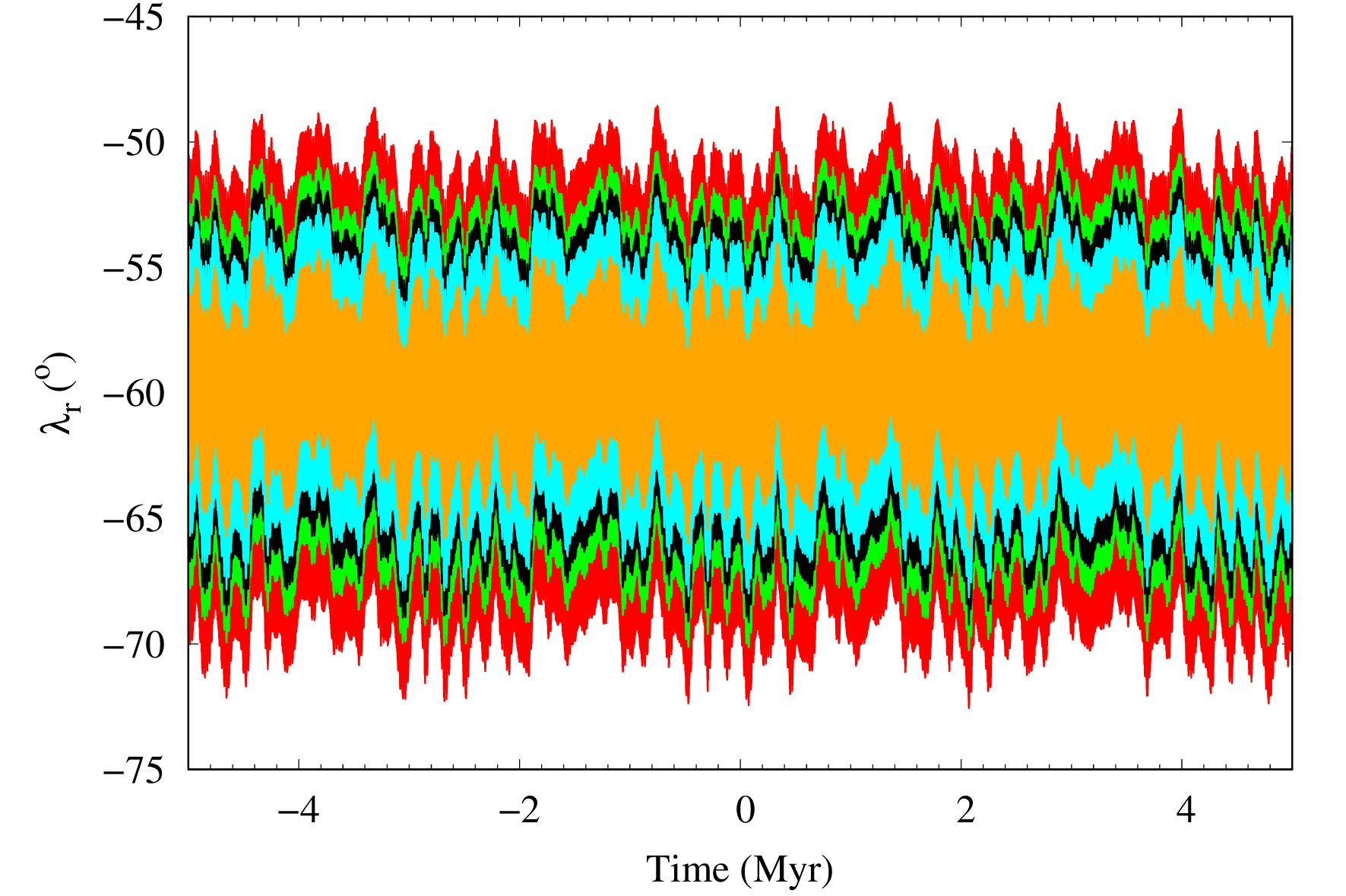}
        \caption{Same as Fig.~\ref{2009SE} but for 2018~FC$_{4}$. As in the case of 2018~EC$_{4}$, even orbits separated $\pm$9$\sigma$ from 
                 the nominal one produce similar evolution on Myr time-scales. Input values from Table~\ref{vector2018FC4}.
                 }
        \label{2018FC4}
     \end{figure}

        {\bf 2018~FC$_\mathbf{4}$}. This Mars Trojan and probable member of the Eureka family was also discovered by the Mt. Lemmon Survey 
        that imaged it on 2018 March 21 although 2018~FC$_{4}$ also appeared in images acquired by the Pan-STARRS~1 telescope system the 
        previous night. Its orbit determination is based on 35 observations spanning a data-arc of 790~d (see Table~\ref{elementsnewtro}). 
        Its absolute magnitude is $H$=21.3~mag (assumed $G=0.15$), which suggests a diameter in the range $\sim$100--800~m for an assumed 
        albedo in the range 0.60--0.01 (as pointed out above, smaller sizes are far more likely). Figure~\ref{Eurekafamily} shows that 
        2018~FC$_{4}$ is perhaps the object closest to Eureka in the $T_{\rm Mars}$-$\lambda_{\rm r}$ plane. If future spectroscopic 
        observations provide consistent results, 2018~FC$_{4}$ could be one of the youngest members of the Eureka family (otherwise, its
        low $T_{\rm Mars}$ relative to Eureka could be mere coincidence). Calculations similar to those performed in the cases of 2009~SE 
        and 2018~EC$_{4}$ show that the orbital evolution of 2018~FC$_{4}$ is as stable as that of 2018~EC$_{4}$ (see Fig.~\ref{2018FC4}) 
        with the value of its relative mean longitude oscillating around $-$60{\degr} with an amplitude close to 20{\degr} and a libration 
        period of nearly 1300~yr.  
        \hfil\par
%
%
%
%

        {\bf 2020~VT$_\mathbf{1}$}. This Mars co-orbital candidate was discovered on 2020 November 10 by the Pan-STARRS~1 telescope system 
        at Haleakala \citep{2020MPEC....V...75W}, although it had been previously observed but not identified as a new discovery by the Mt. 
        Lemmon Survey and Pan-STARRS~1 itself. Its orbit determination is based on 28 observations spanning a data-arc of 24~d and it is in 
        need of improvement (see Table~\ref{elementsnewtro}). Its absolute magnitude is $H$=22.9~mag (assumed $G=0.15$), which suggests a 
        diameter in the range $\sim$40--300~m for an assumed albedo in the range 0.60--0.01 (as pointed out above, smaller sizes are far 
        more likely). Calculations similar to those performed in the cases of 2009~SE, 2018~EC$_{4}$, and 2018~FC$_{4}$ indicate that the 
        evolution of the nominal orbit of 2020~VT$_{1}$ is far less stable than those of the Trojans. Figure~\ref{newtrojans}, bottom panel, 
        shows that it could be a relatively recent capture and that it may follow an extended horseshoe path that encompasses Mars at the 
        heel of one of the branches of the horseshoe. This object is the first of its kind to be found in Mars co-orbital space. A figure 
        for the evolution of 2020~VT$_{1}$ has not been included here because the dispersion for orbits different from the nominal one is 
        much larger than those displayed in Figs~\ref{2009SE}, \ref{2018EC4} and \ref{2018FC4}.
 
     \subsection{New Earth co-orbitals} 
        Our planet shares its orbit with a ring of captured asteroidal debris. The existence of the circumsolar Arjuna asteroid belt was 
        first proposed by \citet{1993Natur.363..704R} and further explored by e.g. \citet{2008MNRAS.386.2031B} and 
        \citet{2013MNRAS.434L...1D}, but its actual properties remain elusive. Most of its members are expected to be temporary co-orbitals 
        of Earth: some of them could be relatively stable, long-term companions, but most are believed to be ephemeral and/or recurrent, 
        short-term visitors that sometimes may become temporary satellites of our planet or minimoons. They tend to be small and dim, and 
        they can only be detected and studied during flybys with Earth, which results in the strong observational bias pointed out above. 
        This so-called Arjuna asteroid belt is a subset of the near-Earth asteroidal and cometary populations (near-Earth objects or NEOs) 
        and includes captured interplanetary dust \citep{2013Icar..226.1550K}. The existence of orbital clusters in NEO space resulting from 
        a combination of mean-motion resonances and catastrophic disruptions (rotation- or impact-induced) has been explored in the 
        literature (see e.g. \citealt{2005Icar..178..434F,2012Icar..220.1050S,2016MNRAS.456.2946D,2019Icar..333..165M,2020MNRAS.494..680J}) 
        using the available but still limited observational evidence. The dynamics of objects part of the Arjuna asteroid belt is mostly 
        controlled by Earth, but in some cases like that of 2020~CD$_{3}$ \citep{2020ApJ...900L..45B,2020MNRAS.494.1089D,
        2020AJ....160..277F}, the Moon also plays a non-negligible role. Our analyses in Sections~4 and 5.2, have uncovered two new 
        temporary Earth co-orbitals, not previously presented in the literature: 2020~PN$_{1}$ and 2020~PP$_{1}$. In addition, we have 
        identified an object that was, until very recently (late November--early December), a quasi-satellite of our planet, 2020~XC. 
        \hfil\par
\begin{figure}
       \centering
        \includegraphics[width=\linewidth]{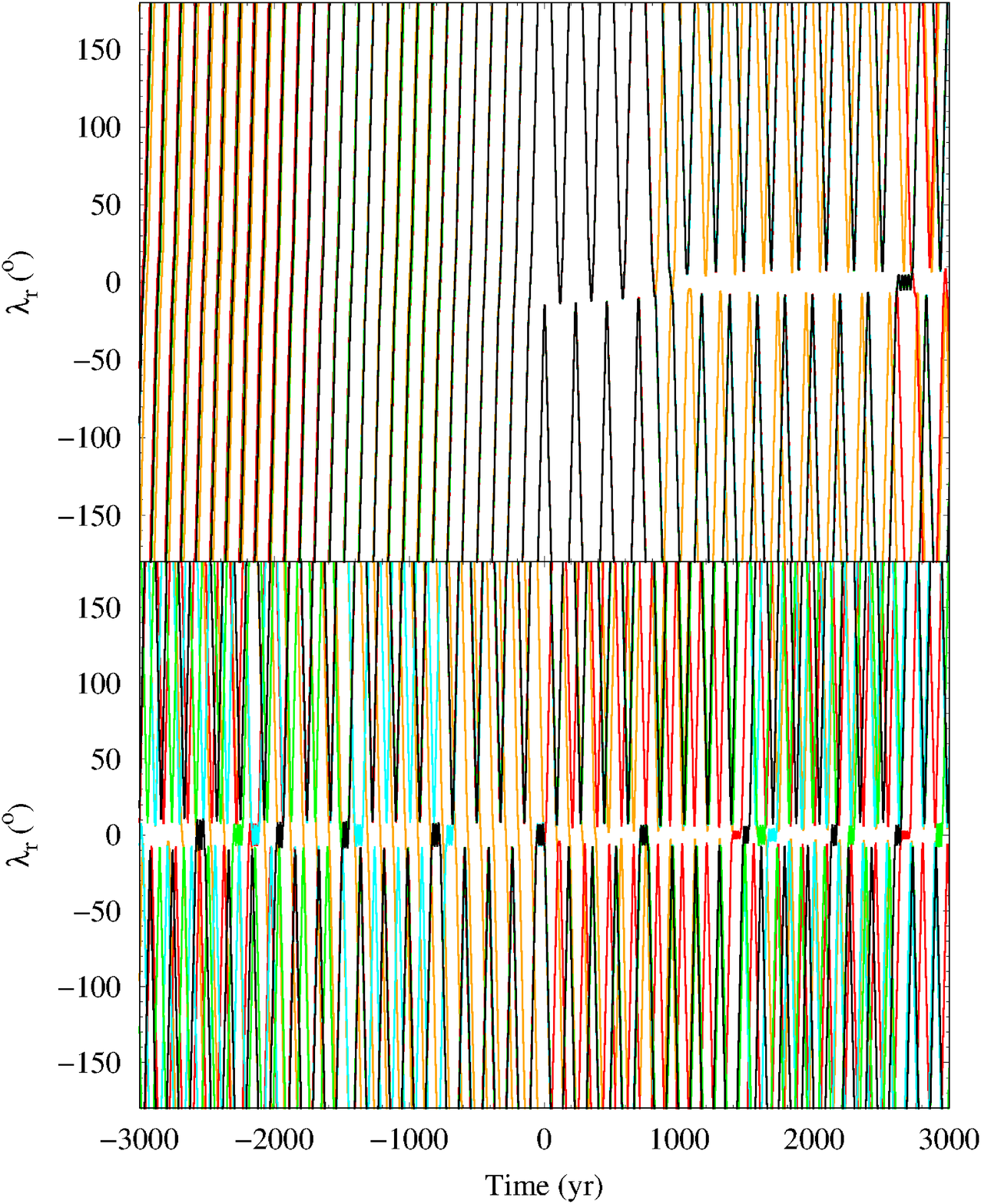}
        \caption{Evolution of the mean longitude difference of 2020~PN$_{1}$ (top panel) and 2020~PP$_{1}$ (bottom panel) and Earth. Our 
                 calculations include the nominal orbits (in black) as described by the orbit determinations in 
                 Table~\ref{elementsnewEarthco} and those of representative control orbits or clones with Cartesian vectors separated 
                 $+$3$\sigma$ (in green), $-$3$\sigma$ (in cyan), $+$9$\sigma$ (in red), and $-$9$\sigma$ (in orange) from the nominal 
                 values in Tables~\ref{vector2020PN1} and \ref{vector2020PP1}. The origin of time is epoch 2459000.5 TDB. 
                }
        \label{newEarthcoorbs}
     \end{figure}

        {\bf 2020~PN$_\mathbf{1}$}. This object was discovered on 2020 August 12 by the Asteroid Terrestrial-impact Last Alert System at 
        Haleakala \citep{2020MPEC....P...66M}.\footnote{\href{https://www.minorplanetcenter.net/mpec/K20/K20P66.html}
        {https://www.minorplanetcenter.net/mpec/K20/K20P66.html}} Its orbit determination (see Table~\ref{elementsnewEarthco}) is robust 
        and based on 41 observations spanning a data-arc of 361~d; it was rapidly improved because multiple precovery observations made by 
        the Pan-STARRS~1 telescope system at Haleakala were found. Its absolute magnitude is $H$=25.5~mag (assumed $G=0.15$), which suggests 
        a diameter in the range $\sim$10--100~m for an assumed albedo in the range 0.60--0.01 (as pointed out above, smaller sizes are far 
        more likely). Figure~\ref{newEarthcoorbs}, top panel, shows the evolution of the mean longitude difference of 2020~PN$_{1}$ for its 
        nominal orbit determination and representative control orbits with Cartesian vectors separated $\pm$3$\sigma$ and $\pm$9$\sigma$ 
        from the nominal values in Table~\ref{vector2020PN1}. For integrations into the past, all the calculations indicate that it was a 
        passing object that was only recently captured as a horseshoe librator to Earth. The evolution into the future predicts that it may 
        continue in this co-orbital configuration and perhaps become a temporary quasi-satellite of our planet. In any case, predictions 
        beyond about 500~yr may not be completely reliable. Horseshoes are thought to be the most numerous group among co-orbitals to Earth 
        (see e.g. \citealt{2010Icar..209..488W,2011MNRAS.414.2965C,2018MNRAS.473.2939D,2020MNRAS.496.4420K}). 
        \hfil\par
        {\bf 2020~PP$_\mathbf{1}$}. This object was discovered on 2020 August 12 by the Pan-STARRS~1 telescope system at Haleakala 
        \citep{2020MPEC....P...68M}.\footnote{\href{https://www.minorplanetcenter.net/mpec/K20/K20P68.html}
        {https://www.minorplanetcenter.net/mpec/K20/K20P68.html}} Its orbit determination (see Table~\ref{elementsnewEarthco}) is in need of
        improvement as it is based on 34 observations spanning a data-arc of 6~d. Its absolute magnitude is $H$=26.9~mag (assumed $G=0.15$), 
        which suggests a diameter in the range $\sim$5--50~m for an assumed albedo in the range 0.60--0.01 (as pointed out above, smaller 
        sizes are far more likely). Figure~\ref{newEarthcoorbs}, bottom panel, displays the evolution of the mean longitude difference of 
        2020~PP$_{1}$ for its nominal orbit and those with Cartesian vectors separated $\pm$3$\sigma$ and $\pm$9$\sigma$ from the nominal 
        values in Table~\ref{vector2020PP1}. For integrations a few hundred years into the past or the future within $\pm$3$\sigma$ from the 
        nominal orbit, predictions are consistent. It is currently leaving a quasi-satellite resonant state (but it is still engaged in it) 
        to become a horseshoe librator. Its overall evolution exhibits a sequence of transitions between the quasi-satellite and horseshoe 
        states. These recurrent transitions are similar to those found for Kamo`oalewa (see Fig.~\ref{ho3fu12}, bottom panel) and several 
        other known Earth co-orbitals \citep{2016Ap&SS.361...16D,2016MNRAS.462.3441D,2018MNRAS.473.3434D}. These episodes correspond to 
        domain III evolution as described by \citet{1999Icar..137..293N}, i.e. horseshoe-retrograde satellite orbit transitions and 
        librations. 
%
%
         \begin{table*}
          \fontsize{8}{11pt}\selectfont
          \tabcolsep 0.10truecm
          \caption{Heliocentric Keplerian orbital elements of 2020~PN$_{1}$, 2020~PP$_{1}$ and 2020~XC used in this study. The orbit 
                   determination of 2020~PN$_{1}$ is based on 41 observations spanning a data-arc of 361~d (solution date, 2020-Aug-28 
                   09:16:12 PDT), the one of 2020~PP$_{1}$ is based on 34 observations spanning a data-arc of 6~d (solution date, 
                   2020-Sep-11 07:48:43 PDT), and that of 2020~XC is based on 51 observations spanning a data-arc of 11~d (solution date,
                   2020-Dec-15 05:27:53 PST). Values include the 1$\sigma$ uncertainty. The orbit determinations have been computed at epoch 
                   JD 2459000.5 that corresponds to 00:00:00.000 TDB on 2020 May 31 (J2000.0 ecliptic and equinox). Source: JPL's SBDB.
                  }
          \begin{tabular}{lcccc}
           \hline
            Orbital parameter                                 &   & 2020~PN$_{1}$                 & 2020~PP$_{1}$           & 2020~XC               \\
           \hline
            Semimajor axis, $a$ (au)                          & = &   0.998105754$\pm$0.000000012 &   1.001715$\pm$0.000012 &   1.00170$\pm$0.00006 \\
            Eccentricity, $e$                                 & = &   0.1269557$\pm$0.0000009     &   0.07384$\pm$0.00007   &   0.10738$\pm$0.00002 \\
            Inclination, $i$ (\degr)                          & = &   4.80807$\pm$0.00003         &   5.827$\pm$0.007       &   0.760$\pm$0.002     \\
            Longitude of the ascending node, $\Omega$ (\degr) & = & 145.63610$\pm$0.00002         & 141.0248$\pm$0.0004     &  71.751$\pm$0.006     \\
            Argument of perihelion, $\omega$ (\degr)          & = &  55.40365$\pm$0.00002         &  44.14$\pm$0.03         & 269.31$\pm$0.03       \\
            Mean anomaly, $M$ (\degr)                         & = &  32.06964$\pm$0.00011         &  56.64$\pm$0.03         & 254.83$\pm$0.02       \\
            Perihelion, $q$ (au)                              & = &   0.8713906$\pm$0.0000009     &   0.92775$\pm$0.00006   &   0.89413$\pm$0.00007 \\
            Aphelion, $Q$ (au)                                & = &   1.124820928$\pm$0.000000013 &   1.075679$\pm$0.000013 &   1.10926$\pm$0.00006 \\
            Absolute magnitude, $H$ (mag)                     & = &  25.5$\pm$0.4                 &  26.9$\pm$0.4           &  29.0$\pm$0.4         \\
           \hline
          \end{tabular}
          \label{elementsnewEarthco}
         \end{table*}
%
%
%
%
%
%
        \hfil\par
\begin{figure}
       \centering
        \includegraphics[width=\linewidth]{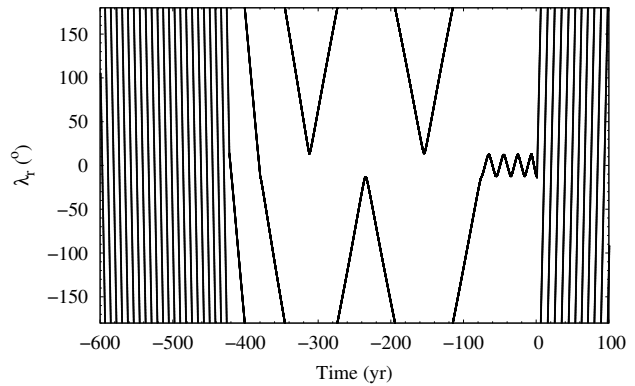}
        \caption{Evolution of the mean longitude difference of 2020~XC and Earth. The output time-step size is 0.01~yr, the origin of time 
                 is epoch 2459000.5 TDB, and only the nominal orbit (see Table~\ref{elementsnewEarthco}) has been displayed.
                }
        \label{2020xc}
     \end{figure}

        {\bf 2020~XC}. This object was discovered by the Mt. Lemmon Survey on 2020 December 4 
        \citep{2020MPEC....X...14R}.\footnote{\href{https://www.minorplanetcenter.net/mpec/K20/K20X14.html}
        {https://www.minorplanetcenter.net/mpec/K20/K20X14.html}} Its orbit determination (see Table~\ref{elementsnewEarthco}) is also in 
        need of improvement as it is based on 51 observations spanning a data-arc of 11~d. Its absolute magnitude is $H$=29.0~mag 
        (assumed $G=0.15$) and it is probably a secondary fragment of a larger object. It experienced a close encounter with the Earth on 
        2020 November 30 at 0.0008~au that altered its co-orbital status dramatically. Figure~\ref{2020xc} shows that prior to this 
        encounter, 2020~XC was a quasi-satellite of our planet, but after the flyby it became a passing body. Therefore, it is no longer
        co-orbital to Earth, but it has been included here because it was one of them at the standard epoch used in this study, 
        2459000.5~TDB. 
%
%
%
%

  \section{Conclusions}
     In this paper, we have applied a novel approach to estimate the importance of rotation-induced YORP break-up events in Earth co-orbital 
     space using data from Mars co-orbitals. Within the framework of the Tisserand's criterion \citep{Tisserand1896}, we assumed that two 
     co-orbital (to a certain planet) minor bodies resulting from a YORP break-up event should have approximately the same values of the 
     Tisserand parameter. The hypothesis is validated using data of Mars co-orbitals with eccentricity below 0.2 and machine-learning 
     techniques. The approach is subsequently applied to an equivalent group of Earth co-orbitals. The main conclusions of our study are:
     \begin{enumerate}[(i)]
        \item We identify three new L$_5$ Mars Trojans: 2009~SE, 2018~EC$_{4}$ and 2018~FC$_{4}$. Two of them, 2018~EC$_{4}$ and 
              2018~FC$_{4}$, are very probably linked to 5261~Eureka (1990~MB). We argue that 2009~SE may be a captured object, not related 
              to Eureka.  
        \item We confirm in the $T_{\rm Mars}$-$\lambda_{\rm r}$ plane the very strong asymmetry between the L$_4$ and L$_5$ Trojan clouds 
              of Mars.
        \item We identify one new Mars co-orbital candidate, 2020~VT$_{1}$, that may follow an extended horseshoe path that encompasses Mars
              at the heel of one of the branches. This object is the first of its kind to be found in Mars co-orbital space.
        \item We identify two new Earth co-orbitals: 2020~PN$_{1}$, that follows a horseshoe path, and 2020~PP$_{1}$, a quasi-satellite.
        \item We identify a former quasi-satellite to Earth, 2020~XC, that was very recently ejected from co-orbital space after a close 
              flyby.
        \item We identify clustering in the $T_{\rm Earth}$-$\lambda_{\rm r}$ plane that could be compatible with the outcome of recent YORP 
              break-up events. The cluster with most members is probably associated with 469219~Kamo`oalewa (2016~HO$_{3}$); 2020~PP$_{1}$
              follows an orbital evolution that closely resembles that of Kamo`oalewa.
        \item Clustering algorithms and numerical simulations both suggest that 2020~KZ$_{2}$ and Kamo`oalewa could be related. 
     \end{enumerate}
     The existence, for both Earth and Mars, of recently captured co-orbitals experiencing similar orbital evolution is confirmed with the
     identification of 2020~PN$_{1}$ (see Fig.~\ref{newEarthcoorbs}, top panel) and 2020~VT$_{1}$ (see Fig.~\ref{newtrojans}, bottom panel).
     Although most known Mars co-orbitals are long-term stable and all known Earth co-orbitals are not, the presence of 2020~VT$_{1}$ 
     suggests that, much like Earth, Mars may host a sizeable population of transient co-orbitals. 

     Although the thermal YORP mechanism may have spun up Eureka, leading to the formation of the Eureka family, Mars Trojans may also be
     the results of impacts \citep{2017NatAs...1E.179P}. A giant impact \citep{2008Natur.453.1216M,2018ApJ...856L..36H} may have led to the 
     formation of the Martian moons Phobos and Deimos as well \citep{2015Icar..252..334C,2017ApJ...845..125H,2017ApJ...851..122H,
     2018ApJ...860..150H,2018MNRAS.475.2452H,2018ApJ...853..118P}. Separate origins for the known Mars Trojans have previously been 
     suggested by \citet{2007Icar..192..434R} and \citet{2007Icar..192..442T}, but see also the discussion in \citet{2021Icar..35413994C}. 
     Our conclusions for the Mars Trojans are however independent of the actual origin of the objects (YORP-induced break-up, impact debris 
     from Mars or collisions between asteroids) as they focus on dynamical evolution and clustering in the $T_{\rm Mars}$-$\lambda_{\rm r}$ 
     plane. A similar statement can be made regarding our conclusions for the Earth co-orbitals that may come from rotational fission 
     events, Lunar impacts or catastrophic NEO collisions (among other sources). 

     The analysis of the structure in the $T_{\rm P}$-$\lambda_{\rm r}$ plane by itself cannot confirm genetic associations between 
     planetary co-orbitals, but it can be used to select candidates worthy of further study (numerical and spectroscopic). In this context,
     the study of the $T_{\rm P}$-$\lambda_{\rm r}$ plane provides evidence akin to that coming from orbital similarity criteria (see e.g.
     \citealt{1991Icar...89...14D,2000Icar..146..453D,1993Icar..106..603J,2016MNRAS.462.2275K}). It can be argued that the use of 
     $T_{\rm P}$ as an orbital similarity criterion has strong limitations that may be absent from other widely used orbital similarity 
     criteria. Figure~\ref{ecceffect} shows that the value of $T_{\rm Mars}$ for Eureka changes by about 0.02 in 10$^{5}$~yr; therefore, 
     small values of ${\Delta}T_{\rm Mars}$ cannot be used to argue for a genetic relationship. This argument is robust in the case of the 
     present-day Mars Trojans if they have remained in their present orbits for time-scales of the order of the age of the Solar system as 
     two initially almost identical $T_{\rm Mars}$ corresponding to two fragments may have had time to randomize parameters sufficiently 
     within the quasi-invariant boundaries (see Fig.~\ref{ecceffect}). This view gains further support if the objects under consideration 
     have very different values of $\lambda_{\rm r}$ and they experience relatively close approaches at high relative speeds. Under these 
     conditions, the existence of small values of ${\Delta}T_{\rm Mars}$ can be attributed to mere coincidence and most if not all pairings 
     could be spurious, placing some of the conclusions of our study in doubt. This line of reasoning however loses most of its credence 
     when considering dynamically young objects like known Earth co-orbitals that may also experience relatively close encounters with both 
     Earth and the Moon. In this case, small values of ${\Delta}T_{\rm Earth}$ can certainly be used to argue for a genetic relationship, 
     unless the size of the population is very large so sampling small values of ${\Delta}T_{\rm Earth}$ is far more probable.

\begin{figure}
       \centering
        \includegraphics[width=\linewidth]{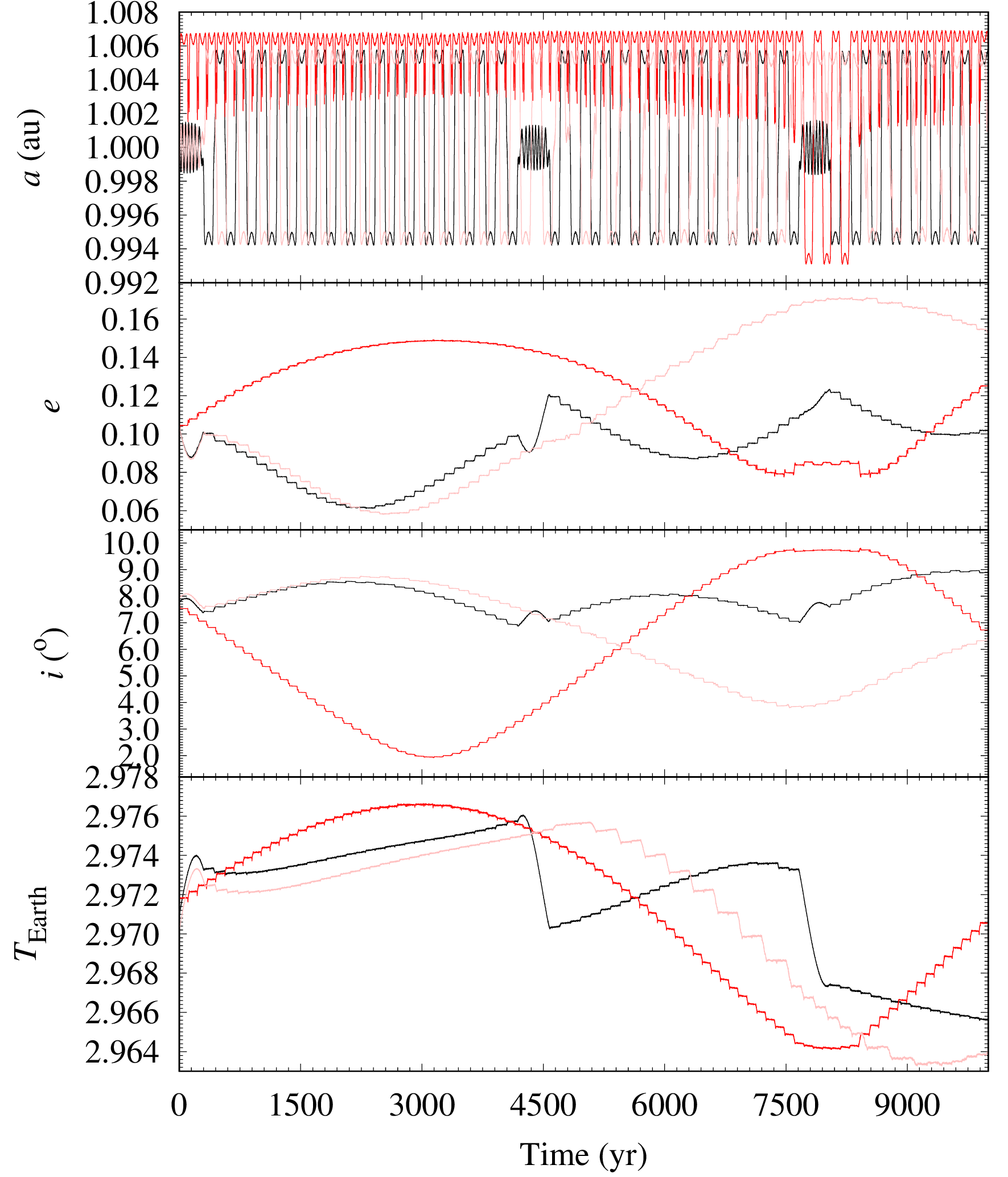}
        \caption{Evolution of the semimajor axis, $a$ (top panel), of the nominal orbit (in black) of 469219~Kamo`oalewa (2016~HO$_{3}$) as 
                 described by the orbit determination in Table~\ref{kamolike} and those of control orbits or clones with Cartesian vectors 
                 separated $+$9000$\sigma$ (in red) and $-$9000$\sigma$ (in pink) in terms of velocity from the nominal values in 
                 Table~\ref{vectorKamo}, i.e. the initial positions are the same, but the starting velocities are different. The second 
                 to top panel shows the evolution of the eccentricity, $e$, for the same sample of control orbits. The second to bottom 
                 panel displays the inclination, $i$. The bottom panel shows the variation of the Tisserand's parameter, $T_{\rm Earth}$.
                 The origin of time is epoch 2459000.5 TDB. 
                }
        \label{exp}
     \end{figure}

     Although we have to admit that the simplest hypothesis regarding the existence of small values of ${\Delta}T_{\rm Earth}$ is that they 
     could be all linked to random orbital alignments, it is possible to design a numerical experiment to confirm or reject the feasibility 
     of scenarios other than mere coincidence for these dynamically young objects. As a test of the implications that a genetic relationship 
     may have on ${\Delta}T_{\rm Earth}$ for Earth co-orbitals, we have performed a numerical experiment with the nominal orbit of 
     Kamo`oalewa and two control orbits with Cartesian vectors separated $+$9000$\sigma$ and $-$9000$\sigma$ in terms of velocity from the 
     nominal values in Table~\ref{vectorKamo} to study the evolution of $a$, $e$, $i$ and $T_{\rm Earth}$ (i.e. the initial positions are 
     the same, but the starting velocities are different). This experiment is equivalent to assuming that two fragments of Kamo`oalewa left 
     its surface with relative velocities close to 100~m~s$^{-1}$. Figure~\ref{exp} shows the results of this experiment: the maximum 
     relative variation in $T_{\rm Earth}$ is about 0.3 per cent while the equivalent results for $e$ and $i$ are 57 per cent and 75 per 
     cent, respectively. For genetically related fragments, the value of ${\Delta}T_{\rm Earth}$ may remain well under 0.014 even after 
     10$^{4}$~yr while the orbits could become rather different in terms of inclination and eccentricity after just 10$^{3}$~yr. In other
     words, Fig.~\ref{exp} shows that the value of ${\Delta}T_{\rm Earth}$ can remain as small as the values reported in Tables~\ref{kamo} 
     and \ref{uv136} for a significant amount of time although the shape and space orientation of the orbit of the fragment drifted 
     noticeably away from those of the parent body.
%
%
%
%

     It has sometimes been argued that planetary co-orbitals are the result of capture events or a side effect of planet formation itself 
     (i.e. the Jovian Trojans), our analysis within the context of the Trojans of Mars provides support to a new paradigm where production 
     of co-orbitals may have more than two active channels. This is particularly interesting in the region of the terrestrial planets where 
     both catastrophic and binary disruptions, and production of debris via planetary impacts, could be possible.   

  \section*{Acknowledgements}
     We thank two anonymous referees for their reports, one of them was particularly incisive, constructive and detailed, and included very 
     helpful suggestions regarding the presentation of this paper and the interpretation of our results. We thank S.~J. Aarseth for 
     providing one of the codes used in this research and A.~I. G\'omez de Castro for providing access to computing facilities. This work 
     was partially supported by the Spanish `Ministerio de Econom\'{\i}a y Competitividad' (MINECO) under grant ESP2017-87813-R. In 
     preparation of this paper, we made use of the NASA Astrophysics Data System and the MPC data server.

  \section*{Data Availability}
     The data underlying this paper were accessed from JPL's SBDB (\href{https://ssd.jpl.nasa.gov/sbdb.cgi}{https://ssd.jpl.nasa.gov/sbdb.cgi}). 
     The derived data generated in this research will be shared on reasonable request to the corresponding author.

  \appendix
  \section{Input data}
     Here, we include the barycentric Cartesian state vectors of the various objects mentioned in the main text of this paper. These vectors 
     and their uncertainties have been used to perform the calculations discussed above and generate those figures that display the time 
     evolution of orbital parameters and the histograms of the close encounters of pairs of objects. For example: a new value of the 
     $X$-component of the state vector is computed as $X_{\rm c} = X + \sigma_X \ r$, where $r$ is an univariate Gaussian random number, and
     $X$ and $\sigma_X$ are the mean value and its 1$\sigma$ uncertainty in the corresponding table.
%
%
     \begin{table}
      \centering
      \fontsize{8}{12pt}\selectfont
      \tabcolsep 0.15truecm
      \caption{Barycentric Cartesian state vector of 2009~SE: components and associated 1$\sigma$ uncertainties. Epoch as in 
               Table~\ref{elementsnewtro}. Source: JPL's SBDB.
              }
      \begin{tabular}{ccc}
       \hline
        Component                         &   &    value$\pm$1$\sigma$ uncertainty                                \\
       \hline
        $X$ (au)                          & = & $-$9.044249670510914$\times10^{-1}$$\pm$3.95127734$\times10^{-5}$ \\
        $Y$ (au)                          & = & $-$1.219519611628439$\times10^{+0}$$\pm$4.43547485$\times10^{-6}$ \\
        $Z$ (au)                          & = & $-$4.181102558282555$\times10^{-1}$$\pm$2.03602531$\times10^{-6}$ \\
        $V_X$ (au d$^{-1}$)               & = &    1.143843538936524$\times10^{-2}$$\pm$3.40882338$\times10^{-7}$ \\
        $V_Y$ (au d$^{-1}$)               & = & $-$6.448195605210014$\times10^{-3}$$\pm$2.27814124$\times10^{-7}$ \\
        $V_Z$ (au d$^{-1}$)               & = & $-$2.919772593230286$\times10^{-3}$$\pm$9.01993401$\times10^{-8}$ \\
       \hline
      \end{tabular}
      \label{vector2009SE}
     \end{table}
%
%
%
%
     \begin{table}
      \centering
      \fontsize{8}{12pt}\selectfont
      \tabcolsep 0.15truecm
      \caption{Barycentric Cartesian state vector of 2018~EC$_{4}$: components and associated 1$\sigma$ uncertainties. Epoch as in
               Table~\ref{elementsnewtro}. Source: JPL's SBDB.
              }
      \begin{tabular}{ccc}
       \hline
        Component                         &   &    value$\pm$1$\sigma$ uncertainty                                \\
       \hline
        $X$ (au)                          & = & $-$9.930212176028788$\times10^{-1}$$\pm$4.53081481$\times10^{-7}$ \\
        $Y$ (au)                          & = & $-$1.261773631639331$\times10^{+0}$$\pm$2.66927453$\times10^{-7}$ \\
        $Z$ (au)                          & = & $-$5.298080705059578$\times10^{-2}$$\pm$1.55233589$\times10^{-7}$ \\
        $V_X$ (au d$^{-1}$)               & = &    9.928799979516468$\times10^{-3}$$\pm$6.94939332$\times10^{-9}$ \\
        $V_Y$ (au d$^{-1}$)               & = & $-$7.155133944288653$\times10^{-3}$$\pm$4.37147830$\times10^{-9}$ \\
        $V_Z$ (au d$^{-1}$)               & = & $-$4.870041988110999$\times10^{-3}$$\pm$3.33948347$\times10^{-9}$ \\
       \hline
      \end{tabular}
      \label{vector2018EC4}
     \end{table}
%
%
%
%
     \begin{table}
      \centering
      \fontsize{8}{12pt}\selectfont
      \tabcolsep 0.15truecm
      \caption{Barycentric Cartesian state vector of 2018~FC$_{4}$: components and associated 1$\sigma$ uncertainties. Epoch as in
               Table~\ref{elementsnewtro}. Source: JPL's SBDB.
              }
      \begin{tabular}{ccc}
       \hline
        Component                         &   &    value$\pm$1$\sigma$ uncertainty                                \\
       \hline
        $X$ (au)                          & = & $-$6.647626395087657$\times10^{-1}$$\pm$5.55753437$\times10^{-6}$ \\
        $Y$ (au)                          & = & $-$1.251996334582715$\times10^{+0}$$\pm$4.96594761$\times10^{-6}$ \\
        $Z$ (au)                          & = &    4.726742793843942$\times10^{-1}$$\pm$7.04561654$\times10^{-6}$ \\
        $V_X$ (au d$^{-1}$)               & = &    1.268921413237325$\times10^{-2}$$\pm$9.38769625$\times10^{-8}$ \\
        $V_Y$ (au d$^{-1}$)               & = & $-$5.581094365762196$\times10^{-3}$$\pm$2.49673168$\times10^{-8}$ \\
        $V_Z$ (au d$^{-1}$)               & = &    2.929559280832830$\times10^{-3}$$\pm$1.95051109$\times10^{-8}$ \\
       \hline
      \end{tabular}
      \label{vector2018FC4}
     \end{table}
%
%
%
%
     \begin{table}
      \centering
      \fontsize{8}{12pt}\selectfont
      \tabcolsep 0.15truecm
      \caption{Barycentric Cartesian state vector of 469219~Kamo`oalewa (2016~HO$_{3}$): components and associated 1$\sigma$ uncertainties. 
               Epoch as in Table~\ref{kamolike}. Source: JPL's SBDB.
              }
      \begin{tabular}{ccc}
       \hline
        Component                         &   &    value$\pm$1$\sigma$ uncertainty                                \\
       \hline
        $X$ (au)                          & = & $-$5.449319912718597$\times10^{-1}$$\pm$4.16666980$\times10^{-7}$ \\
        $Y$ (au)                          & = & $-$9.113174005478253$\times10^{-1}$$\pm$6.84410519$\times10^{-8}$ \\
        $Z$ (au)                          & = &    1.681543176484481$\times10^{-2}$$\pm$6.13612239$\times10^{-8}$ \\
        $V_X$ (au d$^{-1}$)               & = &    1.436478550115864$\times10^{-2}$$\pm$4.18885560$\times10^{-9}$ \\
        $V_Y$ (au d$^{-1}$)               & = & $-$6.974713611012369$\times10^{-3}$$\pm$1.47221869$\times10^{-9}$ \\
        $V_Z$ (au d$^{-1}$)               & = & $-$2.182420313697609$\times10^{-3}$$\pm$4.40382271$\times10^{-9}$ \\
       \hline
      \end{tabular}
      \label{vectorKamo}
     \end{table}
%
%
%
%
     \begin{table}
      \centering
      \fontsize{8}{12pt}\selectfont
      \tabcolsep 0.15truecm
      \caption{Barycentric Cartesian state vector of 2016~FU$_{12}$: components and associated 1$\sigma$ uncertainties. Epoch as in 
               Table~\ref{kamolike}. Source: JPL's SBDB.
              }
      \begin{tabular}{ccc}
       \hline
        Component                         &   &    value$\pm$1$\sigma$ uncertainty                                \\
       \hline
        $X$ (au)                          & = & $-$7.349994733814580$\times10^{-1}$$\pm$8.90684866$\times10^{-3}$ \\
        $Y$ (au)                          & = & $-$9.037930038260460$\times10^{-1}$$\pm$5.26547287$\times10^{-3}$ \\
        $Z$ (au)                          & = &    5.409610013000898$\times10^{-3}$$\pm$4.90283157$\times10^{-4}$ \\
        $V_X$ (au d$^{-1}$)               & = &    1.100523434296072$\times10^{-2}$$\pm$1.18547355$\times10^{-4}$ \\
        $V_Y$ (au d$^{-1}$)               & = & $-$9.543786932245953$\times10^{-3}$$\pm$1.06494665$\times10^{-4}$ \\
        $V_Z$ (au d$^{-1}$)               & = &    5.372770530624185$\times10^{-4}$$\pm$5.47519985$\times10^{-6}$ \\
       \hline
      \end{tabular}
      \label{vector2016FU12}
     \end{table}
%
%
%
%
     \begin{table}
      \centering
      \fontsize{8}{12pt}\selectfont
      \tabcolsep 0.15truecm
      \caption{Barycentric Cartesian state vector of 2020~KZ$_{2}$: components and associated 1$\sigma$ uncertainties. Epoch as in 
               Table~\ref{kamolike}. Source: JPL's SBDB.
              }
      \begin{tabular}{ccc}
       \hline
        Component                         &   &    value$\pm$1$\sigma$ uncertainty                                \\
       \hline
        $X$ (au)                          & = & $-$3.773666435827534$\times10^{-1}$$\pm$3.18219205$\times10^{-5}$ \\
        $Y$ (au)                          & = & $-$9.577429216614861$\times10^{-1}$$\pm$2.04064633$\times10^{-5}$ \\
        $Z$ (au)                          & = & $-$8.960329684632281$\times10^{-3}$$\pm$1.34705032$\times10^{-5}$ \\
        $V_X$ (au d$^{-1}$)               & = &    1.541735728765344$\times10^{-2}$$\pm$7.11067379$\times10^{-7}$ \\
        $V_Y$ (au d$^{-1}$)               & = & $-$5.969582318369662$\times10^{-3}$$\pm$5.08277865$\times10^{-8}$ \\
        $V_Z$ (au d$^{-1}$)               & = & $-$2.093381081617893$\times10^{-3}$$\pm$3.11097775$\times10^{-6}$ \\
       \hline
      \end{tabular}
      \label{vector2020KZ2}
     \end{table}
%
%
%
%
     \begin{table}
      \centering
      \fontsize{8}{12pt}\selectfont
      \tabcolsep 0.15truecm
      \caption{Barycentric Cartesian state vector of 2020~PN$_{1}$: components and associated 1$\sigma$ uncertainties. Epoch as in
               Table~\ref{elementsnewEarthco}. Source: JPL's SBDB.
              }
      \begin{tabular}{ccc}
       \hline
        Component                         &   &    value$\pm$1$\sigma$ uncertainty                                 \\
       \hline
        $X$ (au)                          & = & $-$4.241165537792646$\times10^{-1}$$\pm$4.10976065$\times10^{-7}$  \\
        $Y$ (au)                          & = & $-$7.815068992626035$\times10^{-1}$$\pm$1.25238788$\times10^{-6}$  \\
        $Z$ (au)                          & = &    7.470552943322972$\times10^{-2}$$\pm$4.26388149$\times10^{-7}$  \\
        $V_X$ (au d$^{-1}$)               & = &    1.610713932302730$\times10^{-2}$$\pm$8.18961953$\times10^{-10}$ \\
        $V_Y$ (au d$^{-1}$)               & = & $-$1.021489969786571$\times10^{-2}$$\pm$3.28094580$\times10^{-8}$  \\
        $V_Z$ (au d$^{-1}$)               & = & $-$5.587679529869726$\times10^{-5}$$\pm$1.55793533$\times10^{-9}$  \\
       \hline
      \end{tabular}
      \label{vector2020PN1}
     \end{table}
%
%
%
%
     \begin{table}
      \centering
      \fontsize{8}{12pt}\selectfont
      \tabcolsep 0.15truecm
      \caption{Barycentric Cartesian state vector of 2020~PP$_{1}$: components and associated 1$\sigma$ uncertainties. Epoch as in
               Table~\ref{elementsnewEarthco}. Source: JPL's SBDB.
              }
      \begin{tabular}{ccc}
       \hline
        Component                         &   &    value$\pm$1$\sigma$ uncertainty                                \\
       \hline
        $X$ (au)                          & = & $-$3.443442683083884$\times10^{-1}$$\pm$1.82961900$\times10^{-5}$ \\
        $Y$ (au)                          & = & $-$8.916362488037327$\times10^{-1}$$\pm$5.85976394$\times10^{-5}$ \\
        $Z$ (au)                          & = &    9.314163806080564$\times10^{-2}$$\pm$1.00891106$\times10^{-4}$ \\
        $V_X$ (au d$^{-1}$)               & = &    1.620597866455574$\times10^{-2}$$\pm$3.80871529$\times10^{-7}$ \\
        $V_Y$ (au d$^{-1}$)               & = & $-$7.401387669318608$\times10^{-3}$$\pm$1.46198049$\times10^{-6}$ \\
        $V_Z$ (au d$^{-1}$)               & = & $-$4.536468695716349$\times10^{-4}$$\pm$4.45907592$\times10^{-7}$ \\
       \hline
      \end{tabular}
      \label{vector2020PP1}
     \end{table}
%
%

  \bsp
  \label{lastpage}
\end{document}